\documentclass[a4paper,11pt]{article}
\pdfoutput=1 
\usepackage{jcappub} 
\usepackage{slashed}
\usepackage{graphicx}
\usepackage{subfigure}
\usepackage[normalem]{ulem}
\usepackage{makecell}
\usepackage{xcolor}
\usepackage{upgreek}

\title{Towards a complete scheme of cosmological neutrino self-interactions: Collision term for a wide range of mediator masses}  

\author[a]{Ivan P\'erez-Castro}
\author[a,b]{Josue De-Santiago}
\author[c]{Gabriela Garcia-Arroyo}
\author[a,d]{Jorge Venzor}
\author[a]{Abdel P\'erez-Lorenzana}

\affiliation[a]{Departamento de F\'{\i}sica, Centro de Investigaci\'on y de Estudios Avanzados del Instituto Polit\'ecnico Nacional \\
    Apartado Postal 14-740, 07000, Ciudad de M\'exico, M\'exico.}
\affiliation[b]{Secretar\'ia de Ciencia, Humanidades, Tecnología e Innovación,  Av.   Insurgentes  Sur  1582,  Colonia  Crédito Constructor, Del.  Benito Juárez, 03940, Ciudad de México, México.}
\affiliation[c]{Instituto de Ciencias F\'isicas, Universidad Nacional Aut\' onoma de M\'exico, 62210, Cuernavaca, Morelos, M\'exico.}
\affiliation[d]{Tecnol\'ogico de Monterrey, Escuela de Ingenier\'ia y Ciencias,
Av. Heroico Colegio Militar 4700, 31300, Chihuahua, Chihuahua, M\'exico.}

\emailAdd{ivan.perez.c@cinvestav.mx}
\emailAdd{josue.desantiago@cinvestav.mx}
\emailAdd{arroyo@icf.unam.mx}
\emailAdd{jorge.venzor@cinvestav.mx}
\emailAdd{abdel.perez@cinvestav.mx}

\date{\today}

\abstract{
Neutrino self-interactions (NSI) offer a potential pathway to address anomalies in standard cosmology and explain existing cosmological tensions.
In this work, we present a novel framework to obtain the neutrino--neutrino collision term within the Boltzmann hierarchy, incorporating both neutrino and mediator masses as free parameters. Our calculations encompass both Dirac-like and Majorana neutrinos and distinguish between two neutrino mass eigenstates. This work provides a valuable tool for future analyses, should a NSI signal be detected. 
Remarkably, our results show a smooth transition from the light to the heavy mediator approximation as the Universe cools down for non-resonant cases.
Thus, for the widely studied heavy mediator, our new scheme eliminates the need to approximate at high redshifts when 
the temperature increases above the mediator mass,
and it provides the tools to test the threshold of validity of the heavy mediator paradigm.
While this work focuses on NSI mediated by a scalar particle, the presented framework could be adapted to a broader range of neutrino NSI and possibly to warm dark matter self-interacting scenarios.
}

\arxivnumber{}

\makeatletter
\gdef\@fpheader{}
\makeatother

\begin{document}
\notoc
\maketitle

\tableofcontents

\section{\label{sec:introduction} Introduction}

The standard model of fundamental particles, despite its undeniable success, fails to explain various phenomena related to neutrino physics.
Meanwhile, nonstandard neutrino interactions offer the possible mechanisms required to explain the origin of neutrino masses, accommodate anomalous results in neutrino oscillation experiments, and may be the portal to hypothetical dark matter detection \cite{Gelmini1981left, Chikashige1981there, Miranda_2015, Chao_2019_dm, Bhupal2019status, Babu2020_mass_models, VanLoi2020, ELLIOTT2024}. 

In cosmology, nonstandard neutrino interactions have been extensively tested, especially for low energies compared to typical energies in terrestrial neutrino experiments. 
In particular, neutrino self-interactions (NSI) mediated by a nonstandard particle with mass well below $\sim \mathcal{O}$(GeV)~\cite{Blinov2019, Berryman2023neutrino}.
In this context, some intriguing results have emerged around three different mediator mass regimes: the heavy mediator with mass $\gtrsim \mathcal{O}$(keV) \cite{Bell2006, CyrRacine2014, Archidiacono2014, Oldengott2015, Lancaster2017, Oldengott_2017, Park2019, Kreisch2020, Choudhury2021, brinckmann2020self, mazumdar2022flavour, Kreisch:2022zxp, RoyChoudhury2022, Das:2023npl, Camarena2023, he2024self, camarena2025strong, Poudou:2025qcx, He:2025jwp, Montefalcone:2025ibh, Whitford:2025dmq}, the very light mediator $\lesssim \mathcal{O}$(meV) \cite{Forastieri:2015paa, Forastieri:2019cuf, Venzor2022}, and the resonant case for masses in the range from $\mathcal{O}$(meV) up to $\mathcal{O}$(keV)
\cite{venzor2023resonant, Noriega2025resonant}.
Although, it is important to emphasize that those mass regimes are usually introduced to simplify the analysis by means of using some appropriate approaches for each case. In that sense, the thresholds are an approximation on itself, suggested by the behavior of the interaction rate and the $m/T$ value. A more general analysis would help to understand the differences between those regimes. 

From the data analysis point of view, the main difference between these approximations relies on the cosmological scales being affected,  with the mediator mass having an inverse correlation to the scale.
The very light mediator and resonant regimes affect the largest scales, modifying the CMB.
On the other hand, for the heavy mediator, small-scale data, especially the full shape of the matter power spectrum as seen in the galaxy and Lyman-$\alpha$ forest distributions \cite{Camarena2023, he2024self, camarena2025strong, Poudou:2025qcx, He:2025jwp} (in some scenarios) prefers a large non-null interaction coupling that should be observed in particle detectors unless one proposes or recalls a mechanism to avoid experimental bounds.
Thus, there is an abnormal signal that future data may be able to explore \cite{Beacom2004, Hannestad2005, Heurtier2017, Huang2018, Berryman2018, BLUM2018, Brune2019, Barenboim2019inflation, Schoneberg2019, Escudero2020_CMB_search, Chacko2020, Bustamante2020bounds,  Brdar2020, Deppisch2020, Escudero2020relaxing, 
Shalgar2021, Lyu2021, Creque2021, Venzor2021, Suliga2021,
Barenboim2021invisible,
Das2021, Seto2021, Esteban2021, 
Escrihuela:2021mud, Cerdeno2021, Esteban2021_probing, Ge:2021lur, Medhi:2021wxj, Abellan2022, Chen2022weaker, Kumar_2022, Akita2022, chang2022towards, Fiorillo2022, Dutta:2023fdt, 
cerdeno2023constraints, Doring_2024,
Sandner2023, Fiorillo2024_theoretical,
Fiorillo2024_small_impact,
Bostan_2024,
akita2024limits,
Medhi:2023ebi, dev2025new,
Denton:2024upc, zhang2025neutrino, De2025bounds, foroughi2025enabling, Wang:2025qap,Foroughi-Abari:2025mhj,abellan2026neutrino}.

Theoretical advances are key to studying models against new sets of data.
Roughly a decade ago, I. Oldengott et al.~\cite{Oldengott2015, Oldengott_2017} computed for the first time the Boltzmann hierarchy of self-interacting neutrinos.
They found that effective treatments like viscosity in the energy-momentum tensor did not produce the correct phenomenology
of neutrino self-interactions.
Their work, along with that of Ref. \cite{Kreisch2020}, paved the way for modern treatment of the heavy mediator approximation in recent cosmological studies~\cite{Park2019, Choudhury2021, brinckmann2020self, Kreisch:2022zxp, RoyChoudhury2022, Das:2023npl, Camarena2023, he2024self, camarena2025strong, Poudou:2025qcx, He:2025jwp, Montefalcone:2025ibh, Whitford:2025dmq, Lyu2021, Das2021, Bostan_2024}. 
These studies have ignored the neutrino mass, which is a fundamental parameter in the precision era of cosmology.

In this manuscript, we build on the foundations of Refs. \cite{Oldengott_2017, Kreisch2020} and present a novel theoretical framework for neutrino self-interactions within the thermal approximation.
We construct our scheme systematically from the most general case and seek an answer to the following important questions: Is it possible to compute the collision term of NSI for a range of neutrino temperatures using a free mediator mass? What are the differences in using Dirac or Majorana neutrinos?
What are the effects of neutrino masses?
Is it possible to compute the thermal correction for the light mediator case without relying on further approximations?

The rest of the paper is organized as follows: in Sec.~\ref{sec:Boltzmann and Collision} we revisit the theoretical foundations underlying the Boltzmann hierarchy.
In Sec.~\ref{Sec:nu-nu via scalar mediator}, we focus on the collision term for neutrino--neutrino scattering.
In Sec.~\ref{Sec:Heavy_mediator}, we compare our approach with the well-known massive mediator limit; beyond that, our method provides a more general description of the collision term, valid for any heavy scalar mediator.
In Sec.~\ref{sec:light_mediator}, we study the light mediator based on our present framework as compared to the usual relaxation time approximation.
Finally, in Sec.~\ref{sec:Conclusions}, we give our final conclusions and perspectives. For readers interested in the details of our derivations, appendices are provided following the main text.

\section{Boltzmann hierarchy and the collision term}\label{sec:Boltzmann and Collision}
In this section, we review the fundamental concepts needed to formulate the Boltzmann equation with a collision term for $2 \to 2$ processes. Since the collision term is determined by the local momenta of the particles involved and by its expression in terms of conformal time, we present a coherent and self-contained discussion aimed at resolving potential ambiguities in its formulation, with particular attention to the distribution function, the zero-order collision term (background), the particle degrees of freedom, and the conventions for scattering amplitudes.

To describe these effects in the cosmological context, it is necessary to account for small departures from homogeneity and isotropy. When these deviations remain small, they can be consistently described as linear perturbations around a Friedmann–Lemaître–Robertson –Walker (FLRW) background spacetime.
For scalar perturbations, common choices of
coordinates \cite{Ma1995} are the so called
conformal Newtonian (CN) gauge  and the synchronous gauge, which lead to the corresponding metrics:
    \begin{equation}
        ds^2 = \begin{cases}
        a^2(\tau) \left[ -(1 + 2\Psi) d\tau^2 + (1 - 2\Phi) d {\bf x}^2 \right] & \text{for the CN gauge,} \\
        a^2(\tau)\left[-d\tau^2 + (\eta_{ij}+h_{ij}) dx^idx^j\right] & \text{for the synchronous gauge.}
        \end{cases}  \label{metric-gauge}
    \end{equation}
Here $\tau$ denotes conformal time and $x^i$ are comoving spatial coordinates that remain fixed with respect to the expanding background. The overall conformal factor $a(\tau)$ accounts for the expansion of the Universe. The scalar perturbations are represented in the Newtonian gauge by the potentials $\Psi$ and $\Phi$~\footnote{These quantities have clear physical interpretations: $\Phi$ represents the gravitational potential linked to time dilation effects, while $\Psi$ describes the potential related to distortions in spatial geometry~\cite{Bardeen1980gauge, Durrer20042}.}, while in the synchronous gauge, the perturbation
$h_{ij}$ can be written in terms of the scalar perturbations
$h$ and $\eta$ as shown in Eq.~\eqref{synch_potentials}.

To analyze the 
impact of the particle interactions on the cosmological perturbations, we study the evolution of the distribution function, $f({\bf x}, {\bf P}, \tau)$,
\footnote{In this work, variables in bold represent
3-vectors.} whose phase-space variables are the three positions $x^i$, and their canonical conjugate momenta ${\rm P}^i$, while the function evolves with respect to conformal time $\tau$. By definition, $f$ is a scalar and canonically invariant function that describes the number of particles that occupy a differential volume in phase-space, given by $dN = {\rm g}_s (2\pi)^{-3} f \, d^3x d^3{\rm P}$ where ${\rm g}_s$ is the degeneracy factor per particle.
For any given matter species, the evolution of $f$ is governed by the relativistic Boltzmann equation~\footnote{In some references \cite{Oldengott2015,Barenboim2021invisible}, the relativistic Boltzmann equation is expressed in the following form ${\rm P}^\mu \frac{\partial \tilde{f}}{\partial x^\mu} - \Gamma_{\mu \nu}^\rho {\rm P}^\mu {\rm P}^\nu \frac{\partial \tilde{f}}{\partial {\rm P}^\rho} = Q [\tilde{f}]$.
To satisfy the on-shell condition ${\rm P}_\mu {\rm P}^\mu = - m^2$, one must impose the constraint $f(x^\mu, {\rm P}^j)=\tilde{f}(x^\mu, {\rm P}^j, {\rm P}^0(x^\mu, {\rm P}^j))$, thereby recovering the same physical description.},
    \begin{equation}
        {\rm P}^\mu \frac{\partial f}{\partial x^\mu} - \Gamma_{\mu \nu}^k {\rm P}^\mu {\rm P}^\nu \frac{\partial f}{\partial {\rm P}^k} = Q[f] = \frac{d f}{d \lambda} \,
        .
        \label{RBE-1}
    \end{equation}
Here, $Q[f]$ is the Boltzmann collision term and accounts for scattering or decay processes involving the distribution function $f$, while $\lambda$ is an affine parameter that describes the trajectory of a particle.

To express the relativistic Boltzmann equation on the local energy and momenta $(E,p^i)$ measured by an observer fixed at the coordinates $x^i$, we use the relations
    \begin{eqnarray}
        E = a(1+\Psi) {\rm P}^0\,,\,  
        &p^i = a(1+\Phi) {\rm P}^i \,,&
        \,  
        \text{for the conformal Newtonian gauge, and} \label{conformal Newtonian gauge}
        \\
        E = a {\rm P}^0 \,,
        \qquad
        &p^i =a \left(\delta_j^i+ \frac{1}{2}h^i_j \right) {\rm P}^j \,,&
        \qquad
        \text{for the synchronous gauge.} \label{synchronous gauge}
    \end{eqnarray}
These variables satisfy the dispersion relation
$E^2=p^2+m^2$.
Moreover, separating  the
momenta between its magnitude and direction
${\bf p} = p\hat{\bf p}$, we can write the 
Boltzmann equation for the distribution function
$f(\tau,x^i,p,\hat{p}^i)$ as:
    \begin{equation}
        \frac{\partial f}{\partial \tau} + \frac{d x^j}{d \tau} \frac{\partial f}{\partial x^j} + \frac{d p}{d \tau} \frac{\partial f}{\partial p} + \frac{d \hat{p}^j}{d \tau} \frac{\partial f}{\partial \hat{p}^j} =\frac{d f}{d \tau} = \frac{Q[f]}{{\rm P}^0} \, ,
        \label{Boltzmann-C}
    \end{equation}
First, we consider the FLRW background, where $f=f^{(0)}(\tau,p)$.
Homogeneity of the Universe implies that $\partial f^{(0)}/\partial x^j = 0$ and isotropy that $\partial f^{(0)}/\partial \hat p^j = 0$.    
Using $d p / d\tau = - {\cal H}p$, with ${\cal H} = \dot{a}(\tau)/a(\tau)$ the conformal Hubble expansion rate (here dots represent derivatives with respect to conformal time) leads to
    \begin{equation}
         \frac{\partial f^{(0)}}{\partial \tau} - {\cal H}p \frac{\partial f^{(0)}}{\partial p} = 
         \frac{Q^{(0)}}{{\rm P}^0}\, . 
        \label{Q-zero}
    \end{equation}
The background collision term $Q^{(0)}$ describes the possible number and energy transfers between different components of the
Universe. In this work, we are interested in neutrinos
after nucleosynthesis, where transfer rates with other components of the Universe (like photons and electrons) are negligible due to the smallness of the weak interaction, whereas transfer rates among different neutrino species are functions of their temperature differences~\cite{Abenza2020precision}. However, even though $T_{\nu_e} \neq T_{\nu_\mu}=T_{\nu_\tau}$, the actual difference is small, and  therefore, we can safely assume 
that the neutrino mass eigenstates share the same temperature, and the background collision term cancels out, giving 
$Q^{(0)}=0$.

The equilibrium solution of Eq.~\eqref{Q-zero}, is then
    \begin{equation}
        f^{(0)}_{\alpha}(p) = \frac{1}{e^{p/T_\alpha}-\sigma} = f_\alpha^{(0)}(q) \, , \label{f_background}
    \end{equation}
where the latest equality follows from $p/T_\alpha = q/(a(\tau) T_\alpha)$ with $q=a(\tau) p$ denoting the usual magnitude of the comoving three-momentum, and where $\alpha$ labels the different particle species in the Universe (for neutrinos $\alpha=\nu$), $T_\alpha$ stands for their temperature, and $\sigma=\pm 1$ to distinguish fermions ($-1$) from bosons ($+1$).
Eq.\eqref{f_background} assumes a negligible initial chemical potential, which for Dirac neutrinos occurs if there is no lepton asymmetry.

In the following, we will explicitly write the general collision term at first order in the synchronous and CN gauges, using the previous conventions.
\subsection{Boltzmann hierarchy equations under cosmological perturbations}
By considering our discussion above, at first order in perturbations, where $Q_\alpha= Q_\alpha^{(1)}$ for a null background, the Boltzmann equation takes the form (see Appendix~\ref{appA} for an extended discussion),
    \begin{eqnarray}
        \frac{\partial f_\alpha}{\partial \tau} + \frac{{\bf p}}{E} \cdot \nabla f_\alpha + p \frac{\partial f_\alpha}{\partial p} \left[ - {\cal H} + \frac{\partial \Phi}{\partial \tau} - \frac{E}{p^2} {\bf p} \cdot \nabla \Psi \right] = \frac{Q_\alpha}{{\rm P}^0}  , 
        & \text{for the CN gauge,}
        \label{Q-CN}
        \\
        \frac{\partial f_\alpha}{\partial \tau} + \frac{{\bf p}}{E} \cdot \nabla f_\alpha - p \frac{\partial f_\alpha}{\partial p} \left[ {\cal H} + \frac{1}{2} \hat{p}^i \hat{p}^j \frac{\partial h_{ij}}{\partial \tau} \right] = \frac{ Q_\alpha}{{\rm P}^0} ,
        & \text{for the synchronous gauge.}
        \label{Q-S}
    \end{eqnarray}
It is clear that $Q_\alpha$ affects the evolution of $f_\alpha$, for the $\alpha$ species. Notice that, here, we take ${\rm P}^0=E/a$ at order zero. 
We remark that in certain scenarios, such as decays where $Q_\alpha^{(0)} \neq 0$, the first-order expansion on ${\rm P}^0$ from Eq.~\eqref{conformal Newtonian gauge} must be included in the CN gauge~\cite{Audren2014strongest, Barenboim2021invisible}. In the following, we assume that a possible contribution to zero order is sufficiently small so as not to spoil the background thermal form given by Eq.~\eqref{f_background}.

For a generic scattering process of the form $\alpha(P_1) + \beta(P_2) \rightarrow \alpha(P_3) + \beta(P_4)$, we define the collision term with respect to the conformal time as $C_\alpha = Q_\alpha/E$.
It depends on both the local interactions and the distribution functions of the particles involved. Assuming CP invariance, it has the form
    \begin{equation}
        C_\alpha[{\bf x}, {\bf p}_1, \tau] := \frac{1}{2E_1}\int d \Pi_{2} d\Pi_{3}d\Pi_{4} (2\pi)^4 \delta^{(4)}{(P_1+P_2-P_3-P_4)} \overline{|{\cal M}|}^2 F({\bf x}, {\bf p}_1, {\bf p}_2, {\bf p}_3, {\bf p}_4,\tau) \, ,
        \label{C}
    \end{equation}
where $\overline{|{\cal M}|}^2$ is the invariant squared amplitude averaged over initial states, encoding the quantum interaction~\footnote{For Majorana neutrinos, lepton number is not conserved, so the neutrino chemical potential must vanish in chemical equilibrium, and both chiralities are equally populated. In contrast, for Dirac neutrinos, the lepton asymmetry typically refers to the difference in number densities between neutrinos and antineutrinos.
This possible lepton asymmetry is small at large temperatures, see, for example, ~\cite{kang1992cosmological, Esposito2000standard, Hansen2001constraining, Serpico2005lepton}, we will assume it to be zero.}.
Under these conditions, we average the squared amplitude over the initial neutrino spin states in Eq.~\eqref{C} as in quantum field theory. This assumes that the initial neutrino ensemble is
unpolarized, i.e., both spin states are equally populated. Here, we have used the convention
    \begin{equation}
        d\Pi_j = \frac{ \,d^3p_j}{2E_j(2\pi)^3} \, ,
    \end{equation}
It is worth noticing that in other references~\cite{Oldengott2015, Oldengott_2017}, the phase-space element is defined as $d\Pi_j = {\rm g}_{sj} d^3p_j/[2E_j(2\pi)^3]$, where the  ${\rm g}_{si}$ factors denote the internal degrees of freedom of the particle. In that case, the squared amplitude must also be averaged over the final spin states, in contrast to our convention where one includes a ${\rm g}_{s2}={\rm g}_s$ factor later in the collision term.

In Eq.~\eqref{C}, $F$ stands for the distribution factor for the scattering process
    \begin{equation}    
        F = f_\alpha ({\bf p}_3)f_\beta ({\bf p}_4)[1+ \sigma_1 f_\alpha({\bf p}_1)][1 + \sigma_2 f_\beta ({\bf p}_2)] -
        f_\alpha({\bf p}_1)f_\beta({\bf p}_2)[1 + \sigma_3 f_\alpha({\bf p}_3)][1+\sigma_4 f_\beta ({\bf p}_4)]~,
        \label{F-complete}
    \end{equation}
where $\sigma_i=+1$ ($-1$) for bosons (fermions), and the ${\bf x}$ and $\tau$ dependence should be understood in the distribution factor of the $\alpha$-particle.

This phase-space distribution allows for both inhomogeneities in the neutrinos (through its dependence on ${\bf x}$) and anisotropies (through its dependence on the direction $\hat{p}$).
It is useful to split the phase-space distribution into a background and a perturbed component at linear order as
    \begin{equation}
        f_\alpha({\bf x}, {\bf p}, \tau) = f^{(0)}_\alpha(p) \left[1 + \Theta_\alpha ({\bf x}, {\bf p}, \tau) \right] \, .
        \label{f_def}
    \end{equation}
With this perturbation, we rewrite Eqs.~\eqref{Q-CN} and \eqref{Q-S}, and separate their zero and first-order contributions.
    
In the following, we work in the Fourier transformed
space for $\bf x$. The transformed perturbed distribution function 
$\tilde \Theta_\alpha$ depends on $({\bf k}, {\bf p},\tau)$.
We can expand $\tilde{\Theta}_\alpha$ in a Legendre series in terms of $\upmu = \hat{\bf p} \cdot \hat{\bf k}$ as follows:
    \begin{equation}
        \tilde{\Theta}_\alpha({\bf k}, {\bf p},\tau) = \sum_{\ell=0}^\infty (-i)^\ell(2\ell+1)\vartheta_{\alpha,\ell}
        (k,p,\tau){\cal P}_\ell(\upmu) \,,
        \label{Thetaexp}
    \end{equation}
where the Legendre multipole functions $\vartheta_{\alpha,\ell}$ depend only on the scalar magnitudes $k$ and $p$, whereas $\ell$ is the order of the multipole.

At first order, the collision term in the Fourier space is
    \begin{equation}
        \tilde{C}_\alpha^{(1)}[{\bf k}, {\bf p}_1, \tau] := \frac{1}{2E_1}\int d \Pi_{2} d\Pi_{3}d\Pi_{4} (2\pi)^4 \delta^{(4)}{(P_1+P_2-P_3-P_4)} \overline{|{\cal M}|}^2 \tilde{F}^{(1)}({\bf k},\tau) \, ,
        \label{C-first}
    \end{equation}
where $\tilde{F}^{(1)}({\bf k},\tau) = \tilde{F}^{(1)}({\bf k}, {\bf p}_1, {\bf p}_2, {\bf p}_3, {\bf p}_4,\tau)$ is given by Eq.\eqref{FS2}, which includes the first-order terms that depend linearly on $\Theta_\alpha ({\bf k}, {\bf p}_n, \tau)$, with the labels $n=1,2,3,4$ referring to the four-momenta of the particles involved.

In order to write the Boltzmann equation in terms of the new functions $\vartheta_{\alpha,\ell}$, we substitute \eqref{Thetaexp} into \eqref{f_def} and those in the Fourier transformed version of equations \eqref{Q-CN} and \eqref{Q-S}.
We use the comoving momentum ${\bf q} = a{\bf p}$, where ${\bf p} = {\bf p}_1$ is the local $3$-momentum of the reference particle, and the comoving energy is determined by $\epsilon(q, \tau)=\sqrt{q^2+a(\tau)^2m^2}$. Then we obtain the Boltzmann hierarchy equation at first order in Fourier space:
    \begin{equation}
        f_{\alpha}^{(0)} \left[ \frac{\partial \vartheta_{\alpha, \ell}}{\partial \tau} + \frac{q}{\epsilon} k \left( \frac{(\ell +1)}{(2 \ell +1)} \vartheta_{\alpha, \ell +1} - \frac{\ell}{(2\ell +1)} \vartheta_{\alpha, \ell -1} \right) \right]
        + q \frac{\partial f_{\alpha}^{(0)}}{\partial q} 
        f_{g,\ell} = a \tilde{C}_{\ell} \, ,
        \label{CN-S-ell}
    \end{equation}
where $f_{g,\ell}$ depends on gauge choice as
    \begin{eqnarray}
        f_{g,\ell} =
             \frac{\partial \phi}{\partial \tau} \delta_{0\ell} + \frac{k}{3} \frac{\epsilon}{q} \psi \delta_{1\ell} & \text{for the CN gauge,} \label{fg}\\
        f_{g,\ell} =
             \frac{\dot{h}+ 6\dot{\eta}}{15}\delta_{2\ell} - \frac{\dot{h}}{6}\delta_{0\ell} & \qquad \text{for the synchronous gauge}
         \, ,
        \label{f_g}
    \end{eqnarray}
where dot denotes a partial derivative with respect to $\tau$.
For further details on the derivation of these equations, see
Appendix~\ref{appA}. Here, we have worked under the expansion $C_\alpha = C_\alpha^{(0)} + C_\alpha^{(1)}$, so Eq.~\eqref{CN-S-ell} contains only $C_\alpha^{(1)}= Q_\alpha^{(1)}/E$, and the Legendre coefficients for the collision term are defined as
    \begin{equation}
        \tilde{C}_\ell := \frac{1}{2(-i)^\ell}\int_{-1}^{1}\! d\mu {\cal P}_\ell(\upmu)\tilde{C}_\alpha^{(1)} \, . \label{C_ell}
    \end{equation}
Due to the conservation of CP, energy, and momentum, these
coefficients cancel for the monopole and the dipole ($\tilde{C}_{0,1} = 0$) when we deal with simple scattering processes where the number of particles is conserved, see Appendix~\ref{appE}.

\subsection{Boltzmann hierarchy equations in terms of particle multipole moments}~\label{BHE-MM}
In the most general case of interacting particles, Eqs.~\eqref{CN-S-ell} become a set of integro-differential ones, which are quite difficult to solve exactly~\cite{Ma1995, Oldengott2015}.
Since we aim to describe perturbations around a smooth universe, we introduce a perturbation to the distribution function. This perturbation is characterized by the temperature fluctuation, $\delta T_\alpha ({\bf x}, {\bf p}, \tau)$,
which depends explicitly on the magnitude of the momentum $p$.
In contrast to the standard photon case, whose temperature anisotropies can be treated independently of the momentum magnitude due to tight coupling in the early Universe (see, for example,~\cite{Dodelson2024modern}), other particles, such as massive neutrinos, require the momentum-dependent treatment after decoupling.
Then, expanding around small temperature fluctuations, $\delta T_\alpha$, to first order, we can write 
    \begin{equation}
        f_\alpha({\bf x}, {\bf p}, \tau) = f_\alpha^{(0)}(p) - p \frac{\partial f_\alpha^{(0)}({\bf p})}{\partial p}
        \frac{\delta T_\alpha ({\bf x}, {\bf p},\tau)}
        {T_\alpha(\tau)} \, .
    \end{equation}
Comparing this relation with \eqref{f_def}, we can relate the perturbation $\Theta_\alpha({\bf x}, {\bf p}, \tau)$ with the fractional temperature perturbation, $\delta T_\alpha/T_\alpha$, as 
    \begin{equation}
        \Theta_\alpha({\bf x},{\bf p},\tau) = - \frac{d\ln f^{(0)}(p)}{d\ln p}\frac{\delta T_\alpha({\bf x}, {\bf p},\tau)}{T_\alpha(\tau)} \,.
    \end{equation}
The common approach is to introduce the \emph{temperature fluctuation} variable~\cite{Kreisch2020}, which in Fourier space is 
    \begin{equation}
        \tilde{\Xi}_\alpha({\bf k}, {\bf p}, \tau) = 4\frac{\delta \tilde{T}({\bf k}, {\bf p}, \tau)}{T_\alpha} 
         =
        \sum_{\ell =0}^\infty
        (-i)^\ell (2\ell+1)\chi_{\alpha,\ell}(k, p, \tau){\cal P}_\ell(\mu) \, , \label{Thermal}
    \end{equation}
with the particle multipole moments $\chi_{\alpha,\ell}(k, p, \tau)$.
    
Notice that the factor of $4$ in Eq.~\eqref{Thermal} is a conventional choice within the standard relativistic treatment of the $\ell$-th multipole moment~\cite{CyrRacine2014}, and it also appears in the “separable ansatz” adopted for neutrinos in Refs.~\cite{Oldengott_2017, Chen2022weaker}. This normalization is chosen to reproduce the energy-density–weighted average $\langle d ln f^{(0)}/d \ln q \rangle = -4$, which reflects the fact that the energy density of a relativistic species scales as $a^{-4}$.

In principle, this approach with $\chi_{\alpha,\ell}$ multipole moments remains general and postpones the need to specify species other than neutrinos; however, in what follows, we shall focus on neutrinos.
\subsection{Boltzmann hierarchy equations for massive neutrinos} \label{BHE-neutrinos}
In subsection~\ref{BHE-MM}, we introduced the multipole moments of the particles in the scattering process, which depend on particle momenta. However, for neutrinos, $\alpha = \nu_i$, the absence of neutrino sinks or sources implies that such a momentum dependence should be vanishingly small at early times, when neutrinos behave as a highly relativistic fluid. In this regime, the distribution function can be well approximated by its zeroth-order form, Eq.~\eqref{f_background}, with perturbations characterized primarily by angular (directional) dependence rather than a full momentum dependence. This motivates the expansion of perturbations in terms of Legendre polynomials and the truncation of higher-order momenta contributions when deriving the Boltzmann hierarchy. This procedure is called \emph{thermal approximation}~\cite{Kreisch2020}.

By taking one neutrino species for each $i$ mass eigenstate, such that $\chi_{\alpha,\ell} = \nu_{i,\ell}$, the Legendre multipole moments can be written as
    \begin{equation}
        \vartheta_{\nu_i,\ell}(k, p_i, \tau) = 
        \frac{1}{4} p_i e^{p_i/T_\nu}f_{\nu_i}^{(0)}(p_i) \nu_{i,\ell} \, . \label{vartheta_ell.expand}
    \end{equation}
After neutrino decoupling, all multipole anisotropies with $\ell \geq 2$ are generated by gravitational interactions after a given Fourier mode (with wavenumber $k$) enters the horizon. The gravitational potential induces the same perturbation on all neutrino species at a given spatial location as long as $T_\nu\gg m_i$. This statement leads to the following approximation for the neutrino $\ell$-th multipole moment:
    \begin{equation}
        \nu_{i,\ell} \approx \nu_\ell \quad \text{for all $i$ neutrino mass eigenstates} \, .
        \label{nu-ell}
    \end{equation}
From equation \eqref{CN-S-ell}, the Boltzmann hierarchy for massive neutrinos (indexed by $\alpha=\nu_i$) is expressed in terms of their multipole moments, governed by the following equations without ${\rm g}_s$ degeneracy factors:
    \begin{equation}
         \bigg[ \frac{\partial \nu_{\ell}}{\partial \tau} + \frac{q}{\epsilon} k \left( \frac{(\ell +1)}{(2 \ell +1)} \nu_{\ell +1} - \frac{\ell}{(2\ell +1)} \nu_{\ell -1} \right) \bigg]
        - 4 f_{g,\ell} = -\frac{4 a} {f_{\nu_i}^{(0)}(q)} \frac{q^2}{\epsilon^2}\left[\frac{d \ln f_{\nu_i}^{(0)}(q)}{d \ln q} \right]^{-1} \tilde{C}_\ell \, .
    \label{BH-CN-S}
    \end{equation}
Henceforth, and throughout the remainder of this manuscript, $\epsilon = \sqrt{q^2 + a(\tau) m_i^2}$, is defined using the mass $m_i$ of the reference particle.

In this section, we have presented a self-consistent description of the first-order collision term for a $2 \to 2$ scattering process with $C^{(0)}=0$ formulated in conformal time, and we have established the Boltzmann hierarchy for massive neutrinos using the thermal approximation. We have focused on the main features of this framework, such as CP conservation in the processes, conventions for the scattering amplitudes, and the vanishing chemical potential in the distribution functions, all within the main gauges.

\section{General Boltzmann collision term for neutrino--neutrino elastic scattering via scalar mediator} \label{Sec:nu-nu via scalar mediator}

Hereafter, we will particularize our discussion to the case of self-interacting massive neutrinos with a scalar mediator of mass $m_\varphi$, for which we consider the interaction described by the Lagrangian
    \begin{equation}
        {\cal L}_{\rm int} = \sum_{ij} g_{ij}\varphi \overline{\nu}_i \nu_j \, , \label{L-int-gen}
    \end{equation}
where $g_{ij}$ are the corresponding coupling constants whose labels $i$ and $j$ run over the mass eigenstates. This Lagrangian yields the same behavior as the pseudo-scalar interaction, ${\cal L}_{\rm int}^\prime \sim g_{ij} \varphi \overline{\nu}_i \gamma_5 \nu_j$, when the neutrino masses are negligible, because the squared scattering amplitudes are the same~\cite{Blinov2019}. Viable models of this type have been discussed in Refs.~\cite{Bardin1970nu, Chikashige1981there, Choi199117, Beacom2004}. 

However, here we will conduct the analysis as generally as possible, so that extending the results to other types of mediators or, eventually, any light self-interacting particle (as warm DM candidates; see, for instance, Refs.~\cite {Atrio1997interacting, Yunis2020boltzmann, Yunis2022self}, as well as models of self-interacting dark radiation~\cite{Das2025impostor}) would be straightforward.

Starting from the first-order collision term in Fourier space,
$\tilde{C}^{(1)}_{\alpha} [{\bf k}, {\bf p}_1, \tau]$, given in Eq.~\eqref{C-first}, the distribution factor $\tilde{F}^{(1)} = \tilde{F}^{(1)} ({\bf k},{\bf p}_1, {\bf p}_2, {\bf p}_3, {\bf p}_4,\tau)$ expressed in terms of the Fourier-transformed perturbations $\tilde{\Theta}({\bf k},{\bf p}_j,\tau)$, can be written as
    \begin{equation}
        \tilde{F}^{(1)} =  \left[K_4 \tilde{\Theta}({\bf p}_4) + K_3 \tilde{\Theta} ({\bf p}_3) - K_2 \Theta( {\bf p}_2) - K_1 \tilde{\Theta}({\bf p}_1)\right] \Psi_S \, ,
        \label{F1-eslastic}
    \end{equation}
where we have left the $\tau$ and ${\bf k}$ dependence implicit. We have the following $K_n$-elements for $\nu_i \nu_j \to \nu_k \nu_l$ scattering with all species at same temperature $T_\nu$:
    \begin{align}
        K_1 &= \exp\big\{[ ((E_1+E_2-E_3)^2 - m_l^2)^{1/2} - (p_1 + p_2-p_3)]/T_\nu \big\} + \exp\{-p_1/T_\nu\} \, , \\
        K_2 &= \exp\big\{[((E_1+E_2-E_3)^2 - m_l^2)^{1/2} - (p_1 + p_2-p_3)]/T_\nu \big\} +  \exp\{-p_2 /T_\nu\} \, , \\ 
        K_3 &= 1 + \exp\big\{ [((E_1 + E_2 - E_3)^2 - m_l^2)^{1/2}-(p_1 + p_2)] /T_\nu \big\} \, ,  \\
        K_4 &= 1 +  \exp\big\{ [((E_1 + E_2 - E_4)^2 - m_k^2)^{1/2}-(p_1 + p_2)]/T_\nu\ \big\} \, .
    \end{align} 
where $l,k$ labels denote the mass eigenstates for  neutrinos with energy $E_k = [p_k^2 + m_k^2]^{1/2}$, and the common factor
    \begin{equation} 
        \Psi_S(p_1,p_2,p_3,p_4)  = 
        e^{(p_1+ p_2)/T}f_{\nu_i}^{(0)}(p_1)f_{\nu_j}^{(0)}(p_2)f_{\nu_k}^{(0)}(p_3)f_{\nu_l}^{(0)}(p_4) \, .
        \label{Psi_s}
    \end{equation}

Although many cosmological studies assume Majorana neutrinos to simplify the treatment of scalar interactions~\cite{Oldengott2015, Oldengott_2017, Kreisch2020, Chen2022weaker, RoyChoudhury2022}, restricting the interaction solely to Majorana neutrinos is neither fully general nor necessarily so. Neutrinos may instead be Dirac particles~\cite{Acker1992decaying, Acker1992neutrino}, in which case the interaction with a scalar field involves couplings between active left-handed neutrinos ($\nu_L$) and antineutrinos ($\bar{\nu}_L^c$)~\cite{He_2020} or sterile right-handed states ($\nu_R$) that, while not part of the Standard Model, could exist if physics beyond the Standard Model is realized. Consequently, a complete and general analysis of neutrino–scalar interactions should consider both possible cases, Majorana and Dirac neutrinos.

For simplicity, we consider a toy model for the Dirac-like neutrinos. This means that we assume the existence of light (left)right-handed (anti)neutrinos that are thermally populated. This assumption allows the interaction in Eq.~\eqref{L-int-gen} to be treated as a standard Dirac Yukawa coupling. As we will see below, this simplified setup enables a clear and direct comparison with the Majorana case.

Moreover, the physical and cosmological properties of Dirac-like neutrinos with scalar couplings can differ significantly from those of Majorana neutrinos, affecting thermal abundances, the evolution of the early Universe, and cosmological observables such as $N_{\rm eff}$. In such cases, the toy-model description may no longer be valid and must be discarded~\cite{Blinov2019}.

In the following, we will write the collision terms for the neutrino--neutrino interaction, leaving the scalar mediator mass, $m_\varphi$, as a free parameter.
\subsection{Neutrino--neutrino interaction with different neutrino mass eigenstates} \label{Different-neutrino-mass}

\begin{figure} [!ht]
    \centering
    \subfigure[]{\includegraphics[width=50mm]{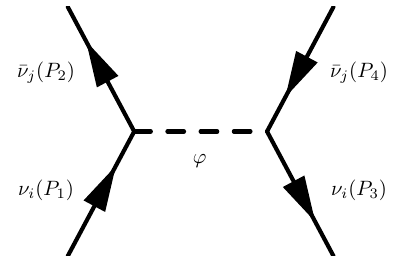}}
    \subfigure[]{\includegraphics[width=50mm]{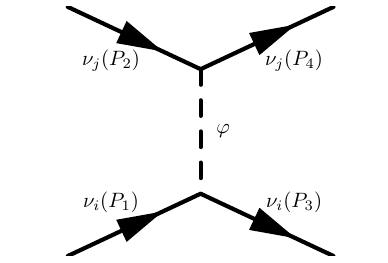}}
    \subfigure[]{\includegraphics[width=50mm]{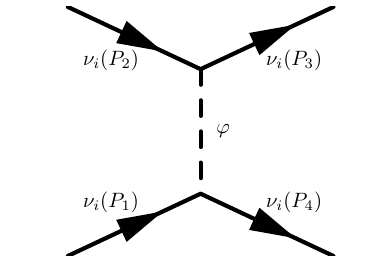}}
    \caption{Feynman diagrams at tree level involved in the scalar mediated $\nu_i\nu_j \rightarrow \nu_i\nu_j$ scattering with: a) s-channel (valid only for Majorana neutrinos and neutrino–antineutrino pairs), b) channel t, c) channel u (only valid for the same mass eigenstate).}
    \label{Scattering with neutrinos}
    \end{figure}

In the general case for elastic scattering: $\nu_i + \nu_j \rightarrow \nu_i + \nu_j$, all possible tree-level $s$, $t$, and $u$ channels via scalar mediator are depicted in Fig.~\ref{Scattering with neutrinos} a), b), and c), respectively.

For the Majorana case, we introduced a factor of $1/2$ to ensure a consistent normalization of the field operators in quantum field theory~\cite{Pal2011dirac}. Thus, the Majorana interaction Lagrangian is written as ${\cal L}_{\rm int}^M = \tfrac{1}{2} g_{ij}\varphi\overline{\nu}_i\nu_j$. It is also well known that the Majorana condition, at the Lagrangian level, forbids certain Dirac structures in the neutrino bilinear, while others lead to two identical contributions (as in the neutral current; see, e.g., Kayser~\cite{Kayser1982distinguishing}).

In the collision term \eqref{C}, we have the dependence on the spin-averaged squared scattering amplitude $\overline{|{\cal M}|}^2$ for the corresponding process. Then, for neutrino--neutrino scattering, we have the general form
    \begin{equation}
        \overline{|{\cal M}|}^2 = g^4 \sum_r \alpha_r T_r(s,t,u;m_i, m_j,m_\varphi)~,
        \label{M2}
    \end{equation}
where $T_r$ is, in general, a dimensionless rational function of the masses and the invariant Mandelstam variables, 
    \begin{eqnarray}
        s &=& (P_1 + P_2)^2 = (P_3 + P_4)^2 \,, \\
        t &=& (P_1 - P_3)^2 = (P_2 - P_4)^2 \,, \\
        u &=& (P_1-P_4)^2 = (P_2 - P_3)^2 \,. 
    \end{eqnarray}
The subindex $r$ runs over the contributions of each independent Feynman diagram in the process and their corresponding crossed interference, while $\alpha_r$ is determined by the ratio between couplings with respect to $|g_{ii}|:=g$. In appendix \ref{Amplitudes} we have written the explicit expressions for the amplitudes of interest at the tree level.

Follow the prescription described in the subsection~\ref{BHE-neutrinos}, $\nu_{i,\ell} = \nu_\ell$ for each mass eigenstate; this approximation reflects the universal gravitational sourcing of anisotropies across all species involved in the interaction. With this in mind, the linear order collision term for one elastic scattering contribution, particularly for scattering $\nu_i-\nu_j$ between massive neutrinos in the Fourier space, is expressed by
    \begin{equation} 
        \tilde{C}^{(1)}_{\nu_i} = \frac{1}{4} \frac{g^4 T_\nu}{128 \pi^3 z_1}
        \frac{d \ln f^{(0)}(p_1)}{d \ln p_1} \sum_{\ell=0 }^\infty (-i)^{\ell}(2\ell+1) \nu_{\ell} {\cal P}_\ell(\mu) \,  {\cal K}_\ell^{ij}(z_1) \, ,
        \label{ctilde_1}
    \end{equation}
where $\upmu = \hat{\bf p}_1 \cdot \hat{\bf k}$ (${\bf p}_1$ being the local momentum of the reference particle). Hereafter, for further convenience, we introduce the dimensionless notation
$z_i = E_i/T_\nu$ and $\mu_i = m_i/T_\nu$ where $T_\nu$ is the effective temperature of the neutrino background, and
    \begin{equation} 
        {\cal K}_\ell^{ij}(z_1) = {\cal A}^{ij}(z_1)+ {\cal B}_\ell^{ij}(z_1) - {\cal D}_\ell^{ij}(z_1) - {\cal E}_\ell^{ij}(z_1) \, . \label{K-ell-ij}
    \end{equation}
The functional coefficients  in ${\cal K}_\ell^{ij}(z)$ above are given by the following integral expressions:
    \begin{align}
        {\cal A}^{ij}(z_1) &= \int_{\mu_j}^\infty\! dz_2\sqrt{z_2^2 -\mu_j^2}
        \int_{\mu_i}^{z_1 +z_2-\mu_j}\,dz_3\sqrt{z_3^2 -\mu_i^2} \Psi_{S,1} \, J_1(z_1,z_2,z_3) \, , 
        \label{calA}\\
        {\cal B}_\ell^{ij}(z_1) &= \frac{1}{q_1} \int_{\mu_j}^\infty dz_2 (z_2^2 - \mu_j^2) \int_{\mu_i}^{z_1+z_2- \mu_j} dz_3 \sqrt{z_3^2 - \mu_i^2} \Psi_{S,2} \, J_{2\ell}(z_1,z_2,z_3) \, , 
        \label{calBl}\\
        {\cal D}_\ell^{ij} (z_1) &= \frac{1}{q_1} \, 
        \int_{\mu_j}^\infty\! dz_2\sqrt{z_2^2 -\mu_j^2} 
        \int_{\mu_i}^{z_1 + z_2 -\mu_j}\! dz_3 \, (z_3^2 - \mu_i^2) \Psi_{S,3} \,  J_{3\ell}(z_1,z_2,z_3) \, , 
        \label{calDl}\\
        {\cal E}_\ell^{ij} (z_1) &= \frac{1}{q_1} \, 
        \int_{\mu_j}^\infty\! dz_2\sqrt{z_2^2 -\mu_j^2}
        \int_{\mu_j}^{z_1 + z_2 -\mu_i}\! dz_4 \, (z_4^2 -\mu_j^2) \, \Psi_{S,4} \, J_{4\ell}(z_1,z_2,z_4) \, , 
        \label{calEl}
    \end{align}
where the inside functions are
    \begin{align}
        \Psi_{S,1} &= \frac{\exp\big\{[(z_1+z_2-z_3)^2 - \mu_j^2]^{1/2} - (q_1 + q_2-q_3) \big\} + \exp\{-q_1\}}{(e^{-q_1} + 1)(e^{-q_2} + 1) (e^{q_3} + 1)(\exp{[(z_1 + z_2 -z_3)^2 - \mu_j^2]^{1/2}} + 1)} \, , \label{Psi-1}\\
        \Psi_{S,2} &= \frac{ \exp\big\{[(z_1+z_2-z_3)^2 - \mu_j^2]^{1/2} - (q_1 + q_2 - q_3) \big\} +  \exp\{-q_2\}}{(e^{-q_2} + 1)^2(e^{q_3} + 1) (\exp{[(z_1 + z_2 -z_3)^2 - \mu_j^2]^{1/2}} + 1)} \, , \label{Psi-2} \\
        \Psi_{S,3} &= \frac{[1 + \exp\big\{ (z_1 + z_2 - z_3)^2 - \mu_j^2]^{1/2}-(q_1 + q_2) \big\}]e^{q_3}}{(e^{-q_2} + 1)(e^{q_3} + 1)^2 (\exp{[(z_1 + z_2 -z_3)^2 - \mu_j^2]^{1/2}} + 1)} \, , \label{Psi-3} \\
        \Psi_{S,4} &= \frac{[1 + \exp\big\{ (z_1 + z_2 - z_4)^2 - \mu_i^2]^{1/2}-(q_1 + q_2) \big\}]e^{q_4}}{(e^{-q_2} + 1)(e^{q_4} + 1)^2(\exp{[(z_1 + z_2 -z_4)^2 - \mu_i^2]^{1/2}} + 1)} \, . \label{Psi-4}
    \end{align}
We have used: $q_1= (z_1^2 - \mu_i^2)^{1/2}$, $q_2= (z_1^2 - \mu_j^2)^{1/2}$, $q_3=(z_1^2 - \mu_i^2)^{1/2}$, $q_4=(z_1^2 - \mu_j^2)^{1/2}$.

Here we provide explicit expressions for the $J$-integrals in Eq.~\eqref{J-integrals-g} as a result of our general strategy outlined in Fig.~\ref{fig:steps}. The particular form of the $J$-integrals is determined as follows: 
\begin{figure}
    \centering
    \includegraphics[width=1\linewidth]{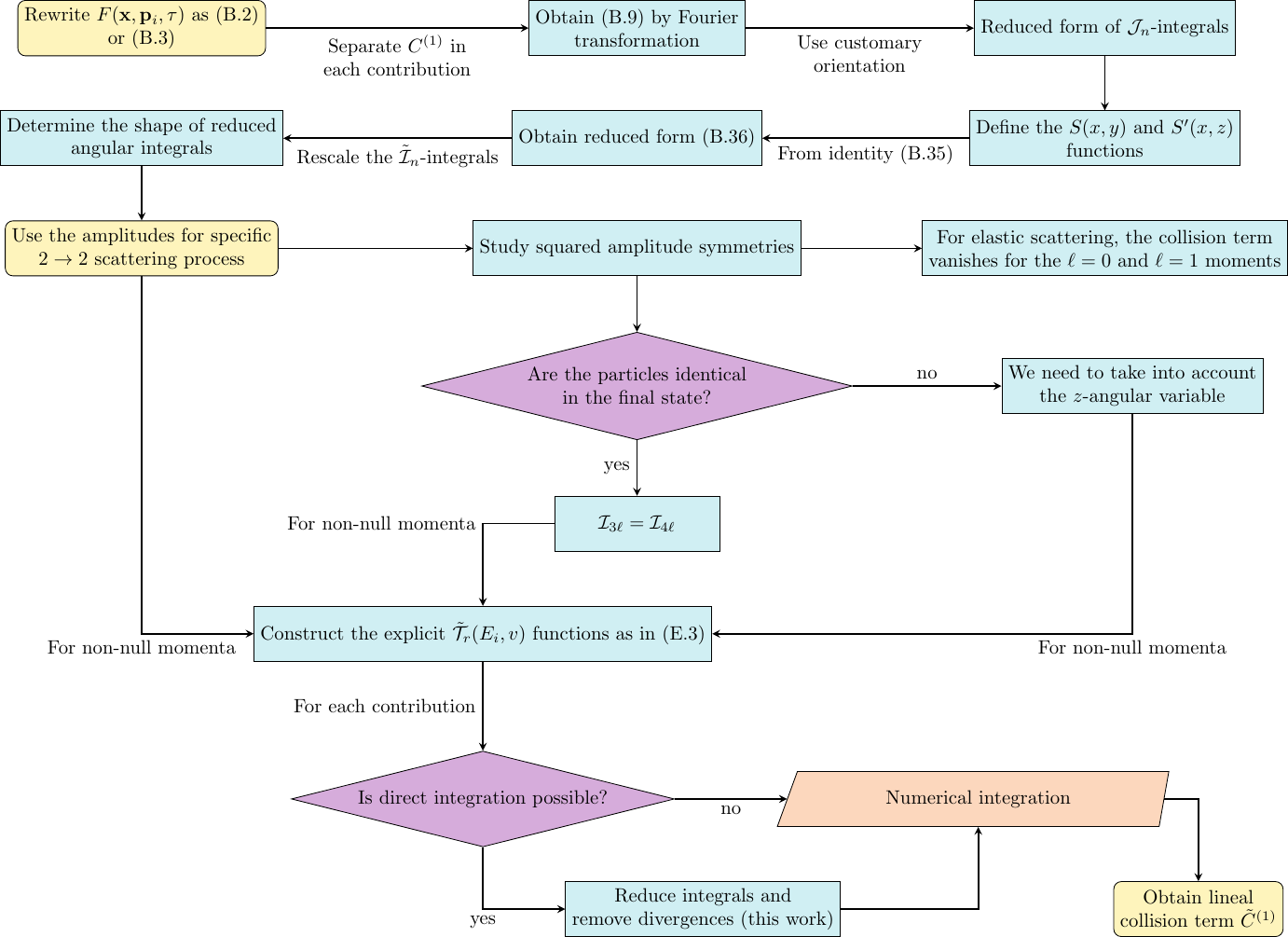}
    \caption{General scheme for computing the collision term in $2 \to 2$ scattering.}
    \label{fig:steps}
\end{figure}
Using the customary orientation of the neutrino spatial momenta relative to ${\bf p}_1$, using the incident angle $\alpha$, and the corresponding scattering angles $\theta$ and $\theta^\prime$, we define the angular variables by $x = \hat{{\bf p}}_1\cdot\hat{{\bf p}}_2 =\cos\alpha$, $y = \hat{{\bf p}}_1\cdot\hat{{\bf p}}_3 =\cos\theta$, and $z = \hat{{\bf p}}_1\cdot\hat{{\bf p}}_4 =\cos\theta^\prime$. Following the prescription presented in the appendix~\ref{appB}, we establish the corresponding second-order polynomials $S(x,y)$ and $S^\prime(x,z)$ whose roots contribute to restricting the limits of the reduced angular integrals.

As all $J$-integrals are linear functions of the angular variables, $x$, $y$, or $z$, it is straightforward to see that any given $T$ term is indeed a rational function of at most order two polynomials on the angular variables, and using the properties of the Heaviside step function and energy conservation, we obtain
    \begin{eqnarray}
        J_1 =
        \int_{\rho_+}^{\rho_{-}} dy \,{\cal T}_y^N \,,
        \qquad
        J_{2\ell} = \int_{\eta_+}^{\eta_{-}} dx \,{\cal T}_x^N {\cal P}_\ell(x) \,,
        \nonumber \\
        J_{3\ell} = \int_{\rho_+}^{\rho_{-}} dy \,{\cal T}_y^N {\cal P}_\ell(y)\,,
        \qquad
        J_{4\ell} =
        \int_{\tau_+}^{\tau_{-}} dz \,{\cal T}_z^N {\cal P}_\ell(z)\,.
        \label{J-integrals}
    \end{eqnarray}
Here, $N = D$ for Dirac-like neutrinos or $N = M$ for Majorana-like neutrinos, while the limits for the angular variables are: $\eta_{-} = {\rm min}(x_{-},1)$, $\eta_{+} = {\rm max}(x_{+}, -1)$, $\rho_{-} = {\rm min}(y_{-},1)$, $\rho_{+} = {\rm max}(y_{+}, -1)$, $\tau_{-} = {\rm min}(z_{-},1)$ and $\tau_{+} = {\rm max}(z_{+},1)$, where the quantities $x_{\pm}$, $y_{\pm}$ and $z_{\pm}$ are determined by the energy conservation constraints:
    \begin{align}
        x_{\pm} &= \frac{z_1 z_2 - z_3 (z_1 + z_2 - z_3) \mp q_3 \sqrt{(z_1 + z_2 - z_3)^2 - \mu_j^2}}{q_1q_2} \, ,
        \label{x-limits} \\
        y_{\pm} &= \frac{z_1 z_3 - z_2 (z_1 + z_2 - z_3) - (\mu_i^2 - \mu_j^2) \mp q_2 \sqrt{(z_1 + z_2 - z_3)^2 - \mu_j^2}}{q_1 q_3} \, ,
        \label{y-limits} \\
        z_{\pm} &= \frac{z_1 z_4 - z_2 (z_1 + z_2 - z_4) \mp q_2 \sqrt{(z_1 + z_2 - z_4)^2 - \mu_i^2}}{q_1 q_4} \, .
        \label{z-limits}
    \end{align}

Following the strategy presented in the Appendix~\ref{appD} the explicit form for ${\cal T}_v^N$ with our short-hand notation are
    \begin{align}
        {\cal T}_y^D &= \frac{\alpha_t}{\sqrt{-a_y}} \frac{(a_1 - B_1y)(a_2 - B_1y)}{(A_{1,i}^\prime - B_1y)^2} \, , \label{Ty_D}\\
        {\cal T}_x^D &= \alpha_t \bigg\{ \frac{1}{\sqrt{-a_x}} - \frac{1}{B_1 \sqrt{-c^{\prime}_{x,i}}} \left[ Q_{ij}  - \frac{Q_i Q_j b_{x,i}^{\prime}}{8 B_1 c_{x,i}^{\prime}} \right] \bigg\} \, , \label{Tx_D}\\
        {\cal T}_z^D &= \alpha_t \bigg\{ \frac{1}{\sqrt{-a_z}} + \frac{1}{B_1^\prime\sqrt{-c_z^\prime}}\left[ Q_{ij} - \frac{Q_i Q_j b_z^\prime}{8 B_1^{\prime}c_z^\prime} \right] h_4(q_z)\bigg\} \, . \label{Tz_D}
    \end{align}

    \begin{align}
        {\cal T}_y^M &= \frac{1}{\sqrt{-a_y}} \left[ 
        \alpha_t \frac{(a_1 - B_1 y)(a_2 - B_1 y)}{(B_{y,i}^\prime)^2} 
        + \alpha_s 
        + \alpha_{st} \frac{B_{y,ij}^{\prime\prime}}{B_{y,i}^\prime B_1^\prime}
        \right] \nonumber \\
        &\quad + \frac{1}{B_1^\prime \sqrt{-c_{y,ij}^{\prime\prime}}} \left[ 
        \alpha_{st} \frac{A_{y,ij}^{\prime\prime} + B_{y,ij}^{\prime\prime} q_{ij}^\prime}{B_{y,i}^\prime}
        - \alpha_s \left( \bar{Q}_{ij} + \frac{\bar{Q}_{ij}^2 b_{y,ij}^{\prime\prime}}{8 B_1^\prime c_{y,ij}^{\prime\prime}} \right)
        \right] h_5(q_{ij}^\prime) \, , \label{Ty_M}\\
        {\cal T}_x^M &= \frac{1}{\sqrt{-a_x}} \left[ 
        \alpha_t 
        + \alpha_s \frac{(A_{4,ij} - B_1^\prime x)^2}{(B_{x,ij}^\prime)^2}
        + \alpha_{st} \frac{B_{x,ij}^{\prime\prime}}{B_{x,ij}^\prime B_1}
        \right] \nonumber \\
        &\quad - \frac{1}{B_1 \sqrt{-c_{x,i}^\prime}} \left[ 
        \alpha_t \left( Q_{ij} - \frac{Q_i Q_j b_{x,i}^\prime}{8 B_1 c_{x,i}^\prime} \right)
        + \alpha_{st} \frac{A_{x,ij}^{\prime\prime} + B_{x,ij}^{\prime\prime} q_i}{B_{x,ij}^\prime} \right] \, , \label{Tx_M} \\
        {\cal T}_z^{M} &= (\alpha_t + \alpha_s + \alpha_{st}) \frac{1}{\sqrt{-a_z}} + \frac{1}{B_1^\prime \sqrt{-c_z^\prime}}\left[ \alpha_t \left( Q_{ij} - \frac{Q_i Q_j b_z^\prime}{8 B_1^{\prime}c_z^\prime} \right) - \alpha_{st} D_{2z} \right] h_4(q_z) \\
        &\quad - \frac{1}{B_1^\prime \sqrt{-c_{z,j}^\prime}} \left[ \alpha_s \left( \bar{Q}_{ij} + \frac{\bar{Q}_{ij}^2 b_{z,j}^\prime}{8B_1^\prime c_{z,j}^\prime} \right) - \alpha_{st} D_{1z} \right] h_6(q_{ij}^\prime) \, .
    \end{align}
Here $\alpha_t = |g_{jj}|^2/g^2$, $\alpha_s = |g_{ij}|^4/g^4$ and $\alpha_{st} = {\rm Re} \{g_{ij}^2 g_{ii}^{*} g_{jj}^{*}\}/g^4$, and the other coefficients that were not defined in Appendix \ref{appD} (see Eqs.~\eqref{short-notation}) are:
    \begin{equation*}
        Q_{ij} = \mu_\varphi^2 - 2(\mu_i^2 + \mu_j^2) \, , \, Q_i=\mu_\varphi^2 - 4\mu_i^2 \, , \, Q_j^2 = \mu_\varphi^2 - 4\mu_j^2 \, , \, q_z = \frac{A_7 + B_1^{\prime\prime}z}{B_1^\prime} \, .
    \end{equation*}
While for the Majorana case:
    \begin{equation*}
    \begin{aligned}
        B_{y,i}^\prime &= A_{1,i}^\prime - B_1 y \, , \, B_{y,ij}^{\prime\prime} = B_{3,ij}^\prime - B_1 B_1^\prime y \, ,\, A_{y,ij}^{\prime\prime} = A_{5,ij} + B_{3,ij}y \, , \, \bar{Q}_{ij} = \mu_\varphi^2 - (\mu_j + \mu_i)^2 \, , \\
        q_{ij}^\prime &= \frac{A_{4,ij}^\prime}{B_1^\prime} \, , \, B_{x,ij}^{\prime} = A_{4,ij}^\prime - B_1^\prime x \, , \, B_{x,ij}^{\prime\prime} = B_{3,ij} - B_1 B_1^\prime x \, , \, A_{x,ij}^{\prime\prime} = A_{5,ij} + B_{3,ij}^\prime x \, , \, q_i = \frac{A_{1,i}^\prime}{B_1} \, , \\
        D_{2z} &= \frac{A_8 + B_4z - (B_4^\prime + B_1^\prime B_1^{\prime\prime}z) q_z + (A_7 + B_1^{\prime\prime}z)^2}{A_{4,ij}^\prime - (A_7 + B_1^{\prime\prime}z)} \, , \, B_1^{\prime\prime} = p_1 p_4 \, , \\
        D_{1z} &= \frac{A_8 + B_4z - (B_4^\prime + B_1^\prime B_1^{\prime\prime}z) q_{ij}^\prime + A_{4,ij}^{\prime 2}}{[A_{4,ij}^\prime - (A_7 + B_1^{\prime\prime}z)]} \, .
    \end{aligned}
    \end{equation*} 
Also, we have the coefficients:
    \begin{equation*}
    \begin{aligned}
        c_{x,i}^\prime&= c_{x,i} + b_{x,i} q_i +a_x q_i^2 \, , \, b_{x,i}^{\prime}= b_{x,i} +2a_x q_i \, , c_z^\prime = c_{z,j} + b_{z,j} q_z+ a_zq_z^2 \, , \, 
        b_z^\prime = b_{z,j} + 2a_z q_z \, , \\
        c_{y,ij}^{\prime\prime} &= c_{y,i} + b_{y,i} q^\prime_{ij} + a_y q^{\prime 2}_{ij} \, , \, b^\prime_{y,ij} = b_{y,i} + 2a_y q^\prime_{ij} \, , \, c_{z,j}^\prime = c_{z,j} + b_{z,j} q_{ij}^\prime + a_z q_{ij}^{\prime 2} \, , \\ b_{z,j}^\prime&= b_{z.j} + 2 a_z q_{ij}^\prime \, .
    \end{aligned}
    \end{equation*}
Moreover, since the propagator behaves as $\sim (r-m_\varphi^2)^{-1}$, with $r=s,t$, collinear infrared singularities as well as kinematic singularities associated with on-shell production may arise. To treat these contributions, we introduce the $h$-functions where the singular terms are regularized by taking their null contribution as in the cases Eqs.~\eqref{intr-1}–\eqref{intr-2} (see an explicit example in~\ref{h-functions}), and we define the $h$-functions as follows:
        \begin{equation}
            h_4(q_z) = \begin{cases}
            1 & \text{for $\tilde{x}_{+}^\prime-q_z > 0$} \\
            -1 & \text{for $\tilde{x}_{-}^\prime-q_z < 0$} \\
            0 & \text{elsewhere}
            \end{cases} \, , \quad 
            h_5(q^\prime_{ij}) = \begin{cases}
            1 & \text{for $\tilde{x}_{+} -q^\prime_{ij} >0$}\\
            -1 & \text{for $\tilde{x}_{-} -q^\prime_{ij}<0$} \\
            0 & \text{elsewhere}
            \end{cases} \, , \label{h4-h5}
    \end{equation}
and 
    \begin{equation}
        h_6(q_{ij}^\prime) = \begin{cases}
        1 & \text{for $\tilde{x}_{+}^\prime-q^\prime_{ij} >0$}\\
        -1 & \text{for $\tilde{x}_{-}^\prime -q^\prime_{ij}<0$} \\
        0 & \text{elsewhere}
        \end{cases} \, , \label{h6}
    \end{equation}
where $\tilde{x}_{\pm}^\prime$ are the roots of $S^\prime(x,z)$ in terms of the angular variable $z$, while $\tilde{x}_{\pm}$ are the roots of $S(x,y)$ in terms of the angular variable $y$ (see Appendix~\ref{appD}).

Using the definition of Eq.~\eqref{C_ell} and Eq.~\eqref{ctilde_1}, we obtain the Boltzmann hierarchy equations for massive neutrinos, for all elastic scattering contributions ($\nu_i \neq \nu_j$): 
    \begin{equation}
    \begin{aligned}
        \frac{\partial \nu_{\ell}}{\partial \tau} & + \frac{q}{\epsilon} k \left[ \frac{(\ell +1)}{(2 \ell +1)} \nu_{\ell +1} - \frac{\ell}{(2\ell +1)} \nu_{\ell -1} \right] - 4 f_{g,\ell} = -\frac{a} {f_{\nu_i}^{(0)}(q)} \frac{g^4 T_\nu}{128 \pi^3} \left( \frac{aT_\nu}{\epsilon}\right) \nu_{\ell} \, {\cal C}^{ij}_\ell(z_1) \, .
        \label{HBE-CN-S}
    \end{aligned}
    \end{equation}
Here, after including a degeneracy factor to account for unpolarized scattering, the elastic contributions are given by the sum of the corresponding collision terms,
    \begin{equation}
       {\cal C}_\ell^{ij}(z_1) = 
           {\rm g}_s \sum_{j \neq i}{\cal K}^{ij}(z_1) \, .
         \label{collision_cases}
    \end{equation}
Here, $j$ denotes the (anti)neutrino mass eigenstate, in the Dirac-like case ${\rm g_s}=2$ refers to the spin polarization, for neutrinos or antineutrinos, and is used to determine their contribution to the energy density, which is constrained by measurements of $N_{\rm eff}$. While ${\rm g}_s=2$ for the Majorana case.

For relativistic neutrinos, it is straightforward to obtain $T_{0, \nu} = aT_\nu$ if we choose $a_0=1$ today. Perturbations from this point on would require the use of numerical analysis to evaluate the integral coefficients and then solve the Boltzmann hierarchy equations Eq.~\eqref{HBE-CN-S}. This analysis is becoming standard for exploring possible bounds to exotic mediators. For instance, the case of self-interacting massless neutrinos with a heavy scalar mediator was discussed in Refs.~\cite{Kreisch2020}, and our formula \eqref{HBE-CN-S} generalizes the ones discussed there. We do not intend to run over that type of analysis here, but rather concentrate on the structure of the collision term. This provides a general method that should allow for exploring all mass mediator ranges without neglecting the neutrino mass, which, in the end, is an important cosmological parameter that one usually intends to constrain from observational data.

Notice that taking $m_i = m_j$ corresponds to a special case in which the final states become indistinguishable. Since the computation of the collision term involves an integration over phase space, we discuss this case in the next subsection.

\subsection{Interactions between neutrinos of the same mass eigenstate}

Consider the scattering between neutrinos in the same mass eigenstate, i.e., mass $m_\nu$, so that all $\alpha_r \to 1$ in the ${\cal T}_v^N$ expressions. If we consider $g$ as a universal coupling, we would have a non-elastic neutrino--neutrino scattering. At the moment, however, we do not consider such a scenario, since the number of particles, in principle, is not conserved in that case.

As discussed in the previous section, the general form of the collision term in Eq.~\eqref{ctilde_1} is already defined up to the reduced angular integrals  $J_{1,n\ell}$ in Eqs.~\eqref{J-integrals}. We shall discuss these in what follows: For the scalar-mediated neutrino--neutrino scattering, the $T$-terms involved in the invariant amplitude presented in \eqref{Am-D} and \eqref{Am-M} for $m_i=m_j=m_\nu$ are reduced to
    \begin{equation}
        T_{r} = \frac{(r-4m_\nu^2)^2}{(r-m_\varphi^2)^2},\qquad \text{for}\quad r=s,t,u~, 
        \label{Tr}
    \end{equation}
whereas for the interference terms, one gets
    \begin{equation}
        T_{tu} = \frac{ut-4m_\nu^2s}{(t-m_\varphi^2)(u-m_\varphi^2)} \quad , \quad T_{st} =\frac{st- 4m_\nu^2 u}{(s-m_\varphi^2)(u-m_\varphi^2)} \, ,
        \label{Tmix}
    \end{equation}
and $T_{su} = T_{st}(u\leftrightarrow t)$ (see Appendix~\ref{Amplitudes} for details). Using the above expressions, we evaluate Eq.~\eqref{M2}. For elastic processes between identical neutrinos, we need to evaluate $r = s,t,u, st,su,tu$ ($r=t,u,tu$) for Majorana (Dirac-like) neutrinos. 

As a side remark, we note that the denominator structure of the above expressions is common to any mediator (scalar, vector, or tensor) since it comes from the propagator of the virtual particle in the diagrams of Fig.~\ref{Scattering with neutrinos}. The numerator, on the other hand, should be expected to have a different form for each case, but since the amplitude is dimensionless, it is at most a polynomial of order four in the momenta, and thus at most a second-order one in Mandelstam variables.

Notice that we have the symmetry between $3 \leftrightarrow 4$ labels, since the outgoing neutrinos are indistinguishable, then ${\cal D}_\ell = {\cal E}_\ell$. Therefore $ {\cal K}_\ell = {\cal A}(z_1) + {\cal B}_\ell(z_1) - 2{\cal D}_\ell(z_1)$ substitutes  ${\cal K}_\ell^{ij}$ in the expression for the collision term ~\eqref{ctilde_1}. The functional forms in Eqs.~\eqref{calA}--\eqref{calEl} take the particular forms~\footnote{We omitted the labels because we work with the same mass eigenstate.}:
    \begin{align}
        {\cal A}(z_1) &= \int_{\mu_\nu}^\infty\! dz_2\sqrt{z_2^2 -\mu_\nu^2}
        \int_{\mu_\nu}^{z_1 +z_2-\mu_\nu}\,dz_3\sqrt{z_3^2 -\mu_\nu^2} \, \Psi_{S,1} \, J_1(z_1,z_2,z_3) \, , 
        \label{calA-red}\\
        {\cal B}_\ell(z_1) &= \frac{1}{q_1} \int_{\mu_\nu}^\infty dz_2 (z_2^2 - \mu_\nu^2) \int_{\mu_\nu}^{z_1 +z_2 - \mu_\nu} dz_3 \sqrt{z_3^2 - \mu_\nu^2} \, \Psi_{S,2} \, J_{2\ell}(z_1,z_2,z_3) \, , 
        \label{calBl-red} \\
        {\cal D}_\ell (z_1) &= \frac{1}{q_1} \, 
        \int_{\mu_\nu}^\infty\! dz_2\sqrt{z_2^2 -\mu_\nu^2} 
        \int_{\mu_\nu}^{z_1 + z_2 -\mu_\nu}\! dz_3 \, (z_3^2 - \mu_\nu^2) \,  \Psi_{S,3} \,  J_{3\ell}(z_1,z_2,z_3) \, ,
        \label{calDl-red}
    \end{align}
where $\mu_\nu = m_\nu/T_\nu$, the $\Psi_{S,n}$ terms in Eqs.~\eqref{Psi-1}-\eqref{Psi-4} are valuated at $\mu_\nu = \mu_i = \mu_j$, and the $J$-integrals in Eq.~\eqref{J-integrals} have the contribution of the channel $u$ (see Fig.~\ref{Scattering with neutrinos}). As in subsection~\ref{Different-neutrino-mass} we follow the strategy and short-hand notation outlined in Appendix~\ref {appD} (see also Fig.~\ref{fig:steps}), then the ${\cal T}_v^N$-forms in the $J$-integrals are
    \begin{align}
        {\cal T}_y^D &=  \frac{1}{\sqrt{-a_y}}\left(1+\frac{c_1^2}{{B_y^\prime}^2}-\frac{B_y}{B_y^\prime B_1^\prime}\right) + \frac{1}{B_1^\prime \sqrt{-c_y^\prime}}\left(
        Q - \frac{Q^2b_y^\prime}{8{B_1^\prime} c_y^\prime}
        -  \frac{A_y+B_y q_y}{B_y^\prime} \right) h_2(q_y) \, , \label{TyD} \\
        {\cal T}_x^D &= \frac{3}{\sqrt{-a_x}} - \left(\frac{Q-C_{1x}}{B_1\sqrt{-c_x^\prime}}-\frac{Q^2b_x^\prime}{8B_1^2c_x^\prime\sqrt{-c_x^\prime}}\right) + 
        \left( \frac{C_{2x} - Q}{B_1\sqrt{-c_x^{\prime\prime}}} -\frac{Q^2b_x^{\prime\prime}}{8B_1^2c_x^{\prime\prime}\sqrt{-c_x^{\prime\prime}}}\right) h_1(q_x) \, , \label{TxD} 
    \end{align}
    \begin{equation}
    \begin{aligned}
        {\cal T}_y^M &= \frac{1}{\sqrt{-a_y}}
        \left(3 +\frac{(A_1 - B_1y)^2}{{B_y^\prime}^2} -\frac{B_y - B_y^{\prime\prime}}{B_y^\prime B_1^\prime} \right) -\frac{h_3(q^\prime)}{B_1^\prime\sqrt{-c_y^{\prime\prime}}}\left( Q - \frac{A_y^{\prime\prime}+B_y^{\prime\prime}q^\prime}{B_y^\prime} + \frac{Q^2b_y^{\prime\prime}}{8B_1^\prime c_y^{\prime\prime}} \right) \\
        &\quad + \frac{1}{B_1^\prime\sqrt{-c_y^\prime}}\left( Q -\frac{A_y+B_yq_y}{B_y^\prime} -\frac{ Q^2b_y^\prime}{8B_1^\prime c_y^\prime} \right) h_2(q_y) + L_y(q^\prime, q_y) \, , \label{TyM}
    \end{aligned}
    \end{equation}
    
    \begin{equation}
    \begin{aligned}
        {\cal T}_x^M &= \frac{1}{\sqrt{-a_x}}\left(3 + \frac{(A_4 -B_1^\prime x)^2}{{B_x^\prime}^2} + \frac{2 B_x^{\prime\prime}}{B_x^\prime B_1} \right) -
        \frac{1}{B_1\sqrt{-c_x^\prime}}\left(Q-C_{1x} + \frac{A_x^{\prime\prime}+B_x^{\prime\prime}q}{B_x^\prime}-\frac{Q^2b_x^\prime}{8B_1c_x^\prime}\right) \\
        &\quad + \frac{1}{B_1\sqrt{-c_x^{\prime\prime}}}\left(
        C_{2x} - Q -\frac{D_x+B_x^{\prime\prime}q_x}{B_x^\prime}
        -\frac{Q^2b_x^{\prime\prime}}{8B_1c_x^{\prime\prime}} \right) h_1(q_x) \, , \label{TxM}
    \end{aligned}
    \end{equation}
where the auxiliary quantities are determined by the Eqs.~\eqref{short-notation}:
    \begin{equation*}
    \begin{aligned}
    c_1 &= A_1 - B_1 y \, , \,  B_y^\prime = A_1^\prime - B_1 y \, , B_y = B_2^\prime + B_1 B_1^\prime y \, , \, Q=\mu_\varphi^2 - 4\mu_\nu^2 \, , \, A_y=A_3+B_2y \, , \\
    q_y &= \frac{A_2^\prime + B_1y}{B_1^\prime} \, , \, A_x = A_3 + B_2^\prime x \, , \, B_x = B_2 + B_1B_1^\prime x \, , \, q = \frac{A_1^\prime}{B_1} \, , \, q_x = \frac{A_x^\prime}{B_1} \, , \, A_x^\prime = A_2^\prime -B_1^\prime x \, , \\
    C_{1x} &= \frac{A_x+B_x\,q-{A_1^\prime}^2}{A_1^\prime+A_x^\prime} \, , \, C_{2x} = \frac{A_x-B_x\,q_x-{A_x^\prime}^2}{A_1^\prime+A_x^\prime} \, , \, B_y^{\prime\prime} = B_3^\prime - B_1 B_1^\prime y \, , \, q^\prime = \frac{A_4^\prime}{B_1^\prime} \, , \\
    C_{1y} &= \frac{A_6 - B_3y + (B_4^\prime + B_1 B_1^\prime y)q^\prime - A_4^{\prime 2}}{A_4^\prime - A_y^\prime} \, , \, C_{2y} = \frac{A_6 - B_3y + (B_4^\prime + B_1 B_1^\prime y)q_y - A_y^{\prime 2}}{A_4^\prime - A_y^\prime} \, , \\
    A_y^{\prime\prime} &= A_5 + B_3y \, , \, B_x^\prime = A_4^\prime-B_1^\prime x \, , \, B_x^{\prime\prime} = B_3-B_1 B_1^\prime x \, , \, A_x^{\prime\prime} = A_5+B_3^\prime x \, ,
    \end{aligned}
    \end{equation*}
are determined with the elements presented in Appendix~\ref{appD}, also, we have
    \begin{equation*}
    \begin{aligned}
        b_y^\prime &= b_y + 2q_y q_y \, , \, c_y^\prime=c_y+ b_yq_y + a_y q_y^2 \, , \, b_x^\prime = b_x +2a_x q \, , \, c_x^\prime = c_x + b_x q+a_x q^2 \, , \\
        b_x^{\prime\prime} &= b_x -2a_x q_x \, , \, c_x^{\prime\prime} = c_x - b_x q_x+a_x q_x^2 \, , 
    \end{aligned}
    \end{equation*}
while in the $h$'s functions, the divergence associated with the propagator $(r-m_\varphi^2)^{-1}$, with $r=s,t,u$, is regularized by taking its null contribution and absorbed into the null contributions (see Appendix~\ref{h-functions}). Consequently, we obtain:
    \begin{equation}
    \begin{aligned}
        h_1(q_x) = \begin{cases}
            1 & \text{for $\tilde{y}_{+} > -q_x $} \\
            -1 & \text{for $\tilde{y}_{-}< - q_x$} \\
            0 & \text{elsewhere}
        \end{cases} , \, h_2(q_y) = \begin{cases}
         - 1 & \text{for $\tilde{x}_{-}<q_y$}\\
         1 & \text{for $\tilde{x}_{+} >q_y$} \\
         0 & \text{elsewhere}
        \end{cases} \, , \, h_3(q^\prime) = \begin{cases}
         1 & \text{for $\tilde{x}_{+}>q^\prime$}\\
         -1 & \text{for $\tilde{x}_{-} <q^\prime$} \\
         0 & \text{elsewhere}
        \end{cases} \, , 
    \end{aligned} \label{h1-h3}
    \end{equation}
and $L_y(q^\prime,q_y)=Ly$ in Eq. (\ref{TyM}) takes the form
    \begin{equation}
        L_y = \begin{cases}
            - \frac{C_{1y}}{B_1^\prime \sqrt{-c_y^{\prime\prime}}} h_3(q^\prime) + \frac{C_{2y}}{B_1^\prime \sqrt{-c_y^\prime}} h_2(q_y) & \text{for $A_4^\prime \neq A_y^\prime$} \\
            \frac{1}{B_1^\prime \sqrt{-c_y^{\prime\prime}}} \bigg[ 2A_4^\prime - (B_4^{\prime\prime} + B_1y) + \frac{A_6 - B_3y + (B_4^{\prime\prime} + B_1 y) A_4^\prime - A_4^{\prime 2}}{ 2B_1^{\prime }} \frac{b_y^{\prime\prime}}{c_y^{\prime\prime}} \bigg] h_3(q^\prime) & \text{for $A_4^\prime = A_y^\prime$}
        \end{cases} \, .
    \end{equation}
Given the previous expressions, the multipoles of the collision are given by
    \begin{equation}
       {\cal C}_\ell(z_1) = 
           \frac{{\rm g}_{s}}{2} {\cal K}(z_1)\, . \label{Collison-Deg}
    \end{equation}
Here, unlike Eq. \eqref{collision_cases}, we have added a $1/2$ factor for indistinguishable particles in the final state, since, as in the case of cross sections, one must integrate only over the range of angles corresponding to physically distinguishable events. 

The collision terms ${\cal C}_\ell(z_1)$ substitute ${\cal C}_\ell^{ij}(z_1)$ in equation \eqref{HBE-CN-S} when the interactions are between the same mass eigenstate except for $\nu \bar{\nu} \to \nu \bar{\nu}$, where the contributions come from the $s$- and $t$-channels, so we need to use the case presented in Subsec.~\ref{Different-neutrino-mass}.
\subsection{\texorpdfstring{$\beta_\ell$}{betaell} coefficients for the effective massless neutrino (\texorpdfstring{$m_i \approx 0$}{mi=0})}
The massless neutrinos possess definite helicity, which alters their spin polarization states compared to the massive case. We can use the amplitudes for the massive case (see App.~\ref{Amplitudes}) for relativistic neutrinos, in which the mass of one of the neutrinos is negligible, $m_i \lesssim 10^{-4} \, {\rm eV}$, for temperatures after neutrino decoupling until recombination. Under this approximation, the Boltzmann hierarchy equations can be reformulated by averaging over the non-perturbed energy density. Following the notation in \cite{Ma1995}, we change the symbols for neutrino multipole moments $\nu_\ell$ to ${\cal F}_\ell$. Then, the hierarchy of Boltzmann equations becomes
    \begin{equation}
        \frac{\partial {\cal F}_{\ell}}{\partial \tau} + k \left[ \frac{(\ell +1)}{(2 \ell +1)} {\cal F}_{\ell +1} - \frac{\ell}{(2\ell +1)} {\cal F}_{\ell -1} \right] - 4 f_{g,\ell} = - a \Gamma_\ell(\mu_\varphi) {\cal F}_{\ell} \, .
        \label{BHE-0-CN-S}
    \end{equation}
In both cases, the rate $\Gamma_\ell(\mu_\varphi)$ contains all possible contributions of $\nu_i-\nu_j$ scattering, then~\footnote{For Dirac-like neutrinos, we do not include the ${\rm g}_s$ factor, since this is a toy model and the realistic case could be modeled by a chiral interaction (see, for example,~\cite{Huang2018, He_2020}).}
    \begin{equation}
        \Gamma_\ell(\mu_\varphi) = \frac{g^4 T_\nu}{16} \begin{cases}
            {\rm g}_s \left[ \frac{1}{2} \beta_\ell (\mu_\varphi) + \sum_{j \neq i} \beta_\ell^j(\mu_\varphi) \right] & \text{for Majorana-like neutrinos} \\
            \frac{1}{2} \beta_\ell (\mu_\varphi) + \sum_{j \neq i} \beta_\ell^j(\mu_\varphi)
            &\text{for Dirac-like (anti)neutrinos}\end{cases} \, ,
            \label{collision-term-MvsD}
    \end{equation}
where we define the coefficients $\beta_\ell^j(\mu_\varphi) = \beta_\ell(\mu_\varphi, \mu_j)$ and $\beta_\ell(\mu_\varphi)$, as
    \begin{equation}
        \beta_\ell^j(\mu_\varphi) := \frac{15}{7\pi^7} \int_0^\infty dq_1 q_1^2 \, {\cal K}_\ell^{ij}(q_1) \, , \label{Beta-ell-j}
    \end{equation}
where the integrant is determined by Eq.~\eqref{K-ell-ij} with $\mu_i \to 0$.
Also, as we take the limit $\mu_\nu \to 0$, recall that for the same mass eigenstate, there are additional contributions, and we define 
    \begin{equation}
        \beta_\ell(\mu_\varphi) := \frac{15}{7\pi^7} \int_0^\infty dq_1 q_1^2 \, {\cal K}_\ell(q_1) \, , \label{Beta-ell}
    \end{equation}
but now ${\cal K}_\ell$ determined by \eqref{calA-red}--\eqref{calDl-red}.

Notice that the functions \eqref{Beta-ell-j} and \eqref{Beta-ell} implicitly depend on the reference temperature, since $\mu_\varphi = m_\varphi/T_\nu$ and $\mu_j = m_j/T_\nu$. Also, given that the monopole and dipole moments do not contribute to the collision term $\beta_{0,1}^{(j)} =0$. This will serve as a benchmark to verify that our numerical implementation using VEGAS algorithm~\cite{PETERLEPAGE1978192, Lepage2021adaptive} is working correctly.
\section{Heavy mediator for Self-interacting neutrinos} \label{Sec:Heavy_mediator}
For neutrino--neutrino scattering via a heavy scalar mediator, when $m_\varphi \gtrsim 10^3 \, {\rm eV}$, from Eq.~\eqref{M2}, the scattering amplitude for the same mass eigenstate ($\alpha_r=1$), with our conventions, gets 
    \begin{equation}
        \overline{|{\cal M}|}^2 =
            \frac{1}{2} G_{\rm eff}^2(s^2+t^2+u^2) \, ,\label{Amp.Heavy} 
    \end{equation}
which is valid for both Majorana and Dirac-like neutrinos, where $G_{\rm eff} = g^2/m_\varphi^2$ is an effective coupling as in Fermi theory~\cite{Kreisch2020}. 

This case has been previously studied (see, for example,~\cite{Archidiacono2014, CyrRacine2014, Poudou:2025qcx, He:2025jwp}, where $G_{\rm eff} \sim (10^{-1}, 10^{-7}) \, {\rm MeV}^{-2}$), and we refer to it as the Heavy Mediator Limit (HML) because the four--fermion interaction emerges in the limit $M \gg q^2$, where $q$ is the transferred momentum.

Even in the HML, we may restrict the mediator mass to the range $m_\varphi \sim (10^3, 10^9) \, {\rm eV}$, to maintain the validity of the perturbative expansion, $g \lesssim 0.3$. Since the HML appears here as a particular case of our general analysis in this regime, we cannot, in principle, distinguish between Dirac-like and Majorana neutrinos~\footnote{This stems from our treatment of the Dirac-like case, in which we include light right-handed neutrinos and left-handed antineutrinos.}. In the following, we provide arguments for this statement.
\subsection{The heavy mediator limit} \label{subsec: HML}
\begin{figure} [!ht]
    \centering
    \subfigure[]{\includegraphics[width=76mm]{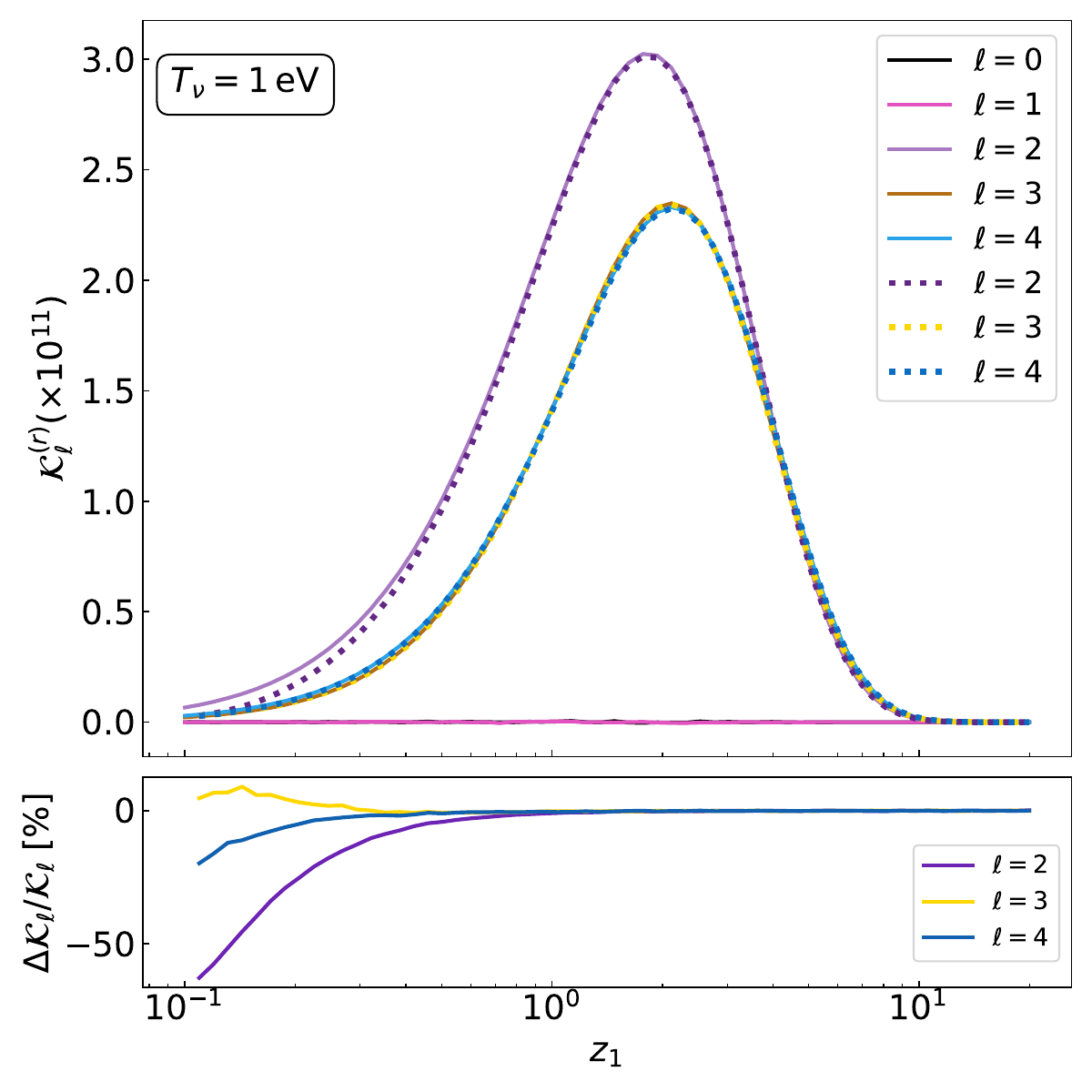}}
    \subfigure[]{\includegraphics[width=76mm]{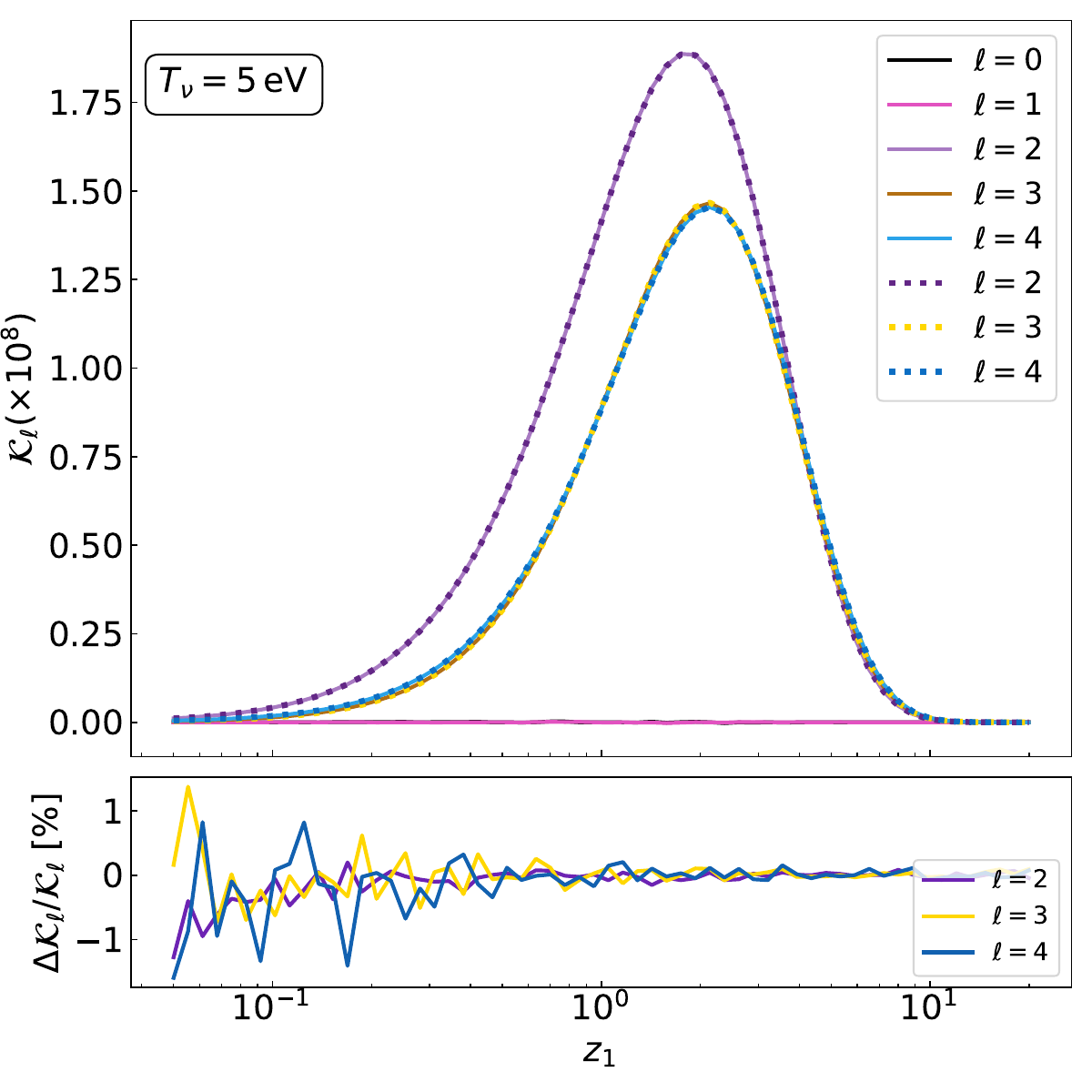}}
    \caption{Massive neutrino effects in the ${\cal K}_\ell(z_1)$ term for elastic scattering involving the same mass eigenstate and a massive mediator ($m_\varphi = 10^3\,\mathrm{eV}$): (a) $T_\nu = 1\,\mathrm{eV}$, (b) $T_\nu = 5\,\mathrm{eV}$. Solid lines correspond to $m_\nu=0$ and dotted lines to $m_\nu=0.1 {\rm eV}$. The percentage deviation with respect to the massless case is also shown for selected non-vanishing multipole contributions. The residual noise arises from the Monte Carlo integration performed with VEGAS.}
    \label{fig:Dirac-massive}
\end{figure}
\begin{figure} [!ht]
    \centering
    \subfigure[]{\includegraphics[width=76mm]{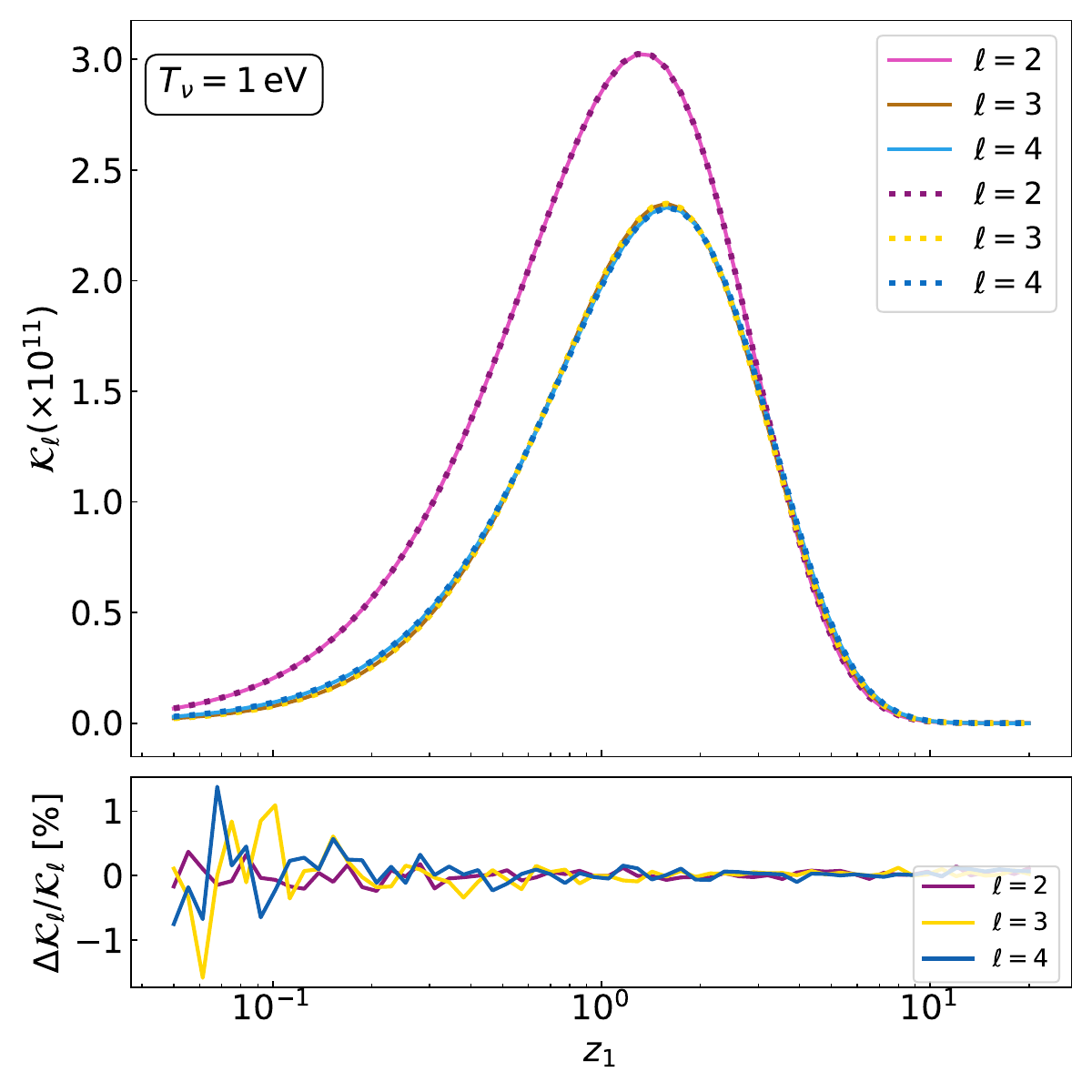}}
    \subfigure[]{\includegraphics[width=76mm]{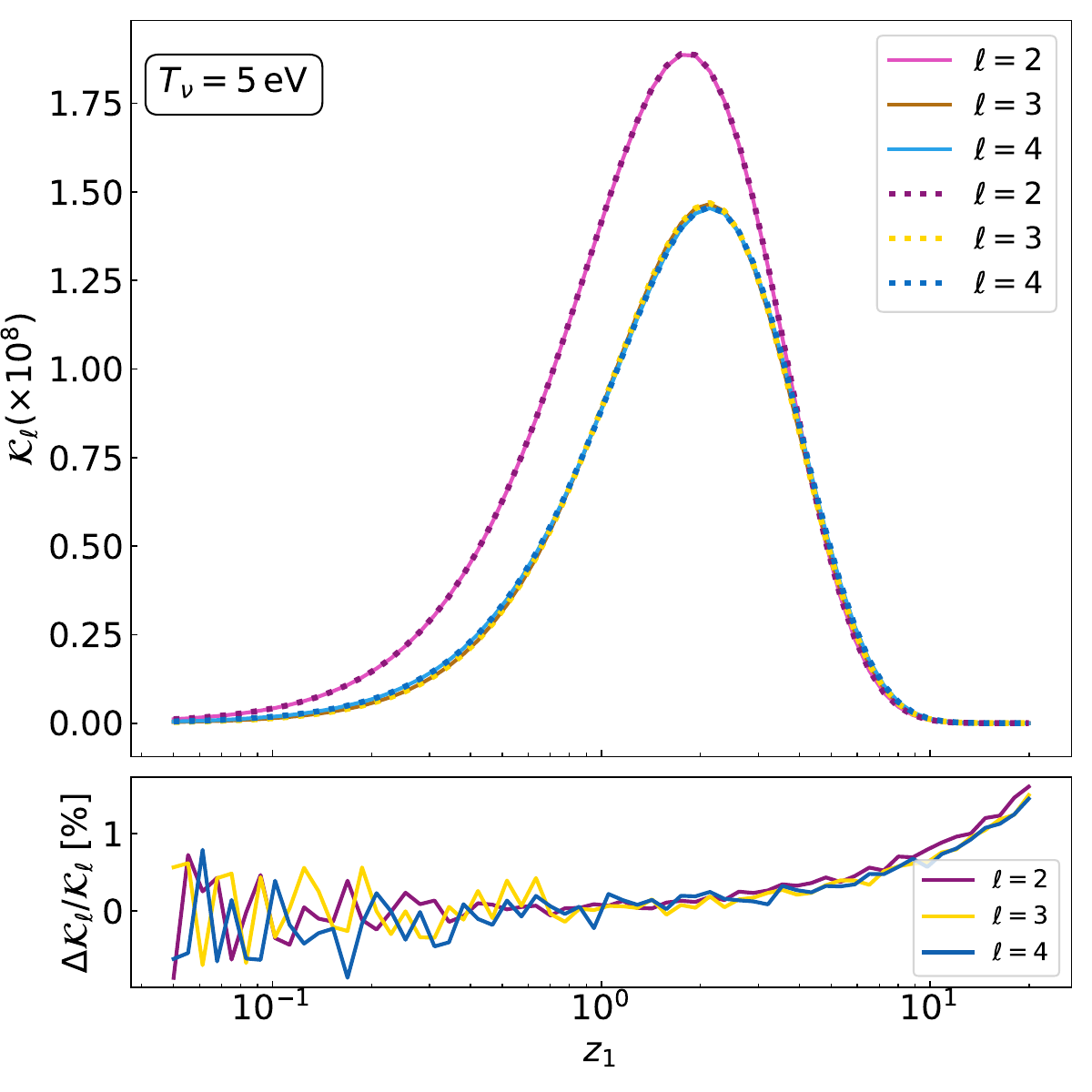}}
    \caption{Neutrino nature effects in the ${\cal K}_\ell(z_1)$-term in elastic scattering for same mass eigenstate, $m_\nu = 0.05 \, {\rm eV}$, where solid lines correspond for Dirac-like and dotted lines for Majorana neutrinos: (a) $T_\nu=1 \, {\rm eV}$, (b) $T_\nu=5 \, {\rm eV}$. The percentage deviation with respect to the Dirac-like case is also shown for selected non-vanishing multipole contributions. The residual noise arises from the Monte Carlo integration performed with VEGAS.}
    \label{fig:Dirac-Majorana-massive}
\end{figure}

Our approach presents a framework where the neutrino masses are free parameters, and their effects are more important for a heavy mediator at low temperatures, because the relevant quantities are the ratios $\mu_\varphi$ and $\mu_\nu$. In Fig.~\ref{fig:Dirac-massive}, we present these possible effects through the term ${\cal K}_\ell(z_1)$. We observe that for neutrino masses close to the observational limits
$m_\nu \lesssim 0.05 \, {\rm eV}$ and even for temperatures as low as
$T_\nu \gtrsim 1 \, {\rm eV}$, with heavy mediator masses ($m_\varphi \gtrsim 10^{3} \, {\rm eV}$), there is no noticeable difference between including or neglecting neutrino masses. Small differences appear for $m_\nu = 0.1 \, {\rm eV}$ in the quadrupole. However, this effect becomes negligible with increasing temperature (see Fig.~\ref{fig:Dirac-massive}). 

Furthermore, in Fig.~\ref{fig:Dirac-Majorana-massive}, we can compare between neutrino nature, notice that in the HML, we cannot distinguish between the neutrino nature, even at low temperatures.
Thus, we can then state that the nature and neutrino mass are irrelevant in the HML.

To assess NSI effects, one typically evolves neutrino perturbations starting from the decoupling of the primordial plasma, $T_\nu\sim 1.5 \, \rm{MeV}$~\cite{Abenza2020precision}, when free streaming would arise in the standard case, while allowing for an NSI-induced tightly-coupled phase that is commonly matched to the HML. In this scenario, the equations that govern the perturbations of density ($\ell = 0$) and velocity ($\ell = 1$) remain unaffected by the interaction; the higher-order moments receive a damping term proportional to the interaction rate $\Gamma$. The effects of the four--fermion interaction occur only within a specific temperature range, 
$T_\nu \lesssim 100 \, {\rm eV}$ ~\cite{Blinov2019}. Before this regime, the system is treated with the tight coupling approximation $\nu_{\ell \geq 2} \to 0$ because the interaction rate satisfies $\Gamma \gg H$ (for instance, $\Gamma \lesssim 10^3 H$ in Ref.~\cite{Kreisch2020}), In contrast, our approach provides a complete description of the neutrino behavior before the HML.

The HML can be determined directly from the expressions~\eqref{calA-red}--\eqref{calDl-red}, neglecting the neutrino mass, only choosing large mediator masses,
or taking the amplitude set in Eq.~\eqref{Amp.Heavy} and compute the new collision term as in Ref.~\cite{Kreisch2020}, but in both cases, we will have the following approximation
    \begin{equation}
        {\cal A}(z_1) \approx \frac{4 T_\nu^4}{m_\varphi^4} 8\pi^3 A(q_1) \, , \quad 
        {\cal B}_\ell(z_1) \approx \frac{4 T_\nu^4}{m_\varphi^4}  8\pi^3 B_\ell(q_1) \quad \text{and} \quad 
        {\cal D}_\ell(z_1) \approx \frac{4 T_\nu^4}{m_\varphi^4} 8\pi^3 D_\ell(q_1) \, .\label{Heavy-approx}
    \end{equation}
The amplitudes take the form \eqref{Heavy-amplitudes-1} and \eqref{Heavy-amplitudes-2}, with the terms $A$, $B_\ell$ and $D_\ell$ given by the integrals summarized in Eqs.~\eqref{A}--\eqref{Dl}. To establish the previous results, we used that $y_-=x_-=1$ in the limit of zero neutrino mass from \eqref{y-limits} and \eqref{x-limits}. Using our approach, we have verified the accuracy of the HML at the level of $\mu_\varphi \gtrsim 10^2$ with desviations $\lesssim 0.6\%$, in this regime, it is a good approximation even for massive neutrinos. 

In order to compare our results with the literature, we define an effective quantity, $\beta_\ell^{\rm HML}$, related directly to the
$\alpha_{\ell}$ from Ref.~\cite{Kreisch2020}~\footnote{Our definition of $\beta_\ell^{HM}$ is equivalent to scaled $\alpha_\ell$, because $q_1 = p_1/T_\nu =  q/T_{0,\nu}$, assuming that $T_{0,\nu}$ is the temperature of background neutrinos and $a_0 = 1$ today.}, as
    \begin{equation}
        \beta_\ell^{\rm HML}(\mu_\varphi) = \frac{4T_\nu^4}{m_\varphi^4} \alpha_\ell \quad \text{with} \quad \alpha_\ell = \frac{120}{7\pi^4} \int_{0}^\infty dq_1 q_1^2 [A(q_1) + B_\ell(q_1)-2D_\ell(q_1)] \, .
    \end{equation}
Note that the authors of~\cite{Kreisch2020} work with massless neutrinos to compute the $\alpha_\ell$ coefficients, since the definition of Eq.~\eqref{ctilde_1} involves an explicit dependence on the energy of the particles participating in the scattering process~\footnote{To reproduce exactly the HML expression used in~\cite{Kreisch2020}, an additional factor ${\rm g}_{s,1}=2$ is introduced on the collision term. However, this extra factor is canceled out after averaging over the neutrino energy density.}. Nonetheless, as previously argued, the effects remain essentially unchanged for massive neutrinos, because to define $\beta_\ell(\mu_\varphi)$ we averaged the collision term at first order over energy density.  

Computing the $\alpha_\ell$-coefficients, we obtain; $\alpha_{0,1}=0$, $\alpha_2 = 0.187$, $\alpha_3 = 0.185$, $\alpha_4 = 0.192$, $\alpha_5 = 0.195$, $\alpha_6 = 0.197$, $\alpha_7 = 0.198$, $\alpha_8=0.198$, $\alpha_9= 0.199$, $\alpha_{\geq 10}= 0.199$. These results differ from those reported in~\cite{Oldengott_2017}, $\alpha_{\ell  \geq 2} \sim (0.4 - 0.48)$, due to their use of the Maxwell–Boltzmann approximation and their absorption of global factors in their definition of $\alpha_\ell$. In the next subsection, we show the self-consistency of our results both within our approach.
\subsection{Beyond the heavy mediator limit} \label{subsec:beyond}

\begin{figure} 
    \centering
    {\includegraphics[width=76mm]{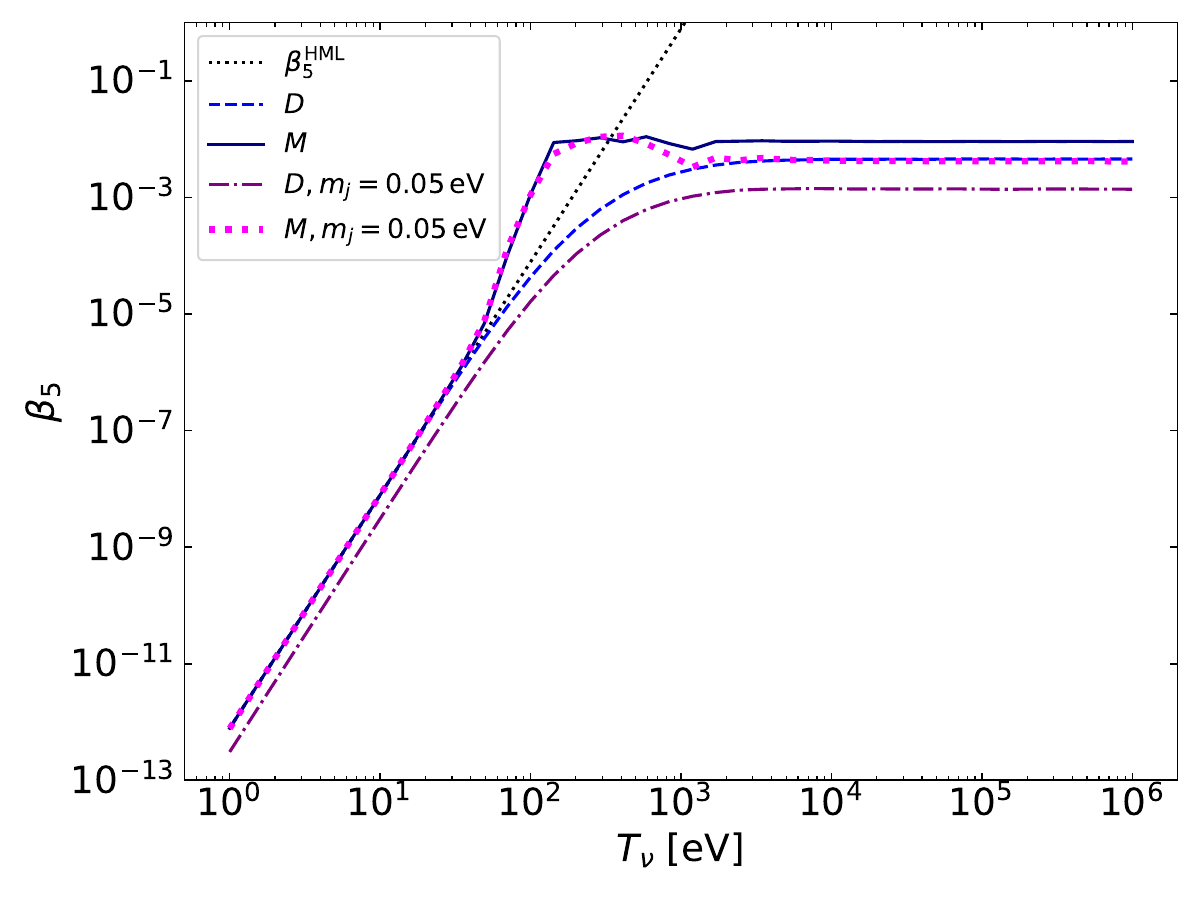}}
    \caption{Relationship between temperature $T_\nu$ and mediator mass 
    $m_\varphi$ for NSI with $\beta_\ell$ computation excluding resonance contribution: Majorana ($M$) vs Dirac-like ($D$) for $m_\varphi = 10^3 \, {\rm eV}$ at $\ell = 5$. In the blue curves, the effectively massless neutrino interacts with itself, while in the purple ones it interacts with another massive neutrino with $m_j=0.05\rm{eV}$.}
    \label{fig:Resonant region}
\end{figure} 

In this subsection, we study the behavior of the rate of interaction of neutrino--neutrino elastic scattering via a heavy mediator, focusing on $\beta_\ell^{(j)}$ elements without taking a hard HML. Specifically, the Dirac-like case involves the scattering of a neutrino ($\nu_i$) with an antineutrino ($\bar{\nu}_j$), exhibiting a behavior similar to $\nu_i-\nu_i$ scattering for Majorana neutrinos. In both cases, there is a deviation that signals the energy range in which the scalar mediator can be produced as a resonance for some values of the incoming energies that contribute to the collision term, but not necessarily for all physically allowed energies; we refer to this as the \emph{resonant regime}. 

Figure~\ref{fig:Resonant region} provides a quantitative insight into the relevance of the resonance. The resonant regime originates from the presence of contributions of the channel $s$; this resonant region, as we preliminarily scanned, lies roughly in the region given by $\mu_\varphi \sim 2\times 10^{-3}- 5\times 10^1$, for effectively massless neutrinos. In Fig.~\ref{fig:Resonant region}, recall that we explicitly subtract the resonant contribution from the $\beta_\ell^{(j)}$ functions by applying the regularization scheme of  Eqs.~\eqref{h1-h3}, \eqref{h6} and \eqref{h4-h5} in which the $s$-channel contribution is set to zero for $B_x^\prime = 0$ (or $B_{x,ij}^\prime =0$).

The resonance can be suppressed by hand in $\nu_i-\nu_j$ scattering by choosing $\alpha_s, \alpha_{st} \ll 1$, or $\alpha_s, \alpha_{st}, \alpha_{su} \ll 1$ for indistinguishable final states. In that limit, the behavior of Majorana neutrinos resembles that of Dirac-like neutrinos. So, the resonance structure and the behavior of the collision integrals are sensitive to the ratio $\mu_\varphi$, and the extremely small neutrino masses through $\mu_\nu, \mu_i, \mu_j$ according to the chosen case~\footnote{However, this sensitivity becomes negligible for the heavy-mediator.}. 

Modeling the resonant region is difficult; however, it has been approximated in Refs. ~\cite{venzor2023resonant, Noriega2025resonant}, and in other contexts for astrophysical neutrinos  ~\cite{Ng2014cosmic, Bustamante2020bounds, Creque2021, Esteban2021_probing}, and has recently gained relevance in scenarios where the lightest neutrino species remains relativistic today~\cite{Wang:2025qap, Machado2025widen}.
Studying in detail the resonant region is beyond the scope of this manuscript, and we expect to give full details in future work.

\begin{figure} 
    \centering
    \subfigure[]{\includegraphics[width=76mm]{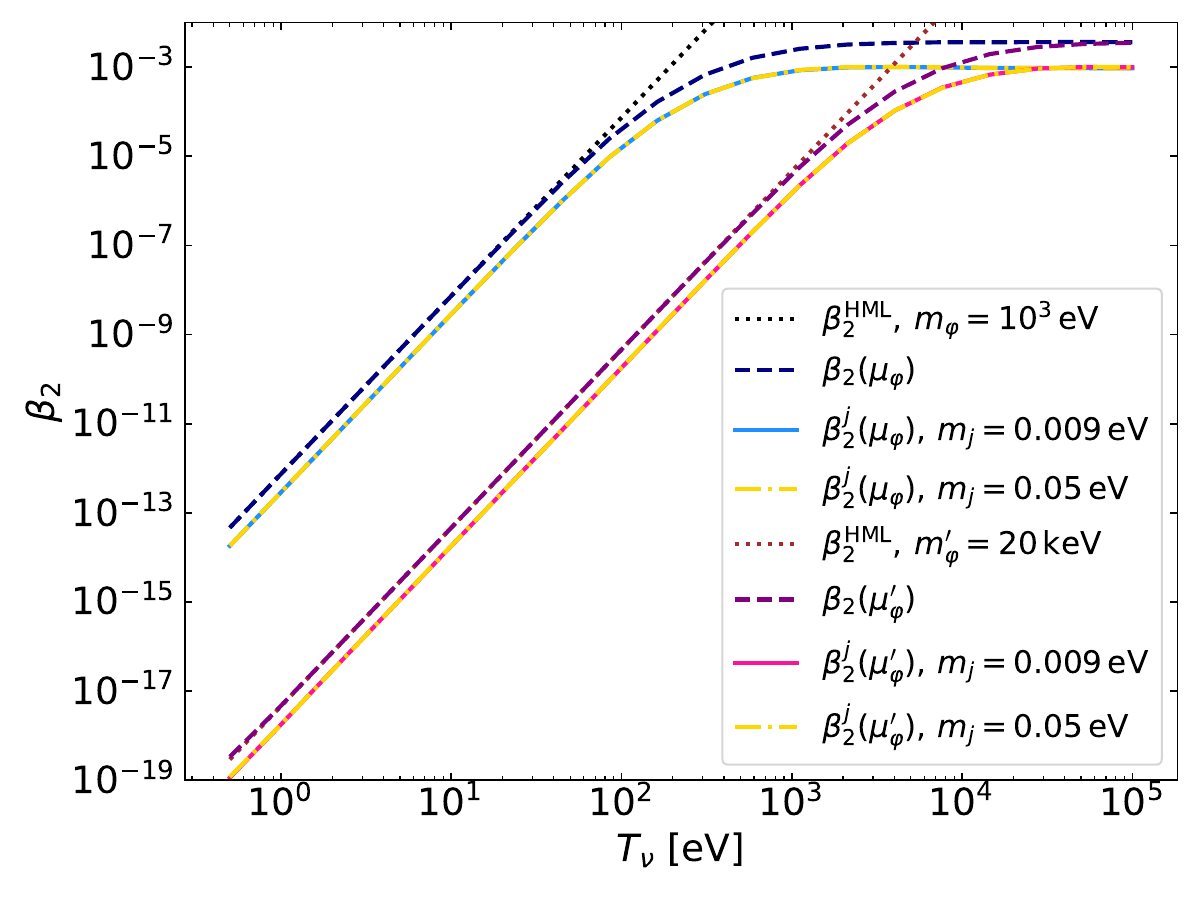}}
    \subfigure[]{\includegraphics[width=76mm]{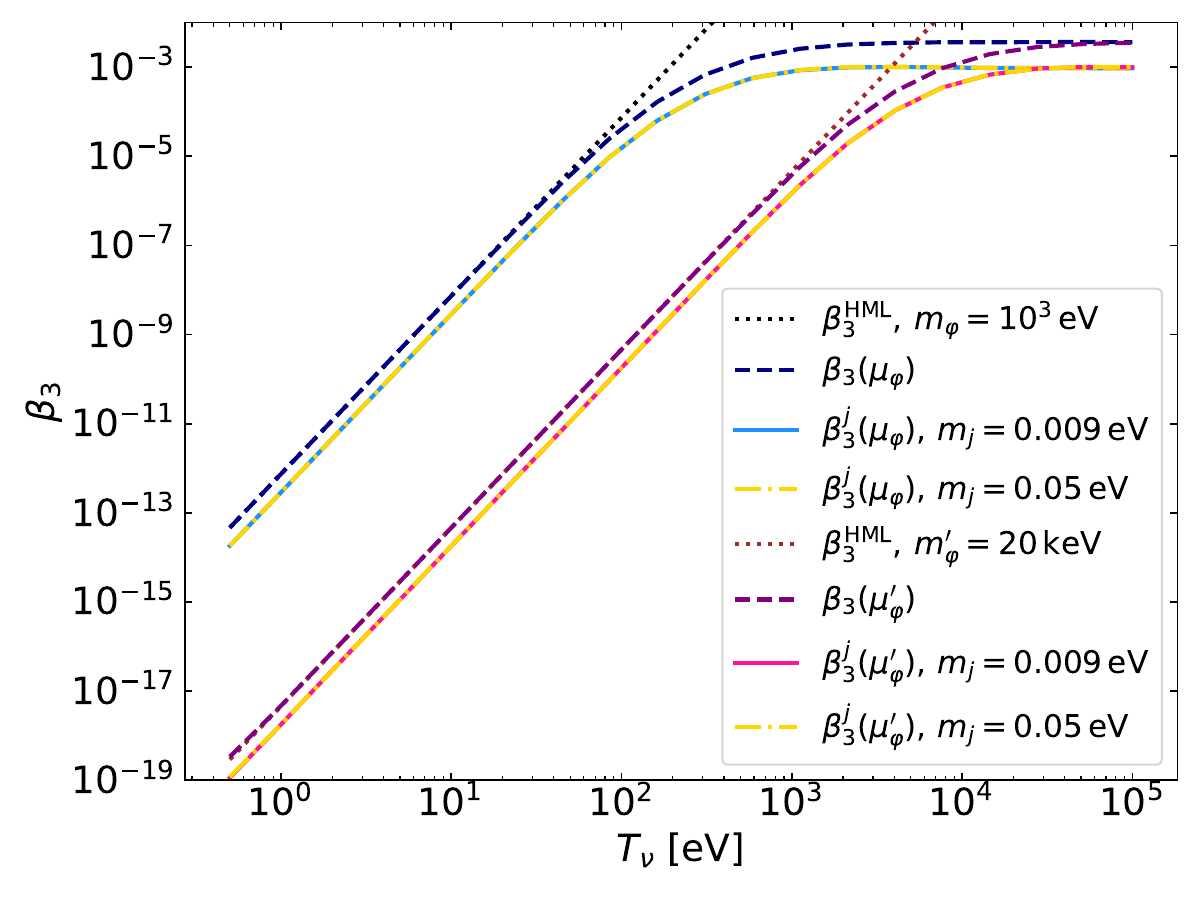}}
    \caption{Comparison between $\beta_\ell(\mu_\varphi)$ for interactions between the same mass eigenstate, $\beta_\ell^{j}(\mu_\varphi)$ for interactions between different mass eigenstates and $\beta_\ell^{\rm HML}(\mu_\varphi)$ in the heavy mediator limit, where $\mu_\varphi=m_\varphi/T_\nu$ and $\mu_\varphi^\prime= m_\varphi^\prime/T_\nu$. For the massive mediator scenario for neutrino--neutrino elastic scattering for Dirac-like neutrinos: (a) $\ell=2$, (b) $\ell=3$.}
    \label{fig:Dirac-betas}
\end{figure}

Now, we focus our discussion on what happens for heavy mediators at high temperatures.
In Fig.~\ref{fig:Dirac-betas}, we compare results of coefficients $\beta_\ell^{(j)}$ and $\beta_\ell^{\rm HML}$ for several temperatures, where we choose representative masses $m_\nu \in \{0.009, 0.05\} \, {\rm eV}$~\footnote{We consider for normal hierarchy $m_1 \lesssim 10^{-4} \, {\rm eV}$, then $m_2 \approx 8.66 \times 10^{-3} \, {\rm eV}$ and $m_3 \approx 5.050 \times 10^{-2} \, {\rm eV}$, while for inverted hierarchy $m_3 \lesssim 10^{-4} \, {\rm eV}$, then $m_1 \approx 4.950 \times 10^{-2} \, {\rm eV}$ and $m_2 \approx 5.025 \times 10^{-2} \, {\rm eV}$ from the values~\cite{de20212020}.}.
We observe a couple of interesting behaviors.
First, we can see that for Dirac-like neutrinos, there is a smooth transition from the HML for low temperatures to a light mediator case where $\beta_\ell^{(j)}$ is effectively a constant at high temperatures.
Second, massive neutrinos are indistinguishable from massless ones. In fact, the visible difference when considering a single mass eigenstate is not due to the value of the mass itself, but rather to the reduced number of available channels that contribute to the scattering process in $\beta_\ell^j$. This simplification suppresses part of the dynamics, making the interaction appear less efficient, even though the underlying phase space remains mostly unaffected by the small mass values.
Furthermore, in figure \ref{fig:Resonant region}, we can see that for different mass eigenstates ($\nu_i \neq \nu_j$), assuming a scenario of universal coupling, we observe a significant difference between Majorana and Dirac cases due to the difference in available scattering channels.

We want to remark that the usual approximation $\Gamma \sim G_{\rm eff}^2 T_\nu^5$ is valid at temperatures of the order $T_\nu \lesssim 10^{-2}m_\varphi$. However, it is worth noting that in the exact computation, the dimensionless quantities $\beta_\ell$ tend toward a constant of order ${\cal O}(10^{-3})$ for temperatures around $T_\nu \gtrsim (0.5 \times 10^3) \, m_\varphi$ for Majorana neutrinos~\footnote{This behavior occurs at temperatures above the resonant region, which spans $T_\nu \in (1/30,500) \, m_\varphi$.}, and $T_\nu \gtrsim 10 \, m_\varphi$ for Dirac-like neutrinos. Where differences from the constant are smaller than $1\%$ in this regime. This behavior indicates a saturation in the interaction rate, reflecting the breakdown of the simple power-law scaling at high energies and highlighting the importance of using full numerical evaluations for a complete study of the heavy mediator. 

The light mediator regime starts when $\mu_\varphi$ drops to a value where $\beta_\ell{(j)}$ becomes constant.
Note that what matters for this behavior is the dimensionless ratio $\mu_\varphi=m_\varphi/T_\nu$, and the constant value itself depends on $\ell$. To arrive at this regime before matter radiation equality (redshift close to $3 \times 10^3$ or $T_\nu\sim 0.5  \, \rm{eV}$) the mediator needs to be lighter than $10^{-3} \rm{eV}$ for Majorana ($m_\varphi < 0.05 \, \rm{eV}$ for Dirac-like neutrinos) according to the magnitudes obtained above.
Remarkably, $m_\varphi \lesssim 10^{-3} \, {\rm eV}$ is the usual assumption reported in the literature as light mediator~\cite{Forastieri:2019cuf, Venzor2022}; this will be discussed further in the next section.

\section{Light mediator limit vs relaxation time approximation}\label{sec:light_mediator}
In the early works of kinetic theory~\cite{Gross1951plasma, Bhatnagar1954model}, a simplified kinetic model was introduced to overcome the mathematical difficulties posed by the Boltzmann equation in describing collision processes in gases. They proposed replacing the complex collision integral with a relaxation term, which reflects the
physical idea that collisions tend to relax the system toward equilibrium over a characteristic
time scale, today known as the relaxation time approximation (RTA).

For the light mediator, $m_\varphi \lesssim 10^{-3} \, {\rm eV}$, studies of $\nu-\nu$ scattering~\cite{Hannestad2005} have used the RTA, where the effect of interactions is modeled by driving the distribution function $f_{\nu_i}$ toward its equilibrium form $f_{\nu_i}^{(0)}$ over a characteristic timescale $t_c$. In this approximation, the collision term takes the simple form:
    \begin{equation}
        \left.\frac{df_{\nu_i}}{d\tau}\right|_{\text{coll}} = - a\frac{f_{\nu_i} - f_{\nu_i}^{(0)}}{t_c} = - a \frac{f_{\nu_i}^{(0)}\Theta_{\nu_i}}{t_c} \, .
        \label{RTA}
    \end{equation}
Here, $t_c$ encodes the microphysics of the interaction and is typically related to the interaction rate $\Gamma \sim 1/t_c$~\footnote{The scale factor, $a(\tau)$, appears in the RHS of Eq.~\eqref{RTA} because we are using conformal time $\tau$.}. This approach allows us to track the implementation in the Boltzmann hierarchy, where damping of higher multipoles reflects the suppression of anisotropic stress due to efficient scattering. A commonly used approximation is $\Gamma \sim n \langle \sigma v\rangle$ ~\cite{Hannestad2000self, Forastieri:2019cuf},
where $n$ is the number density of the particles that scatter with the species that we are currently studying, and $\langle \sigma v\rangle$ is their thermally averaged cross-section. This expression provides a simple but powerful estimate for comparing interaction rates with the Hubble expansion rate, and is widely used in the context of early Universe processes such as decoupling, freeze-out, and thermalization.

Following the discussion presented in~\ref{appE}, the approximation is only valid for $\ell \geq 2$, and then the right side of Eq.~\eqref{BHE-0-CN-S} is equal to $-a \Gamma_{\nu\nu \to \nu\nu} {\cal F}_\ell$, according to Forasteri et al.~\cite{Forastieri:2015paa,Forastieri:2019cuf}:
    \begin{equation}
        \Gamma_{\nu\nu \to \nu\nu} = \frac{{\rm g}_s}{2 \pi^2} \frac{3}{2} \zeta(3) g_{\rm eff}^4 T_\nu \, ,
    \end{equation}
where $g_{\rm eff} = \xi^{1/4} g$, $\xi$ accounts for the lack of knowledge about the collision computation and/or thermal effects. We can compare this approximation with our case in Eqs.~\eqref{BHE-0-CN-S} for a light mediator, within the thermal approximation approach that we adopted. 
This procedure can lead us to determine the previously unknown value of $\xi$~\footnote{Here, we do not include the indistinguishability factor $1/2$ in the Majorana case.}. We focus on the region where resonances do not appear for $T_\nu \gtrsim 0.5 \, {\rm eV}$,
    \begin{equation}
        \Gamma_{\nu\nu \to \nu\nu} \to {\rm g}_s \frac{g^4}{16} \beta_\ell(\mu_\varphi) T_\nu \, .
    \end{equation}
In the case of effectively massless neutrinos, $m_\nu \lesssim {\cal O} (10^{-4} \, {\rm eV})$, this suggests that we need to replace $\xi \to \xi_\ell$.
All $\beta_\ell$'s have constant values but are $\ell$-dependent; then, a detailed study of this regime with our approach could, in principle, slightly modify the current constraints on the coupling~\cite{Forastieri:2019cuf, Venzor2022}. Therefore, we have the following correspondence:
    \begin{equation} 
        \xi_\ell \equiv \frac{\pi^2}{12 \zeta(3)} \beta_\ell  \, . \label{Xi-ell}
    \end{equation}
\begin{table}
\centering
\caption{Values of $\xi_\ell$ for $\ell$-multipoles in $\nu-\nu$ scattering.}
\begin{tabular}{|c|c|c|c}
\hline
$\ell$ & $\xi_\ell$ $[10^{-3}]$ (Majorana) & $\xi_\ell [10^{-3}]$ (Dirac-like) \\
\hline
2  & $3.454$ & $1.728$ \\
3  & $4.979$ & $2.491$ \\
4  & $5.765$ & $2.883$ \\
5  & $6.216$ & $3.106$ \\
6  & $6.480$ & $3.242$ \\
7  & $6.655$ & $3.329$ \\
8  & $6.782$ & $3.389$ \\
9  & $6.867$ & $3.432$ \\
10 & $6.925$ & $3.466$ \\
\hline
\end{tabular}
\label{tab:beta-xi}
\end{table}

In Table~\ref{tab:beta-xi}, we show the first $\xi_\ell$ values up to $\ell=10$, we observe a saturation for $\ell \geq 20$ and differences are smaller than $1 \, \%$.
Furthermore, with the convention in Dirac-like neutrinos (degenerate in neutrino/antineutrino mass), we have $\xi_\ell \, ({\rm Majorana})\approx 2\xi_\ell \,$(Dirac-like). This factor is due to $s$-channel suppression.
\begin{table}
\centering
\caption{Values of $\beta_\ell^j$ for $\ell$-multipoles in $\nu_i-\nu_j$ scattering with $i\neq j$.}
\begin{tabular}{|c|c|c|c}
\hline
$\ell$ & $\beta_\ell^j$ $[10^{-3}]$ (Majorana) & $\beta_\ell^j [10^{-3}]$ (Dirac-like) \\
\hline
2  & $2.530$ & $0.843$ \\
3  & $3.638$ & $1.213$ \\
4  & $4.213$ & $1.405$ \\
5  & $4.536$ & $1.513$ \\
6  & $4.734$ & $1.579$ \\
7  & $4.862$ & $1.621$ \\
8  & $4.951$ & $1.650$ \\
9  & $5.014$ & $1.671$ \\
10 & $5.062$ & $1.687$ \\
\hline
\end{tabular}
\label{tab:beta-ell}
\end{table}

The current constraints on a light mediator for Majorana neutrinos in $\nu-\nu$ scattering require an effective coupling 
$g_{\rm eff} < 1.94 \times 10^{-7}$, based on Planck 2018 data~\cite{Venzor2022}, which translate for a single process to $g < 9.52 \times 10^{-7}$ for $\ell=2$ or $g < 8 \times 10^{-7}$ for $\ell=20$, reflecting that 
$\xi_\ell^{1/4} \sim \mathcal{O}(10^{-1})$. However, these bounds can be updated by including a complete collision term that accounts for additional processes, as $\nu_i \nu_j \to \nu_i \nu_j$, as shown in Eq.~\eqref{collision-term-MvsD}; moreover, we showed that the contributions are $\ell$-dependent. 

When we incorporate the scattering between different neutrino mass eigenstates $\nu_j \neq \nu_i$ under the assumption of universal coupling, new contributions arise that modify the scattering rate, because $\beta_\ell^j \, ({\rm Majorana}) \approx 3\beta_\ell^j \,$(Dirac-like), see Table~\ref{tab:beta-ell}, where the neutrino mass of the non-reference neutrino, $m_j \lesssim 0.06 \, {\rm eV}$, is negligible for temperatures $T_\nu \gtrsim 1\, {\rm eV}$. 

Furthermore, the available channels play a critical role in the collision term. From Eq.~\eqref{collision-term-MvsD}, we do not take into account processes such as $\nu_i\nu_j \to \nu_k \nu_l$. However, in the regime where neutrino masses do not contribute significantly, their contribution is the same as in the process $\nu_i\nu_j \to \nu_i \nu_j$. This highlights the importance of a comprehensive treatment that includes mass–eigenstate mixing and all relevant processes to constrain light-mediator scenarios in neutrino self-interactions accurately.
\section{Summary and conclusions} \label{sec:Conclusions}
Throughout this paper, we have presented a novel calculation of the neutrino collision term for neutrino--(anti)neutrino scattering mediated by a scalar particle at first order in cosmological perturbation theory, within the thermal approximation. Our framework is versatile and can be extended to describe any pair of massive self-interacting particles whose collision term arises from elastic (and, in some cases, inelastic) scattering processes.

Along the present paper, we have revisited Boltzmann hierarchy equations for massive neutrinos and computed our most general form of the collision term, which in general distinguishes between mass eigenstates and whether neutrinos are Dirac-like or Majorana particles.
Furthermore, our framework includes both the neutrino mass and the mediator mass as free parameters.
From there, we systematically distill the three approximations used commonly in the literature: the light, heavy, and resonant mediators. 
In each case, we compared our results with previous results and showed the advantages and limitations of our approach.

As it comes forward from our constructions and findings, the series of initial questions we posed along the introduction to frame the scope of our work do find answers in our present approach. First of all, we have shown along Subsection~\ref{subsec:beyond} that it is indeed possible to properly compute the collision term of NSI for a wide range of neutrino temperatures using a free mediator mass, at least for the non-resonant interaction regimes, since, in resonant cases, it seems numerically challenging to get reliable results. 
On the other hand, as already noted, the main differences that Dirac or Majorana-like neutrino nature implies are primarily determined by the available NSI interaction channels (see Figure~\ref{Scattering with neutrinos}). 
We have, therefore, adopted a Dirac-like model, since neutrino--neutrino scattering between the same mass state reproduces the HML in the same way
as in the Majorana case; nevertheless, one should keep in mind that the neutrino nature should become important when resonances are considered, which lies beyond the scope of this paper.
From our results, we would also like to underline that, as expected, neutrino masses can be neglected at sufficiently high temperatures; however, as they become relevant at lower temperatures, they suppress the collision term. 
Moreover, we should remark that in this work, we have succeeded in providing a careful calculation for  thermal corrections on the light mediator case, without using 
additional approximations, as demonstrated in Sec. 5.

The commonly used HML is a popular framework that has the potential to address cosmological tensions. However, this approach has previously relied on the effective coupling $G_{\rm eff}$ combined with a tight coupling approximation at high temperatures. The latter is particularly well motivated, as previous computations did not permit a systematic exploration of the coupling strength and the mediator mass. In contrast, our work allows exploration of the parameter space using the fundamental coupling $g$ directly, enabling explicit tests of the mediator mass $m_\varphi$. We believe this is the most important result of our analysis, and plan to present results with cosmological observables in the future.

Here, we showed that the neutrino mass does not affect the Boltzmann collision term for $m_\nu \lesssim 0.06 \, {\rm eV}$ for redshifts larger than $1.8 \times 10^4$. We also relaxed the heavy mediator validity condition down to $m_\varphi \geq 10^{2} \, {\rm eV}$, and showed that Dirac-like neutrinos do not exhibit resonance production in $\nu-\nu$ scattering, but they do in the $\nu_i - \bar{\nu}_j$ scattering. Modeling the resonant regime for Majorana neutrinos remains challenging since it requires the inclusion of one-loop corrections for a complete amplitude description. We anticipate that a full understanding of these corrections will allow for a thorough exploration of heavy mediators in NSI in future work.

In Sec.~\ref{sec:light_mediator}, we presented an exact treatment for the light mediator regime, establishing that for $m_\varphi \lesssim 10^{-3} \, {\rm eV}$ the mediator mass and neutrino mass ($m_j \lesssim 0.06 \, {\rm eV}$) can be safely neglected according to our numerical analysis for $z \gtrsim 6 \times 10^3$ ($T_\nu \gtrsim 1\rm{eV}$).
To conclude, we were able to find the correct value of $\xi_l$, which is typically codified into $g_{\rm eff}=\xi^{1/4} g$ that was not previously computed, and find the multipolar dependence on $\ell$. All of our results would be useful for future cosmological data analysis of neutrino self-interactions.

Another important feature is that interactions such as \eqref{L-int-gen} can produce scalar particles through the process $\varphi$, via $\nu\nu \to \varphi \varphi$ and $\nu\nu \to \varphi$ when the temperature is comparable to $m_\varphi$. This may invalidate the assumption $C^{(0)}=0$ (equivalently, $Q^{(0)} = 0$), or at least render its contribution subdominant rather than negligible. We plan to address this effect and provide a more accurate description in future work.

Looking ahead, our future research will focus on improving the precision of neutrino self-interaction constraints by implementing this formalism. Additionally, connecting this theoretical development with upcoming cosmological and astrophysical observations will be crucial to test the viability of mediator models and potentially distinguish between Majorana and Dirac neutrinos.
New sets of data, which will be available in the coming years, will play an important role in distinguishing the signal from statistical fluctuations (see, for instance, \cite{Pal_2025, Libanore2025}).
Furthermore, if a self-interaction signal is to be detected, then it would be necessary to find the correct model and the other phenomenological implications that may arise in experiments \cite{He_2020, Berbig2020, Lyu2021}.
Such advancements will deepen our understanding of neutrino physics and its role in the evolution of the universe.

\acknowledgments
The authors thankfully acknowledge Cluster de Supercómputo Xiuhcoatl (Cinvestav) for allocating computer resources. I.P.C. thanks SECIHTI for the scholarship as ``Ayudante de Investigador SNII III'', also thanks Pablo Roig for their valuable discussions during this work.
Work partially supported by Conacyt, Mexico, under FORDECYT-PRONACES grant No. 490769.  G.G.A. and JV acknowledge support from SECIHTI post-doctoral fellowships.

\appendix
\section{The relativistic Boltzmann equation: perturbation of the distribution function near the thermal equilibrium} \label{appA}
The metric perturbation $h_{ij}$ can be decomposed into a trace part and a traceless part; The trace part is defined as $h = {h^i}_{i}$, and the traceless part consists of three components: $h^{(k)}_{ij}$, $h^{(\perp)}_{ij}$, and $h^{(T)}_{ij}$. Thus, we write $h_{ij} = h \, \delta_{ij}/3 + h^{(k)}_{ij} + h^{(\perp)}_{ij} + h^{(T)}_{ij}$. By definition, the divergences of $h^{(k)}_{ij}$ and $h^{(\perp)}_{ij}$ correspond to longitudinal and transverse vectors, respectively, while $h^{(T)}_{ij}$ is fully transverse. The components of $h_{ij}$ satisfy the following conditions: $\epsilon_{ijk} \, \partial_j \partial^l h^{(k)}_{lk} = 0$, $ \partial^i \partial^j h^{(\perp)}_{ij} = 0 \, , \partial^i h^{(T)}_{ij} = 0 \, .$
It follows that $h^{(k)}_{ij}$ can be expressed in terms of a scalar field $\chi$, and $h^{(\perp)}_{ij}$ in terms of a divergenceless vector field $A_i$, as
    \begin{equation}
        h^{(k)}_{ij} = \left( \partial_i \partial_j - \frac{1}{3} \delta_{ij} \nabla^2 \right) \chi \, , \quad 
        h^{(\perp)}_{ij} = \partial_i A_j + \partial_j A_i \, , \quad \partial^i A_i = 0 \, .
    \end{equation}
The two scalar fields $h$ and $\chi$ (or $h^{(k)}_{ij}$) characterize the scalar modes of the metric perturbations. The vector $A_i$ (or $h^{(\perp)}_{ij}$) describes the vector modes, and the transverse tensor $h^{(T)}_{ij}$ encodes the tensor modes (i.e., gravitational waves). With the scalar part of the metric perturbation, $h_{ij}$, is described in the Fourier space through two fields, $h=h({\bf k},\tau)$ and $\eta=\eta({\bf k},\tau)$,
    \begin{equation}
        h_{ij}({\bf x}, \tau)= \int d^3k e^{i{\bf k}\cdot{\bf x}} \left\{ \hat{k}_i \hat{k}_j h + \left(\hat{k}_i \hat{k}_j-\frac{1}{3}\delta_{ij}\right)6 \eta\right\} \, , \quad {\bf k}=k\hat{k} \, . \label{synch_potentials}
    \end{equation}
This last represents local gravitational potentials coupled to matter through Einstein's equations, which bind the evolution of both.

Using the perturbation given in Eq.~\eqref{f_def}, we can divide the Relativistic Boltzmann equation into parts: the zero order given in Eq.~\eqref{Q-zero}, and the first order that depends on the gauge choice, for CN gauge from Eq.~\eqref{Q-CN},

    \begin{equation}
        f^{(0)}_\alpha \left[ \frac{\partial \Theta_\alpha}{\partial \tau}  + \frac{{\bf p}}{E} \cdot \nabla \Theta_\alpha \right] + p \frac{\partial  f^{(0)}_\alpha}{\partial p} \left[ \frac{\partial \Phi}{\partial \tau} - \frac{E}{p^2} {\bf p} \cdot \nabla \Psi \right] = \frac{Q^{(1)}_\alpha}{{\rm P}^0} \, .
    \end{equation}
While the relativistic Boltzmann equation in the Synchronous gauge is
    \begin{equation}
        f^{(0)}_\alpha \left[ \frac{\partial \Theta_\alpha}{\partial \tau}  + \frac{{\bf p}}{E} \cdot \nabla \Theta_\alpha \right] - \frac{1}{2} \hat{p}^j \hat{p}^l p \frac{\partial  f^{(0)}_\alpha}{\partial p} \left( \frac{\partial h_{jl}}{\partial \tau} \right) =  \frac{Q^{(1)}_\alpha}{{\rm P}^0} \, .
    \end{equation}
In both cases $f^{(0)}= f^{(0)}(E(p))$. We neglected terms like $\frac{\partial \Theta_\alpha}{\partial p}$, and consider
    \begin{equation}
        \left[\frac{\partial}{\partial \tau} f^{(0)} - {\cal H} p \frac{\partial}{\partial p} f^{(0)}\right] \Theta_\alpha = \frac{Q_\alpha^{(0)}}{{\rm P}^0} \Theta_\alpha = 0 \, .
    \end{equation}
To solve the RBE for massive particles, work in Fourier space, taking the Fourier transform with the convention $\tilde{\chi}({\bf k}, \tau) = (2\pi)^{-3} \int d^3 x e^{-i {\bf k} \cdot {\bf x}} \chi({\bf x}, \tau)$ and we can write in the CN gauge
    \begin{equation}
        f_{\alpha}^{(0)}(q, \tau) 
        \left[ \frac{\partial \tilde{\Theta}_\alpha}{\partial \tau} + i \frac{q}{\epsilon} \upmu k \tilde{\Theta}_\alpha \right]
        + q \frac{\partial f_{\alpha}^{(0)}(q, \tau)}{\partial q} 
        \left[ \frac{\partial \phi}{\partial \tau} - i \frac{\epsilon}{q} \upmu k \psi \right] = a \tilde{C}^{(1)}_{\alpha}[q] \, .
        \label{C1-CN}
    \end{equation}
Here $\upmu = \hat{p} \cdot \hat{k}$ encodes the angular dependence of $\tilde{\Theta}_\alpha$ and $\tilde{C}^{(1)}_\alpha$ is the rescaled collision term to linear order on $\tilde{\Theta}$ in the Fourier space. Again, as in the CN gauge, we introduce the perturbations as \eqref{f_def}. As a consequence, the RBE, in Fourier space, becomes
    \begin{equation}
        f_\alpha^{(0)}(q, \tau) 
        \left[ \frac{\partial \tilde{\Theta}_\alpha}{\partial \tau} + i \frac{q}{\epsilon} \upmu k \tilde{\Theta}_\alpha \right]
        + q \frac{\partial f_\alpha^{(0)}(q, \tau)}{\partial q} 
        \left[ \dot{\eta} - \frac{\dot{h} + 6\dot{\eta}}{2} \upmu^2 \right] = a\tilde{C}_\alpha^{(1)}[q] \, ,
        \label{C1-S}
    \end{equation}
where $\dot{\eta} = \frac{\partial \eta}{\partial \tau} ({\bf k}, \tau)$, $\dot{h} = \frac{\partial h}{\partial \tau} ({\bf k}, \tau)$, and $\tilde{C}^{(1)}$ is the collision term to linear order on $\tilde{\Theta}$.
\section{Collision integral reduction for \texorpdfstring{$\alpha - \beta$}{nunu} elastic scattering processes} \label{appB}
Collision term from scattering of the form $\alpha(P_1) + \beta(P_2)\rightarrow \alpha(P_3) + \beta(P_4)$ 
have a generic distribution factor written as
    \begin{equation}
    \begin{aligned}
        F({\bf x}, {\bf p}_1, {\bf p}_2, {\bf p}_3,{\bf p}_4,\tau) &= f_\alpha({\bf p}_3) f_\beta({\bf p}_4)[1+ \sigma_1 f_\alpha({\bf p}_1)][1+\sigma_2 f_\beta({\bf p}_2)] \\
        &\quad - f_\alpha({\bf p}_1)f_\beta({\bf p}_2)[1+ \sigma_3 f_\alpha({\bf p}_3)][1 + \sigma_4 f_\beta({\bf p}_4)] \, ,
        \label{FS}
    \end{aligned}
    \end{equation}
where $\sigma_{j}$ stands for $+1$ ($-1$) for bosons (fermions). By assuming that the phase space distribution function deviates slightly from
the equilibrium ones, Eq.~\eqref{f_def}, to linear order on the perturbations one gets 
    \begin{equation}
        F^{(1)} = \Psi_S 
        \left[2K_3\Theta({\bf p}_3) - K_1\Theta({\bf p}_1) - K_2\Theta({\bf p}_2)\right] \,,
        \label{FS2p}
        \end{equation}
for identical particles, while
\begin{equation}
        F^{(1)} = \Psi_S
         [K_3\Theta({\bf p}_3) + K_4\Theta({\bf p}_4) - K_1\Theta({\bf p}_1) 
        \, - K_2\Theta({\bf p}_2)] \,,
        \label{FS2}
    \end{equation}
for different particles in the final state.
We have dropped the implicit $\tau$ and ${\bf x}$ dependence; here, the $K_n$ terms are determined by

    \begin{align}
        K_1 &= \exp\big\{[(p_3 + p_4) - (p_1 + p_2)]/T \big\} -  \sigma_1 e^{-p_1 /T} \, ,\\
        K_2 &= \exp\big\{[(p_3 + p_4) - (p_1 + p_2)]/T \big\} -  \sigma_2 e^{-p_2 /T} \, , \\
        K_3 &= 1 - \sigma_3 \exp\big\{  [p_4 -(p_1 + p_2)]/T \big\} \, , \\
        K_4 &= 1 - \sigma_4 \exp\big\{  [p_3 -(p_1 + p_2)]/T \big\} \, ,
    \end{align}
where $p_4$ and $p_3$ are determined by energy conservation, while $\Psi_S$ is given by
    \begin{equation}
        \Psi_S = \Psi_S(p_1,p_2,p_3,p_4)  = 
        e^{(p_1+ p_2)/T}f_\alpha^{(0)}(p_1)f_\beta^{(0)}(p_2)f_\alpha^{(0)}(p_3)f_\beta^{(0)}(p_4) \, .
        \label{Psi_S}
    \end{equation}
The general form of $\Psi_S$ is valid for any elastic scattering processes
$\alpha(P_1) + \beta(P_2)\rightarrow \alpha(P_4) + \beta(P_3)$ and 
annihilation processes as $\alpha + \alpha \to \beta + \beta$, since the necessary and sufficient condition to establish this expression is $\sigma_1 \sigma_2 = \sigma_3 \sigma_4$, though, we will not analyze inelastic processes here~\footnote{In certain inelastic scattering processes, the same condition on sigmas applies, such as in $\nu_i \nu_j \to \varphi\varphi$ or $\nu_i \nu_j \to \nu_k \nu_l$.}. Note that the double contribution of $\Theta({\bf p}_3)$ in Eq.\eqref{FS2p} comes from
$\Theta({\bf p}_4)$ after the explicit use of the ${\bf p}_4 \leftrightarrow {\bf p}_3$ symmetry for identical particles in the final state. The corresponding rescaled collision term in Fourier space then behaves as follows
    \begin{equation}
        \tilde{C}^{(1)}_{\alpha}[{\bf k}, {\bf p}_1, \tau] = {\cal J}_3({\bf k}, p_1, \tau) + {\cal J}_4({\bf k}, p_1, \tau) - {\cal J}_2({\bf k}, p_1, \tau) - {\cal J}_1({\bf k}, p_1, \tau) \, .
    \end{equation}
Here, we have the general integral form
    \begin{equation}
        {\cal J}_n ({\bf k}, p_1, \tau) = 
        \frac{1}{2 (2\pi)^5 E_1 }\int\!\frac{d^3p_2}{ 2E_2}\frac{d^3p_3}{ 2E_3}\frac{d^3p_4}{2E_4} \delta^{(4)}(P_1+P_2 - P_3 - P_4) G_n(p_n)~,
        \label{Gen.Jn}
    \end{equation}
where $G_n = \overline{|{\cal M}|}^2 \Psi_S K(E_n) \tilde{\Theta}({\bf k},{\bf p}_n,\tau)$ is a function of all individual momenta ${\bf p}_n$, for $n= 1,\dots,4$ and $\overline{|{\cal M}|}^2$ standing for the squared invariant amplitude (averaged over initial spin states and summed over final spin states, one must include symmetry factors that arise due to identical particles in the final states) of the process and the distribution factor to first order $\tilde{F}^{(1)} = \tilde{F}^{(1)}({\bf k},{\bf p}_1, {\bf p}_2, {\bf p}_3, {\bf p}_4,\tau)$ is now defined in terms of the Fourier transformed
perturbations $\tilde{\Theta}({\bf k},{\bf p}_n,\tau)$. Notice that the perturbation can be extracted from the ${\cal J}_n$ integral.

\begin{figure} [ht!]
    \centering
    \subfigure[]{\includegraphics[width=55mm]{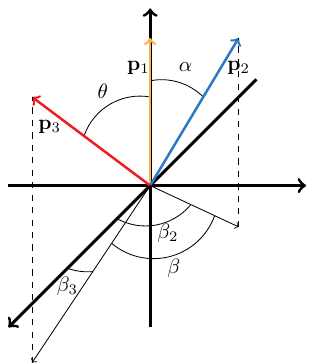}}
    \subfigure[]{\includegraphics[width=55mm]{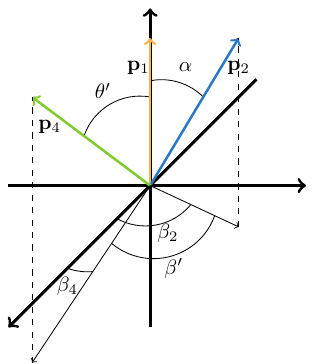}}
    \caption{Customary orientation according to \eqref{angles}: (a) for first identity, (b) for second identity.}
    \label{fig: Customary orientation}
\end{figure}

Next, we must correctly identify the ordering and orientation of the particle momenta for each term in the amplitude. The following steps remain general, including for bosons; however, we focus on analyzing the specific case of two neutrino eigenstates labeled $i$ and $j$. We start by using the mass on-shell identities for $\nu_i - \nu_j$ elastic scattering~\footnote{A similar approach was adopted previously in~\cite{Yueh1976scattering} for use in supernova calculations and for neutrino decoupling in the early universe~\cite{ Hannestad1995neutrino}.}, 
    \begin{equation}
        \frac{d^3p_4}{2E_4} = d^4 P_4\delta(P_4^2-m_j^2)H(E_4 - m_j)\,, \quad \text{and} \quad \frac{d^3p_3}{2E_3} = d^4 P_3\delta(P_3^2 - m_i^2)H(E_3 - m_i) \, . \label{identities}
    \end{equation}
Furthermore, we take ${\bf p}_1$ as a fixing orientation direction, taken along the $\mathrm{z}$-axis, such that 
    \begin{equation}
    \begin{array}{cc}
        \hat{{\bf p}}_1 \cdot \hat{{\bf p}}_2 = \cos\alpha = x \, , & \hat{{\bf p}}_1 \cdot \hat{{\bf p}}_4 = \cos\theta^\prime = z\,, \\
        \hat{{\bf p}}_1 \cdot \hat{{\bf p}}_3 = \cos\theta = y \, , & \hat{{\bf p}}_2 \cdot \hat{{\bf p}}_4 = \cos\alpha\cos\theta^\prime + \sin\alpha\sin\theta^\prime \cos\beta^\prime \, , \\
        \hat{{\bf p}}_2 \cdot \hat{{\bf p}}_3 = \cos\alpha\cos\theta + \sin\alpha\sin\theta\cos\beta \, . &
    \end{array} \label{angles}
    \end{equation}
with $\beta$ ($\beta^\prime$) the ${\bf p}_2$ azimuth angle, measured with respect to the projection of ${\bf p}_3$ ($p_4$) onto the $\mathrm{x}-\mathrm{y}$ plane.

Hence, we take $d^3p_2 = p_2^2dp_2dx d\beta$ ($d^3p_2 = p_2^2dp_2dx d\beta^\prime$), whereas
$d^3p_3 = p_3^2dp_3dy d\phi$ ($d^3p_4 = p_4^2dp_4dz d\phi^\prime$), and rewrite
    \begin{equation}
    \begin{aligned}
        {\cal J}_n ({\bf k}, p_1, \tau) &=
        \frac{1}{8(2\pi)^5 E_1 }\int\!\frac{p_2^2dp_2\,p_3^2dp_3\,d x d y\,d\beta d\phi}{ E_2E_3} \delta\left[(P_1 + P_2 - P_3)^2 - m_j^2\right]\times \\
        &\quad G_n \big|_{P_4 = P_1+P_2 - P_3}\, H(E_1+E_2-E_3 -m_j) \quad \text{for $n=1,2,3$}\, . \\
        {\cal J}_4 ({\bf k}, p_1, \tau) &=
        \frac{1}{8(2\pi)^5 E_1 }\int\!\frac{p_2^2dp_2\,p_3^2dp_3\,d x d y\,d\beta d\phi}{ E_2 E_3} \delta\left[(P_1 + P_2 - P_4)^2 - m_i^2\right] \times \\
        &\quad   G_n \big|_{P_3 = P_1+P_2 - P_4}\, H(E_1+E_2-E_4 -m_i) \, .
    \end{aligned}
    \end{equation}
Next, we proceed with the integration over the azimuth angle $\beta$ ($\beta^\prime$), for which we note that  the argument of the delta function 
    \begin{align*}
        (P_1 + P_2 - P_3)^2-m_j^2
        &= 2 \left(P_1 \cdot P_2 - P_1 \cdot P_3 - P_2 \cdot P_3 + m_i^2 \right) \, , \\
        (P_1 + P_2 - P_4)^2 - m_i^2
        &= 2 \left( P_1 \cdot P_2 - P_1 \cdot P_4 - P_2 \cdot P_4 + m_j^2 \right) \, ,
    \end{align*}
because of Eq.~(\ref{angles}), fixes the angle to the value $\beta_0$ ($\beta_0^\prime$) that comply with the (respective)condition
    \begin{align}
        \cos\beta_0 &= \frac{E_2E_3 - E_1(E_2-E_3)-m_i^2 + p_1p_2\cos\alpha-p_1p_3\cos\theta-p_2p_3\cos\alpha\cos\theta}{p_2p_3\sin\alpha\sin\theta} \, , \label{cosbeta} \\
        \cos\beta_0^\prime &= \frac{E_2E_4 - E_1(E_2-E_4)-m_j^2 + p_1p_2\cos\alpha-p_1p_4\cos\theta^\prime -p_2p_4\cos\alpha\cos\theta^\prime}{p_2p_4\sin\alpha\sin\theta^\prime} \, , \label{cosbetap}
    \end{align}
that we should accordingly for the corresponding case.

\noindent Therefore, using the Delta function properties, we can write
    \begin{align}
        \delta\left[(P_1+P_2 - P_3)^2-m_j^2\right] &= \frac{\delta(\cos\beta-\cos\beta_0)}{|2p_2p_3\sin\alpha\sin\theta|} \, , \\
        \delta\left[(P_1+P_2 - P_4)^2-m_i^2\right] &= \frac{\delta(\cos\beta^\prime - \cos\beta_0^\prime)}{|2p_2p_4\sin\alpha\sin\theta^\prime|} \, ,
    \end{align}
whereas for the integral, we use the convenient change of variables
    \begin{equation}
        \int_{-\pi}^\pi d\beta = 2\int_{0}^\pi d\beta =  2\int_{-1}^{1}\frac{d\cos\beta}{\sin\beta} H(\sin\beta) \, ,
    \end{equation}
where the Heaviside function states that $0\leq \sin\beta\leq 1$~. From Eq.~\eqref{cosbeta} and Eq.~\eqref{cosbetap}, after some algebra we can write, respectively
    \begin{equation}
        \sin^2\beta_0 =\frac{\tilde{S}(x,y)}{|p_2p_3\sin\alpha\sin\theta|^2} \quad \text{and} \quad \sin^2\beta_0^\prime =\frac{\tilde{S}^\prime(x,z)}{|p_2p_4\sin\alpha\sin\theta^\prime|^2} \, ,
    \end{equation}
where $\tilde{S}(x,y) = T^4 S(x,y)$ and $\tilde{S}^\prime(x,z) = T^4 S^\prime (x,z)$ are second-order polynomial expressions which, for the purpose of integration, can be conveniently expressed either as 
\begin{itemize}
    \item $S(x,y) = a_xy^2 + b_{x,i}y +c_{x,i}$, where
    \begin{align}
        a_x &= -q_3^2(q_1^2+q_2^2+2q_1q_2 x)~,\label{ax}\\
        b_{x,i} &= 2q_3(q_1+q_2x) \left[z_2z_3 + z_1(z_3-z_2)-\mu_i^2+q_1q_2x\right]~, \label{bx.i}\\
        c_{x,i} &= -[z_2z_3 + z_1(z_3-z_2)-\mu_i^2+q_1q_2x]^2 + 
        q_2^2q_3^2(1-x^2)~. \label{cx.i}
    \end{align}
    \item Alternatively, $S(x,y) = a_yx^2 + b_{y,i}x +c_{y,i}$, where
    \begin{align}
        a_y &= -q_2^2(q_1^2+q_3^2-2q_1q_3 y) \, , \label{ay}\\
        b_{y,i} &= 2q_2(q_1-q_3y)\left[z_1z_2 - z_3(z_1+z_2) + \mu_i^2 + q_1q_3y \right] \, , \label{by.i}\\
        c_{y,i} &= -\left[z_1z_2 - z_3(z_1+z_2) + \mu_i^2+q_1q_3y\right]^2 + 
        q_2^2q_3^2(1-y^2) \, .  \label{cy.i}
    \end{align}
\end{itemize}
Similarly, choosing the other scattering angle, 
    \begin{itemize}
    \item $S^\prime(x,z) = a_x^\prime z^2 + b_{x,j}^\prime z +c_{x,j}^\prime$, where
    \begin{align}
        a_x^\prime &= -q_4^2(q_1^2+q_2^2+2q_1q_2 x)~,\label{axp}\\
        b_{x,j}^\prime &= 2q_4(q_1+q_2x)[z_2z_4 + z_1(z_4-z_2)-\mu_j^2 + q_1q_2x]~, \label{bxp.j} \\
        c_{x,j}^\prime &= -[z_2z_4 + z_1(z_4-z_2)-\mu_j^2+q_1q_2x]^2 +   
        q_2^2q_4^2(1-x^2)~. \label{cxp}
    \end{align}
    \item Alternatively, $S^\prime(x,z) = a_z x^2 + b_{z,j} x + c_{z,j}$, where
    \begin{align}
        a_z &= -q_2^2(q_1^2 + q_4^2-2q_1q_4 z)~, \label{az}\\
        b_{z,j} &= 2q_2(q_1 - q_4 z)[z_1z_2 - z_4(z_1+z_2) + \mu_j^2 + q_1q_4z]~, \label{bz.j}\\
        c_{z,j} &= -[z_1z_2 - z_4(z_1+z_2) + \mu_j^2 + q_1q_4z]^2 + 
        q_2^2q_4^2(1-z^2)~.  \label{cz.j}
    \end{align}
    \end{itemize}
Clearly, in any case, the integration limits of either $x$ or $y$ ($z$) are restricted to the interval bounded by the roots of $S$ ($S^\prime$). Therefore, by putting all this back together into the integral, one can express 
    \begin{equation}
        {\cal J}_n ({\bf k}, p_1,\tau) =
        \frac{1}{8 (2\pi)^5 E_1} \times \begin{cases}
        \int \frac{p_2^2dp_2\,p_3^2dp_3\,}{E_2E_3}\,
        \Psi_S K_n \, I_n \,H(E_4-m_j) \quad \text{for $n=1,2,3$} \\
        \int \frac{p_2^2dp_2\,p_4^2dp_4\,}{E_2E_4}\,
        \Psi_S K_4 \, I_4 \,H(E_3 - m_i) \quad \text{for $n=4$}
        \end{cases} \, ,
    \end{equation}
where $I_n$ are the angular integrals
    \begin{equation}
        I_n = \begin{cases}
            
        I_n (\mu_n,p_1,p_2,p_3, \tau) = \frac{1}{T^2}\int \frac{dx dy\, d\phi} {\sqrt{S(x,y)}} \Theta_n({\bf k}, {\bf p}_n,\tau) \overline{|{\cal M}|}^2 H[S(x,y)] \quad \text{for $n=1,2,3$} \\
        I_4 (\mu_4,p_1,p_2,p_4, \tau) = \frac{1}{T^2} \int \frac{dx dz\, d\phi^\prime} {\sqrt{S^\prime(x,z)}} \Theta_4({\bf k}, {\bf p}_4,\tau) \overline{|{\cal M}|}^2 H[S^\prime(x,z)] \quad \text{for $n=4$}
        \end{cases} \, .
    \end{equation}
Both expressions were evaluated under the momentum conservation condition $P_4 = P_1+P_2-P_3$. After inserting the Legendre expansion (\ref{Thetaexp}), where $k=|{\bf k}|$, and $P_\ell$ is the Legendre polynomial of order $\ell$,
    \begin{equation}
    \begin{aligned}
        \hat{\bf p}_1 \cdot\hat{{\bf k}} &= \cos\gamma = \upmu \, , \quad
        \hat{\bf p}_2 \cdot\hat{{\bf k}} = 
        \cos\alpha\cos\gamma + \sin\alpha\sin\gamma\cos(\phi-\beta)\quad\text{and}\\
        \hat{\bf p}_3 \cdot \hat{\mathbf{k}} &=
        \cos\theta\cos\gamma + \sin\theta\sin\gamma\cos\phi \, , \quad \hat{\bf p}_4 \cdot \hat{\mathbf{k}} =
        \cos\theta^\prime\cos\gamma + \sin\theta^\prime\sin\gamma\cos\phi^\prime \, .
    \end{aligned}
    \end{equation}
In the chosen coordinate system, we have used the identity
    \begin{equation}
        \int_{0}^{2\pi} d\phi P_{\ell} (\cos \varphi \cos \gamma + \sin\varphi \sin\gamma \cos\phi) = 2\pi P_\ell (\cos\varphi) P_\ell(\cos\gamma) \, .
    \end{equation}
Then, the elements of the collision term in Fourier space are written as
    \begin{equation}
    \begin{aligned}
        {\cal J}_0 (p_1, \tau) &= \frac{1}{8 (2\pi)^4 E_1}
        \sum_{\ell=0 }^\infty (-i)^{\ell}(2\ell+1) \vartheta_l (p_1) {\cal P}_\ell(\upmu)\,
        \int\! \frac{p_2^2dp_2}{E_2} \frac{p_3^2dp_3}{E_3} \Psi_S K_1 H(E_4 - m_j) \tilde{\cal I}_1 \, , \\
        {\cal J}_n (p_1, \tau) &= \frac{1}{8 (2\pi)^4 E_1}
        \sum_{\ell=0 }^\infty (-i)^{\ell}(2\ell+1){\cal P}_\ell(\upmu)\,
        \int\! \frac{p_2^2dp_2}{E_2} \frac{p_3^2dp_3}{E_3} \Psi_S K_n H(E_4 - m_j) \vartheta_l (p_n) \tilde{\cal I}_{nl} \, , \\
        {\cal J}_4 (p_1, \tau) &= \frac{1}{8 (2\pi)^4 E_1}
        \sum_{\ell=0 }^\infty (-i)^{\ell}(2\ell+1){\cal P}_\ell(\upmu)\,
        \int\! \frac{p_2^2dp_2}{E_2} \frac{p_4^2dp_4}{E_4} \Psi_S K_4 H(E_3 - m_i) \vartheta_l (p_4) \tilde{\cal I}_{4l} \, ,
    \end{aligned} \label{J.red}
    \end{equation}
where $E_4 = E_1 + E_2 - E_3$, and the reduced angular integrals are given by
    \begin{equation}
    \begin{aligned}
        {\cal I}_{1}(q_1,q_2,q_3) &=  \int \frac{dxdy}{\sqrt{S}}
        \overline{|{\cal M}|}^2\,H(S) \, , \\
        {\cal I}_{n\ell} (q_1,q_2,q_3) &= \int \frac{dxdy}{\sqrt{S}}
        \overline{|{\cal M}|}^2\,H(S)\,  
        \times \begin{cases}
        P_\ell(x) & \text{for} \quad n = 2\\
        P_\ell(y) & \text{for} \quad n = 3
        \end{cases} \, , \\
        {\cal I}_{4\ell} (q_1,q_2,q_4) &= \int \frac{dxdz}{\sqrt{S^\prime}}
        \overline{|{\cal M}|}^2\,H(S^\prime)\,{\cal P}_{\ell}(z) \, ,
        \label{reduced angular integrals}
    \end{aligned}
    \end{equation}
with $\tilde{\cal I}_{1, n\ell} = \frac{1}{T^2} {\cal I}_{1, n\ell}$. So we can rewrite the rescaled collision term as
    \begin{equation}
        \tilde{C}^{(1)} [{\bf k}, p_1, \tau] = \frac{1}{4\pi} \frac{T_\nu}{128 \pi^3 z_1}
        \frac{d \ln f^{(0)}(p_1)}{d \ln p_1} \sum_{\ell=0 }^\infty (-i)^{\ell}(2\ell+1) \nu_{\ell} {\cal P}_\ell(\upmu) \tilde{{\cal K}}(z_1) \,,
        \label{C.general}
    \end{equation}
where $\tilde{K}(z_1) = \tilde{{\cal A}}(z_1)+ \tilde{{\cal B}}_\ell(z_1) - \tilde{{\cal D}}_\ell(z_1) - \tilde{{\cal E}}_\ell(z_1)$ and  
    \begin{equation}
    \begin{aligned}
        \tilde{{\cal A}}(z_1) &= \int_{\mu_j}^\infty\! dz_2\sqrt{z_2^2 -\mu_j^2} 
        \int_{\mu_i}^{z_1 +z_2-\mu_j}\,dz_3\sqrt{z_3^2 -\mu_i^2} \, \Psi_{S,1} {\cal I}_1(z_1,z_2,z_3) \, , \\
        \tilde{{\cal B}}_\ell(z_1) &= \frac{1}{q_1}\, \int_{\mu_j}^\infty\! dz_2 \, (z_2^2 -\mu_j^2)
        \int_{\mu_i}^{z_1 + z_2 - \mu_j }\!dz_3\sqrt{z_3^2 -\mu_i^2} \, \Psi_{S,2} {\cal I}_{2\ell}(z_1,z_2,z_3) \, , \\
        \tilde{{\cal D}}_\ell (z_1) &= \frac{1}{q_1} \, 
        \int_{\mu_j}^\infty\! dz_2\sqrt{z_2^2 -\mu_j^2}
        \int_{\mu_i}^{z_1 + z_2 -\mu_j}\! dz_3 \, (z_3^2 - \mu_i^2) \, \Psi_{S,3} \, {\cal I}_{3\ell}(z_1,z_2,z_3) \, , \\
        \tilde{{\cal E}}_\ell (z_1) &=  \frac{1}{q_1}\, 
        \int_{\mu_j}^\infty\! dz_2\sqrt{z_2^2 -\mu_j^2} 
        \int_{\mu_j}^{z_1 + z_2 -\mu_i}\! dz_4 \, (z_4^2 -\mu_j^2) \, \Psi_{S,4} \, {\cal I}_{4\ell}(z_1,z_2,z_4) \, ,
    \end{aligned}
    \end{equation}
where ${\cal I}_{1,n\ell}$ are the reduced angular integrals of the invariant amplitude, and for neutrinos $T=T_\nu$. And the functions $\Psi_{S,n}$ are determined by
    \begin{equation}
    \begin{aligned}
        \Psi_{S,1} &= \Psi_S K_1 \,, 
        &\qquad
        \Psi_{S,2} = \frac{e^{q_2} f^{(0)}(q_2)}{e^{q_1} f^{(0)}(q_1)} \, \Psi_S K_2 \,, \\
        \Psi_{S,3} &= \frac{e^{q_3} f^{(0)}(q_3)}{e^{q_1} f^{(0)}(q_1)} \, \Psi_S K_3 \,, 
        &\qquad
        \Psi_{S,4} &= \frac{e^{q_4} f^{(0)}(q_4)}{e^{q_1} f^{(0)}(q_1)} \, \Psi_S K_4 \,.
    \end{aligned}
    \end{equation}

\section{On \texorpdfstring{$\ell=0,1$}{ell01} moments} \label{appE}
Consider the general expression for the collision term, as in Eq.~\eqref{C}, where the distribution function factor is written as $F = F^{(0)} + F^{(1)}$, with
    \begin{equation}
    \begin{aligned}
        F &= f_\alpha^{(0)}(p_3)f_\beta^{(0)}(p_4)\left[1 + \sigma_1 f_\alpha^{(0)}(p_1)\right]\left[1 + \sigma_2 f_\beta^{(0)}(p_2)\right] \\
        &\quad - f_\alpha^{(0)}(p_1)f_\beta^{(0)}(p_2)\left[1 + \sigma_3 f_\alpha^{(0)}(p_3)\right]\left[1 + \sigma_4 f_\beta^{(0)}(p_4)\right]~,
    \end{aligned}
    \end{equation}
the in-equilibrium factor, and $F^{(1)} = F^{(1)}({\bf x}, {\bf p}_1, {\bf p}_2, {\bf p}_3, {\bf p}_4, \tau)$,
    \begin{equation}
        F^{(1)} = \Psi_S\big[K_3\Theta({\bf x},{\bf p}_3) +K_4\Theta({\bf x},{\bf p}_4) -K_1\Theta({\bf x},{\bf p}_1)-K_2\Theta({\bf x},{\bf p}_2)\big] \, .
    \end{equation}
The linear order contribution from the perturbation of the distribution functions, $\Theta$'s, and we have omitted the $\tau$-dependence. The functions involved here are given in the App.~\ref{appA}. Thus, one can write the collision integral as $ C_\alpha = C_\alpha^{(0)} + C_\alpha^{(1)}$. Only the linear perturbation $ C_\alpha^{(1)}$ has relevance for the evolution of the neutrino perturbations. 

Next, we observe that the squared invariant amplitude $|{\cal M}|^2$ (see App.~\ref{Amplitudes}) is invariant under any of the following momentum exchanges
    \begin{equation}
    \begin{array}{l l l l}
        \text{(i)}\; P_3 \leftrightarrow P_4 \, ,
        &\quad \text{(ii)}\; P_1 \leftrightarrow P_2 \, , &
        \text{(iii)}\; P_1 \leftrightarrow P_3 \ \text{and}\  P_2 \leftrightarrow P_4 \, , &\quad 
        \text{(iv)}\; P_1 \leftrightarrow P_4 \ \text{and}\  P_2 \leftrightarrow P_3 \, .
    \end{array}
    \end{equation}
Under (i) and (ii), the functional forms in the amplitudes are preserved in order to maintain the same interpretations for the Mandelstam variables. However, (iii) and (iv) are true symmetries for the amplitude. Clearly, these are also symmetries of the momentum conservation factor within the integral that defines $ C$~\footnote{So far, we have omitted the label $\alpha$.}.
In contrast, both terms of the distribution function factor $F$ remain invariant under (i) and (ii), but they are antisymmetric under (iii) and (iv). Furthermore, consider the integral,
    \begin{equation}
        \int d^3 p_1 C[{\bf x},{\bf p}_1] = (2\pi)^3
        \int d\Pi_{1} d\Pi_{2} d\Pi_{3}d\Pi_{4} (2\pi)^4 \delta^{(4)}{(P_1+P_2-P_3-P_4)} \overline{|{\cal M}|}^2 F \, .
    \end{equation}
By renaming the momentum variables in the above integral according to either (iii) and (iv), and by using that under such exchanges $F\rightarrow -F$, we get that $\int d^3 p_1 C[{\bf x},{\bf p}_1,\tau] = 0$. Since the antisymmetry of $F$ holds independently for $F^{(0)}$ and $F^{(1)}$, same conclusion arises for each independent collisional term, i.e. $\int d^3 p_1 C^{(0)}[{\bf x},{\bf p}_1] = 0 = \int d^3 p_1 C^{(1)}[{\bf x},{\bf p}_1]$, where first equality reflects the number conservation,
as so does for the Fourier-transformed term    \begin{equation}
        \int d^3 p_1 \tilde{C}^{(1)}[{\bf k},{\bf p}_1] = 0~.
        \label{intC0}
    \end{equation}
On the other hand, any integral of the form $\int d^3p_1\,d\Omega\, A\, C^{(1)}[{\bf x},{\bf p}_1]$ is equal to zero if $A$ is a collisional invariant.  Then, the expression
    \begin{equation}
        \int d^3p_1\,d\Omega\,p_1\, \tilde{C}^{(1)}[{\bf k},{\bf p}_1] = 0~,
        \label{intC0.f}
    \end{equation}
reflects momentum conservation and is also valid for energy conservation; ${\bf p}_i \to E_i$.  
Similarly, by considering the (dipole) integral in Fourier space and with the use of the transformation properties under (ii) and (iii), we can write
    \begin{equation}
    \begin{aligned}
        \int d^3 p_1 ({\bf p}_1 \cdot \hat{\bf k}) \tilde{C}^{(1)}[{\bf k},{\bf p}_1] \propto
        \int d\Pi_{1} d\Pi_{2} d\Pi_{3}d\Pi_{4} & (2\pi)^4 \delta^{(4)}{(P_1+P_2-P_3-P_4)} \overline{|{\cal M}|}^2 \\
        & \times ({\bf p}_1 + {\bf p}_2 - {\bf p}_3 - {\bf p}_4)\cdot {\bf \hat{k}} \,\tilde{F}^{(1)} \, .
    \end{aligned}
    \end{equation}
Therefore, due to the conservation of momentum, we conclude that
    \begin{equation}
        \int d^3 p_1 ({\bf p}_1\cdot \hat{\bf k}) \tilde{C}^{(1)}[\mathbf{k},\mathbf{p_1}] = 0 \, .
        \label{intC1}
    \end{equation}
Eqs. \eqref{intC0} \eqref{intC0.f} 
and \eqref{intC1} have important implications for the $\ell = 0,1$ weights. In general terms, the Legendre expansion of the collisional term in Fourier space goes as
    \begin{equation}
        \tilde{C}^{(1)}[{\bf k},{\bf p}_1] = 
        \sum_{\ell = 0}^{\infty} (-i)^\ell (2\ell +1) \tilde{C}_\ell(p_1)\,{\cal P}_\ell(\upmu) \,,
    \end{equation}
where, as already defined in the main text, $\upmu=\hat{\bf p}_1\cdot\hat{\bf k}$.
Nonetheless, by definition, the last is given in Eq.\eqref{C_ell}. As $d^3p_1 = p_1^2dp_1d\varphi_1d\upmu$, we can express for $\ell =0$ and $\ell=1$, respectively (since $P_0(\upmu) = 1$ and $P_1(\upmu) = \upmu$)
    \begin{equation}
        \int_0^\infty dp_1 p_1^2 C_0 = \frac{1}{4\pi} \int d^3p_1\tilde{C}^{(1)}=0 \quad \text{and} \quad \int_0^\infty dp_1 p_1^3 C_1 = \frac{1}{4\pi} \int\! d^3p_1 p_1(\mathbf{\hat{p}_1}\cdot\mathbf{\hat{k}})\tilde{C}^{(1)}=0 .
    \end{equation}
Obviously, as $\tilde{C}_\ell$ has no angular dependencies, then $\tilde{C}_{0,1} = 0$. Note that this result is valid as long as the amplitude has the symmetries (i)-(iv), which, in principle, are valid for any elastic scattering process. In our case of interest, the coefficient $\tilde{C}_\ell$ can be read out from the formal integral expression given in Eq.~\eqref{ctilde_1} as
    \begin{equation}
        \tilde{C}_{\ell} = \frac{1}{4} \frac{g^4 T_\nu}{128 \pi^3 z_1}
        \frac{d \ln f^{(0)}(p_1)}{d \ln p_1} \nu_\ell \, {\cal K}_\ell^{ij}(z_1)  \, .
    \end{equation}
Therefore the natural solution to the above constraints is that ${\cal K}_{0,1}^{ij} = 0$ (also ${\cal K}_{0,1}=0$).
\section{Elastic Scattering Amplitudes for \texorpdfstring{$\nu_i-\nu_j$}{nuinuj} Via Scalar Mediator} \label{Amplitudes}
With the Feynman rules obtained from the Lagrangian with Dirac-like neutrinos, 
    \begin{equation}
        {\cal L}_{int}^{D} = \frac{1}{2} \sum_{i, j} \varphi \left( g_{ij} \overline{\nu}_i \nu_j +  g_{ji}\overline{\nu}_j \nu_i \right) \, .
        \label{L.Dirac}
    \end{equation}
With the condition $g_{ij} = g_{ji}^{*}$ that allows the interaction to be consistently expressed in the flavor basis as ${\cal L}_{int} \sim g_{\alpha\beta} \overline{\nu}_\alpha \nu_\beta $ for $g_{\alpha\beta} \in \mathbb{R}$. We can establish the following scattering amplitudes:
\begin{itemize}
    \item {\bf For Dirac-like neutrinos}. In the same mass eigenstate, both t and u channels contribute, while for different mass eigenstates, only the t channel contributes, so the amplitudes are (averaged over the initial spin-polarizations of neutrinos)
    \begin{equation}
        \overline{\left|{\cal M}_D^{(ij)} \right|}^2 = \begin{cases}
            |g_{ii}|^4 \left[T_t^{(i)} + T_u^{(i)} + T_{tu}^{(i)} \right] & \text{for $m_i = m_j$}\\
            |g_{ii}|^2 |g_{jj}|^2 T_t^{(ij)} & \text{for $m_i \neq m_j$}
        \end{cases} \, ,
        \label{Am-D}
    \end{equation}
    with $T_t^{(i)}=T_t^{(ii)}$ and
        \begin{equation}
            T_t^{(ij)}= \frac{(4m_i^2 - t) (4m_j^2 - t)}{(m_\varphi^2 - t)^2} \, , \quad T_u^{(j)}= \frac{(4m_j^2 - u)^2}{(m_\varphi^2 - u)^2} \, , \quad T_{tu}^{(j)} = \frac{ tu - 4m_j^2 s}{(m_\varphi^2-t)(m_\varphi^2-u)} \, .
        \end{equation}
    \item {\bf For Majorana-like neutrinos}. We can rescale the interaction Eq.~\eqref{L.Dirac} due to Majorana fields' properties (see Subsec.~\ref{Different-neutrino-mass}), so we can use the same couplings for our calculations, the Lagrangian remains the form \eqref{L.Dirac} and using the consistent Feynman rules for Majorana neutrinos (see, e.g., Ref.~\cite{Denner1992feynman}).
    \begin{equation}
        \overline{\left|{\cal M}_M^{(ij)}\right|}^2 = \begin{cases}
            |g_{ii}|^4 \left[T_s^{(i)} + T_t^{(i)} + T_u^{(i)} + T_{st}^{(i)} + T_{su}^{(i)} + T_{tu}^{(i)}\right] & \text{for $m_i = m_j$}\\
            |g_{ij}|^4 T_s^{(ij)} + |g_{ii}|^2 |g_{jj}|^2 T_t^{(ij)} + {\rm Re}\{g_{ij}^2 g_{ii}^{*} g_{jj}^{*}\} T_{st}^{(ij)} & \text{for $m_i \neq m_j$}
        \end{cases} \, ,
        \label{Am-M}
    \end{equation}
    where $T_s^{(i)}=T_s^{(ii)}$, $T_{st}^{(i)} = T_{st}^{(ii)}$, $T_{su}^{(i)} = T_{s( t \to u)}^{(i)}$
    \begin{equation}
    \begin{aligned}
        T_s^{(ij)} &= \frac{(s - (m_i + m_j)^2)^2}{(m_\varphi^2 - s)^2} \, , \\
        T_{st}^{(ij)} &= -\frac{2 \bigg\{ (4m_i^2 -t)(4m_j^2 - t) - [(m_i+m_j)^2 -u]^2 + [s-(m_i+m_j)^2]^2 \bigg\}}{(m_\varphi^2 -s)(m_\varphi^2 - t)}\, .
    \end{aligned}
    \end{equation}
And the other terms are the same as those appearing in the case of Dirac-like neutrinos.
\end{itemize}

\noindent Note that the previous results can be formulated for an arbitrary $n$-eigenstates of mass; however, in practice, we work with the three neutrino mass eigenstates of the SM extended to include massive neutrinos. By assuming universal couplings, $|g_{ii}| = g$, we can compare our analysis with approximations known in the literature, heavy or light mediators (see Subsec.~\ref{subsec: HML} and Sec.~\ref{sec:light_mediator}). In this limit, the amplitudes that take the general form given in Eq.~\eqref{M2}, after some algebra, reduce to Eq.~\eqref{Amp.Heavy}.
\section{First reduced angular integrals}
\label{appD}
Here, we discuss the general strategy to perform the first reduced angular integrals as \eqref{reduced angular integrals}, focusing on an exact example of neutrino--neutrino elastic scattering process with the same mass eigenstate.
To compute the reduced angular integrals, we notice that the roots of $S(x,y)$ are given by
    \begin{equation}
        \tilde{x}_{\pm} = \frac{-b_y \pm \sqrt{\Delta_y}}{2a_y} \quad \text{and} \quad \tilde{y}_{\pm} = \frac{-b_x \pm \sqrt{\Delta_x}}{2a_x} \, , \label{S-roots}
    \end{equation}
and for non-zero moments satisfy the condition
    \begin{equation}
        |\tilde{x}_{\pm}| \leq 1 \quad \text{and} \quad |\tilde{y}_{\pm}| \leq 1 \, .
    \end{equation}
This result is valid for any elastic scattering process; this means, $|\tilde{x}^\prime_{\pm}| \leq 1$ and $|\tilde{z}_{\pm}| \leq 1$ in our notation presented in Appendix \ref{appB} for $S(x^\prime, z)$.
Thus, we consider the general form
    \begin{equation}
        \tilde{\cal T}_{r} (E_i,v) = \int_{-1}^1\frac{dw}{\sqrt{S(w,v)}}\, T_r(w,v)\,H[S(w,v)] \, ,
        \label{tilde-T-integrals}
    \end{equation}
where $w = x$  ($w=y$)  and $v=y$ ($v=x$) for $\tilde{\Theta}({\bf p}_{1,3})$ [$\tilde{\Theta}({\bf p}_2)$] and $S(w,v)=a_vw^2+b_vw+c_v$. Note that the actual limits would correspond to the interval where $S(w, v)>0$, as indicated by the Heaviside function, which is bounded by the roots of $S(w,v)$ in terms of $v$.
Note that the integral $\tilde{\cal T}_r(E_i,v)$ is dimensionless on the energy units since $S(w,v)$ is dimensionless and $T$-terms in the amplitude are also dimensionless. The reduced angular integrals for our case of interest (see Appendix~\ref{Amplitudes}) become
    \begin{equation}
    \begin{aligned}
        {\cal I}_{1} &= g^4 \int_{-1}^{1} dv \sum_r \tilde{\cal T}_r (E_i,v) = \pi g^4 \int_{-1}^{1} dv H(\Delta_v) {\cal T}_v^{N} = \pi g^4 J_1 \, , \\
        {\cal I}_{n \ell} &= g^4 \int_{-1}^{1} dv \sum_r \tilde{\cal T}_r (E_i,v) {\cal P}_\ell(v) = \pi g^4 \int_{-1}^{1} dv H(\Delta_v) {\cal T}_v^N {\cal P}_\ell(v) = \pi g^4 J_{n\ell} \, .
        \label{RAI.Gen}
    \end{aligned}
    \end{equation}
Here ${\cal T}_v^{N} = \sum_r \alpha_r {\cal T}_r (E_i,v)$ with $r \in \{t, u, tu\}$ for Dirac-like neutrinos ($N=D$) or in another case $r \in \{s, t, u, tu, st, su\}$ for Majorana-like neutrinos ($N=M$) according to the contributions presented in Fig.~\ref{Scattering with neutrinos}. Also, we introduce adimensional $\alpha_r$ coefficients that determine the ratio between couplings respect to $|g_{ii}| =g$, for example, for Dirac-like neutrinos with $m_i \neq m_j$: $\alpha_t = \frac{|g_{jj}|^2}{g^2} = \frac{g_{jj}^2}{g^2}$, from Eq.~\eqref{Am-D}, for the same mass eigenstate or universal coupling all $\alpha_r =1$. Here, the $J$-integral take the forms
    \begin{equation}
        J_1 = \int_{-1}^{1} dv H(\Delta_v) {\cal T}_v^{N} \quad \text{and} \quad J_{n\ell} = \int_{-1}^{1} dv H(\Delta_v) {\cal T}_v^N {\cal P}_\ell(v) \, .
        \label{J-integrals-g}
    \end{equation}
Where $n$-labels are related directly to the labels in the momenta of the involved particles, if the final states are identical, $J_{3\ell} = J_{4\ell}$.

As discussed in the main text, any given $T$-term for the neutrino--neutrino interaction goes as a rational function of up to second-order polynomials on the angular variable $w=x$ or $y$. As a matter of fact, by expanding the corresponding expressions in Eq.~\eqref{Tr} and \eqref{Tmix}, they can be  explicitly written down as
    \begin{equation}
        T_t^{(ij)} = \begin{cases}
            \frac{(a_1 - B_1y)(a_2 - B_1y)}{(A^\prime_{1,i} - B_1y)^2} & \text{for $\mu_i \neq \mu_j$} \\
            \frac{(A_1 - B_1y)^2}{(A^\prime_1 - B_1y)^2} & \text{for $\mu_i = \mu_j$}
        \end{cases}  \quad \text{or} \quad T_t^{(ij)} = \frac{[a_3-B_1^\prime x + B_1^{\prime\prime} z][a_4-B_1^\prime x + B_1^{\prime\prime}z]}{(A_7-B_1^\prime x + B_1^{\prime\prime} z)^2} \, , \label{Tt}
    \end{equation}
    \begin{equation}
        T_u = \frac{(A_2 -B_1^\prime x + B_1y)^2}{(A^\prime_2 - B_1^\prime x + B_1y)^2} \, , \, T_s^{(ij)} = \begin{cases}
        \frac{(A_{4,ij}-B_1^\prime x)^2}{(A_{4,ij}^\prime -B_1^\prime x)^2} & \text{for $\mu_i \neq \mu_j$} \\
        \frac{(A_4 -B_1^\prime x)^2}{(A^\prime_4 - B_1^\prime x)^2} & \text{for $\mu_i = \mu_j$}
        \end{cases} , \label{Tu}
    \end{equation}
whereas the interference terms are
    \begin{equation}
        T_{tu} = \frac{A_3 +B_{2}^\prime x + B_{2}y + B_1B_1^\prime xy - B_1^2y^2}{(A^\prime_1 - B_1y)(A^\prime_2 - B_1^\prime x + B_1y)} \, ,  \, T_{st}^{(ij)} = \begin{cases}
        - \frac{A_{5,ij} + B_{3,ij}^\prime x+B_{3,ij} y - B_1B_1^\prime xy}{(A_{4,ij}^\prime - B_1^\prime x) (A_{1,i}^\prime - B_1 y)} \\ 
        \frac{A_8 - B_4^\prime x + B_4 z + B_1^{\prime 2} x^2 - B_1^\prime B_1^{\prime\prime} xz}{(A_{4,ij}^\prime - B_1^\prime x)(A_7 - B_1^\prime x + B_4z)}
        \end{cases} ,
    \end{equation}
and
    \begin{equation}
        T_{su} =-\frac{A_6 + B_4^\prime x - B_3 y + B_1B_1^\prime x y - {B_1^\prime}^2 x^2}{(A_4^\prime-B_1^\prime x)(A_2^\prime - B_1^\prime x + B_1 y)}~,
    \end{equation}
where we used the short-hand notation for the elements in $T$-terms:
    \begin{equation}
    \begin{aligned}
        a_1 &= z_1z_3 + \mu_i^2 \, , \, a_2 = z_1z_3 + 2\mu_j^2 - \mu_i^2 \, , \, a_3 = z_1(z_2-z_4) + \frac{1}{2} (\mu_j^2+3\mu_i^2) \, , \\
        a_4 &= z_1(z_2-z_4) + \frac{1}{2} (5\mu_j^2-\mu_i^2) \, , \, A_1 = z_1z_3 + \mu_\nu^2 \, , \, A_{1,i}^\prime = z_1z_3 + \frac{1}{2}\mu_\varphi^2 - \mu_i^2 \, , \\
        A_1^\prime &= A_{1,i}^\prime |_{\mu_i = \mu_j=\mu_\nu} \, , \, A_2 = z_1(z_2-z_3) + 2\mu_\nu^2 \, , \, A_2^\prime = z_1(z_2 - z_3) + \frac{1}{2}\mu_\varphi^2 \, , \\
        A_3 &= z_1(z_2 - z_3)(z_1z_3-\mu_\nu^2)-2\mu_\nu^2(\mu_\nu^2 +z_1z_2) \, , \, A_{4,ij} = z_1z_2-\mu_i \mu_j \, , \\
        A_4 &= A_{4,ij} |_{\mu_i=\mu_j=\mu\nu} \, , \, A^\prime_{4,ij} = z_1z_2 -\frac{1}{2} [\mu_\varphi^2 - (\mu_i^2 + \mu_j^2)] \, , \, A_4^\prime = A^\prime_{4,ij} |_{\mu_i = \mu_j=\mu_\nu} \, , \\
        A_8 &= 2\mu_i \mu_j(\mu_i \mu_j - z_1 z_4) + (\mu_i^2 + \mu_j^2 - \mu_i \mu_j + z_1z_2)[z_1(z_2-z_4)] \, , \, \\
        A_6 &= (z_1z_2 + 3\mu_\nu^2)(z_1z_3-\mu_\nu^2) - (z_1^2 z_2^2 - \mu_\nu^4) \, , \\
        A_{5,ij} &= [z_1z_2 + (\mu_i^2 + \mu_j^2 + \mu_i\mu_j)] [(\mu_i^2 + 2\mu_i\mu_j) - z_1z_3] - 2\mu_i\mu_j (\mu_i + \mu_j)^2 \, , \\
        A_7 &= z_1(z_2-z_4) + \frac{1}{2}\mu_\varphi^2 \, , \, B_1 = q_1q_3 \, , \, B_1^\prime = q_1q_2 \, , \, B_1^{\prime\prime} = q_1q_4 \, , \\
        B_2 &= (z_1(2z_3-z_2)-\mu_\nu^2) B_1 \, , \, B_2^\prime = (3\mu_\nu^2 -z_1z_3)B_1^\prime \, , \\
        B_{3,ij} &= [z_1 z_2 + (\mu_i^2 + \mu_j^2 + \mu_i \mu_j)]B_1 \, , \, B_3 = B_{3,ij} |_{\mu_i = \mu_j=\mu_\nu} \, , \\
        B_{3,ij}^\prime &= [z_1 z_3 - (\mu_i^2 + 2\mu_i \mu_j)] B_1^\prime \, , \, B_3^\prime = B_{3,ij}^\prime |_{\mu_i = \mu_j=\mu_\nu} \, , \\
        B_4 &= (z_1z_2 + \mu_i^2 + \mu_j^2 - \mu_i \mu_j) B_1^{\prime\prime} \quad \text{and} \quad B_4^\prime = [z_1(2z_2-z_4) + \mu_j^2 + \mu_i^2 - \mu_i\mu_j] B_1^\prime \, ,
    \end{aligned} \label{short-notation}
    \end{equation}
where the rescaled energy is, as usual, defined as $z_{1,3} = \sqrt{q_{1,3}^2 + \mu_i^2}$ and $z_{2,4} = \sqrt{q_{2,4}^2 + \mu_j^2}$.

By direct inspection, it is straightforward to see that the above $T$-terms have  generic forms that  can be reduced into partial fractions according to the following four general rules 
    \begin{equation}
        \frac{(a - B w)(b - B w)}{(A^\prime - B w)^2} = 1 + \frac{ (a + b - 2A^\prime)}{A^\prime - B w} + \frac{(A^\prime -a)(A^\prime - b)}{(A^\prime - Bw)^2} \, , \label{frac1}
    \end{equation}
or its particular form, $B \to - B$ and $a_1=a_2$, so $A = \frac{1}{2}(a + b)$;
    \begin{equation}
        \frac{(A + Bw)^2}{(A^\prime + Bw)^2} = 
        1 +\frac{2(A -A^\prime)}{A^\prime + Bw} + 
        \frac{(A - A^\prime)^2}{(A^\prime + Bw)^2}~,
        \label{frac1.a}
    \end{equation}
    \begin{equation}
        \frac{a + bw - B^2w^2}{(A - Bw)(A^\prime + Bw)} = 
        1 +\frac{C_1}{A- Bw} + 
        \frac{C_2}{A^\prime + Bw}~,
        \label{frac2}
    \end{equation}
where 
$$
C_1 = \frac{a+bA/B -A^2}{A+A^\prime}~,\quad\text{and}\quad
C_2 = \frac{a-bA^\prime/B -(A^\prime)^2}{A+A^\prime}~,
$$
Another useful partial fractions are:
    \begin{equation}
        \frac{a-bw + B^2w^2}{(A-Bw)(A^\prime + Bw)} = -1 + \frac{D_1}{(A-Bw)} + \frac{D_2}{(A+Bw)}~,
        \label{frac6}
    \end{equation}
    \begin{equation}
        \frac{a + bw }{(A - Bw)(A^\prime + Bw)} = 
        \frac{C_3}{A- Bw} + 
        \frac{C_4}{A^\prime + Bw}~,
        \label{frac4}
    \end{equation}
    \begin{equation}
        \frac{a+bw}{A-Bw} = -\frac{b}{B} + \frac{a +bA/B}{A-Bw}~,
        \label{frac3}   
    \end{equation}
where 
    \begin{equation*}
        D_1 = \frac{a-Ab/B +A^2}{A+ A^\prime} \, ,\,
        D_2 = \frac{a+A^\prime b/B +A^{\prime 2}}{A+ A^\prime} \, , \, C_3 = \frac{a+bA/B}{A+A^\prime} \quad \text{and} \quad  C_4 = \frac{a-bA^\prime/B}{A+A^\prime} \, .
    \end{equation*}
Thus, as any given term on the  RHS of the above equations is either a constant or goes as $(w\pm q)^{-1}$ or $(w\pm q)^{-2}$, for a properly defined $q$, we are left to consider the following integrals, the majority of which are obtained from the general formulas given in Ref.~\cite{tables} for $a < 0$,
    \begin{eqnarray}
        \int_{-1}^{1}\frac{dw}{\sqrt{S(w)}}H[S(w)] &=& \frac{\pi}{\sqrt{-a}}H(\Delta_v)~, \label{int1} \\
        \int_{-1}^{1} \frac{dw}{(w\pm q)\sqrt{S(w)}}H[S(w)] &=& \frac{\pi}{\sqrt{-c^\prime}}H(\Delta_v) \begin{cases}
            1 & \text{for $w_{+} \pm q >0$}\\
            - 1 & \text{for $w_{-} \pm q < 0$}
             \end{cases}~, \label{int2}\\
        \int_{-1}^{1}\frac{dw}{(w\pm q)^2\sqrt{S(w)}}H(S(w)) &=& -
        \frac{\pi b^\prime}{2c^\prime\sqrt{-c^\prime}}H(\Delta_v) 
        \begin{cases}
            1 & \text{for $w_{+} \pm q >0$}\\
            - 1 & \text{for $w_{-} \pm q < 0$}
             \end{cases}~, \label{int3}
    \end{eqnarray}
where  $b^\prime = b \mp 2aq$, $c^\prime = c\mp bq +aq^2$ and $\Delta_v = b^2 -4ac$ and for $c^\prime \geq 0$ \eqref{int2} and \eqref{int3} are zero, because
    \begin{align}
        \int_{-1}^{1} \frac{dw}{(w \pm q)\sqrt{S(w)}} H[S(w)] &= - \frac{2\sqrt{S(w)}}{b^\prime (w \pm q)} \bigg|_{w_{+}}^{w_{-}} H(\Delta_v) =0 \quad \text{for $c^\prime =0$} \, , \label{intr-1} \\
        \int_{-1}^{1} \frac{dw}{(w \pm q)\sqrt{S(w)}} H[S(w)] &= - \frac{1}{\sqrt{c^\prime}}\ln\left| \frac{2c^\prime+b^\prime (w \pm q) + 2\sqrt{c^\prime S(w)}}{(w \pm q)}\right| \bigg|_{w_{+}}^{w_{-}} H(\Delta_v) = 0 \, , \, c^\prime > 0 \, . \label{intr-2}
    \end{align}
The resulting Heaviside function restricts the angular integrals to regions where the discriminant $\Delta_v$ of $S(w)$ is positive; further integrations must be performed numerically.
\subsection{The \texorpdfstring{$h$}{h}-functions} \label{h-functions}
The integrals \eqref{int1}-\eqref{intr-2} fail for $w\pm q =0$, which corresponds to either on-shell scalar production or an infrared divergence. In the following, we illustrate how to regularize these cases in our numerical implementation.

For the $u$-channel we can write Eq.~\eqref{Tu} in partial fractions as
    \begin{equation}
        T_u(y) = 1 - \frac{Q}{B_1(y+q_x)} +\frac{Q^2}{4{B_1}^2(y+q_x)^2} \quad \text{with} \quad Q= \mu_\varphi^2 - 4\mu_\nu^2 \,  , \, q_x = \frac{A_2^\prime - B_1^\prime x}{B_1} \,.
    \end{equation}
The integration over the $y$ variable required to construct the $J_{2\ell}$ integral for \eqref{calBl-red}, with $\tilde{y}_{\pm}$ defined in Eq.~\eqref{S-roots} for $c_x^{\prime\prime} = c_x - b_xq_x + a_x q_x^2 < 0$, as obtained from Eq.~\eqref{int2}, is given by
    \begin{equation}
        \int_{-1}^1 \frac{dy}{(y+q_x)\sqrt{S(y)}} H[S(y)] = \frac{\pi}{\sqrt{-c_x^{\prime\prime}}} H(\Delta_x) \begin{cases}
            1 & \text{for $\tilde{y}_{+} > -q_x $} \\
            -1 & \text{for $\tilde{y}_{-}< - q_x$}
        \end{cases} 
        \label{Int Sy2-qx}
    \end{equation}
while for $c_x^{\prime\prime} \geq 0$ from Eq.~\eqref{intr-1} and Eq.~\eqref{intr-2}, one finds
    \begin{equation}
        \int_{-1}^1 \frac{dy}{(y+q_x)\sqrt{S(y)}} H[S(y)] = 0 \, .
    \end{equation}
We take a null contribution for the infrared divergence, which is absorbed in the case $c_x^{\prime\prime} \geq 0$, as follows:
    \begin{equation}
        \int_{-1}^1 \frac{dy}{(y+q_x)\sqrt{S(y)}} H[S(y)] = \frac{\pi}{\sqrt{-c_x^{\prime\prime}}} H(\Delta_x) h_1(q_x) \quad \text{where} \quad h_1(q_x) = \begin{cases}
            1 & \text{for $\tilde{y}_{+} > -q_x $} \\
            -1 & \text{for $\tilde{y}_{-}< - q_x$} \\
            0 & \text{elsewhere}
        \end{cases} \, .
    \end{equation}
Similarly, from Eqs.~\eqref{int3}–\eqref{intr-2}, and using our regularization prescription, we obtain
    \begin{align}
        \int_{-1}^1 \frac{dy}{(y+q_x)^2\sqrt{S(y)}} H[S(y)]&= -\frac{\sqrt{S(y)}}{c_x^{\prime\prime} (y+q_x)} \bigg|_{\tilde{y}_{+}}^{\tilde{y}_{-}} H(\Delta_x)-
        \frac{b_x^{\prime\prime}}{2c_x^{\prime\prime}}\int_{-1}^1 \frac{dy}{(y+q_x)\sqrt{S(y)}} H[S(y)] \nonumber \\
        &= - \pi H(\Delta_x) \frac{b_x^{\prime\prime}}{2 c_x^{\prime\prime} \sqrt{-c_x^{\prime\prime}}} h_1(q_x) \, ,
    \end{align}
where $b_x^{\prime\prime}= b_x - 2a_x q_x$.

We note that the infrared divergence in the $u$-channel of the previous example may be neglected, as it depends on whether the relevant energies lie within the integration domain. Nevertheless, it is useful for our code to check whether it is necessary to include these contributions. Similarly, we have constructed the remaining functions in Eqs.~\eqref{h4-h5}, \eqref{h6}, and \eqref{h1-h3}. Notice that this procedure is useful for excluding the resonant production and for identifying the range in which it appears.
\subsection{Massive mediator limit} \label{appD-1}
The amplitudes for Majorana or Dirac-like neutrinos in the HML take the form
    \begin{equation}
        |{\cal M}_y|^2  =  \Delta_2(y)\left(\frac{3b_y^2 - 4a_yc_y}{8a_y^2}\right) - \Delta_1(y)\left(\frac{b_y}{2a_y}\right) + \Delta_0(y) \, ,
        \label{Heavy-amplitudes-1}
    \end{equation}
    \begin{equation}
        |{\cal M}_x|^2  =  \Delta_2(x)\left(\frac{3b_x^2 - 4a_xc_x}{8a_x^2}\right) - \Delta_1(x)\left(\frac{b_x}{2a_x} \right) + \Delta_0(x)~, \label{Heavy-amplitudes-2}
    \end{equation}
where
    \begin{align*}
    \begin{aligned}
        \Delta_2(y) &=  q_1^2 q_2^2~, \\
        \Delta_1(y) &= q_1^2 q_2 \left[q_3(1-y) - 2q_2\right]~, \\
        \Delta_0(y) &= q_1^2 \left[q_2^2 - q_2 q_3(1-y) + q_3^2(1-y)^2\right]~.
    \end{aligned}
    \quad
    \begin{aligned}
        \Delta_2(x) &= q_1^2 q_3^2~, \\
        \Delta_1(x) &= q_1^2 q_3 \left[q_2(1-x) - 2q_3\right]~, \\
        \Delta_0(x) &= q_1^2 \left[q_3^2 - q_2 q_3(1-x) + q_2^2(1-x)^2\right]~,
    \end{aligned} 
\end{align*}
and the integral forms become
    \begin{align}
        A(q_1) &= \frac{1}{8\pi^3}\int_0^\infty\frac{e^{q_2}\,dq_2}{e^{q_2}+1}\int_0^{q_1+q_2}\frac{q_3\,dq_3}{(e^{q_3}+1)(e^{(q_1+q_2-q_3)}+1)}\int_{\rho_{+}}^{1}\,dy\frac{{|\cal M}_y(q_i)|^2}{\sqrt{q_1^2 + q_3^2 -2q_1 q_3 y}} ~, \label{A}\\
        B_\ell (q_1)&= \frac{1}{8\pi^3 q_1}\int_0^\infty\frac{e^{q_2}q_2^2\,dq_2}{e^{q_2}+1}\int_0^{q_1+q_2}\frac{dq_3}{(e^{q_3}+1)(e^{(q_1+q_2-q_3)}+1)}\int_{\eta_{+}}^{1}\,dx\frac{{|\cal M}_x(q_i)|^2{\cal P}_\ell(x)}{\sqrt{q_1^2 + q_2^2 +2q_1 q_2 x}} \, , \label{Bl}\\
        D_\ell(q_1)&= \frac{e^{-q_1}}{8\pi^3 q_1}\int_0^\infty\frac{dq_2}{e^{q_2}+1}\int_0^{q_1+q_2}\frac{e^{q_3}q_3^2\,dq_3}{(e^{q_3}+1)(e^{-(q_1+q_2-q_3)}+1)}\int_{\rho_{+}}^{1}\,dy\frac{{|\cal M}_y(q_i)|^2{\cal P}_\ell(y)}{\sqrt{q_1^2 + q_3^2 -2q_1 q_3 x}} \, . \label{Dl}
    \end{align}
\section{Symbols} \label{appF}
In this appendix, we present some relevant symbols to clarify the notation used in this paper.
\begin{table}[!ht]
\centering
\caption{Guided symbols in this paper.}
\begin{tabular}{|c|l|c|}
\hline
\textbf{Symbol} & \textbf{Meaning} & \textbf{Reference Eqn} \\
\hline
\hline
$a$ & scale factor & \eqref{metric-gauge}, \eqref{conformal Newtonian gauge}, \eqref{synchronous gauge}, \eqref{Q-zero} \\
\hline
$\tau$ & conformal time & \eqref{metric-gauge}, \eqref{Boltzmann-C}, \eqref{Q-zero}
\\
\hline
\makecell[c]{$\Phi$ \\ $\Psi$} & conformal Newtonian
metric perturbations & \eqref{metric-gauge}, \eqref{conformal Newtonian gauge}, \eqref{Q-CN}, \eqref{f_g} \\
\hline
$h_{ij}$ & \makecell[l]{metric perturbations in \\ the synchronous gauge} & \eqref{metric-gauge}, \eqref{synchronous gauge}
\\
\hline
$f$ & phase space distribution function & \eqref{RBE-1}, \eqref{Boltzmann-C} \\
\hline
${\rm P}^i$ & \makecell[l]{canonical conjugate momenta of \\
(comovil) position $x^i$} & \eqref{RBE-1}, \eqref{conformal Newtonian gauge}, \eqref{synchronous gauge} \\
\hline
${\rm P}^0$ & \makecell[l]{time component of 4-momentum in \\ conformal coordinates (gauge-dependent)} & \eqref{RBE-1}, \eqref{conformal Newtonian gauge}, \eqref{synchronous gauge}\\
\hline
$\Gamma_{\mu \nu}^k$ & second-rank Christoffel symbols & \eqref{RBE-1} \\
\hline
$p$ & magnitude of the local three-momentum & \eqref{Boltzmann-C} \\
\hline
$\hat{p}_j$ & \makecell[l]{unit vector along the momentum direction \\ (also for any symbol with hat)} & \eqref{Boltzmann-C}, \eqref{Q-S} \\
\hline
$Q[f]$ & Boltzmann collision term & \eqref{RBE-1}, \eqref{Q-zero}, \eqref{Q-zero} \\
\hline
$f_\alpha^{(0)}$ & unperturbed phase space distribution function & \eqref{f_background}, \eqref{f_def} \\
\hline
${\bf p}$ & local three momentum of reference particle & \eqref{Q-CN}, \eqref{Q-S} \\
\hline
$f_\alpha$ & phase space distribution function of $\alpha$-particle & \eqref{Q-CN}, \eqref{Q-S}, \eqref{F-complete}, \eqref{f_def} \\
\hline
$C_\alpha[{\bf x}, {\bf p}_1, \tau]$ & collision term & \eqref{C} \\
\hline
$\Theta_\alpha$ & perturbation to $f_\alpha^{(0)}$ & \eqref{f_def}, \eqref{Thetaexp} \\
\hline
$\tilde{C}_\alpha^{(1)}$ & transformed collision term at first order & \eqref{C-first} \\
\hline 
$\upmu = \hat{\bf p} \cdot \hat{\bf k}$ & \makecell[l]{projection of the momentum unit vector onto \\ the Fourier wavevector associated with ${\bf x}$} & \eqref{Thetaexp}, \eqref{C_ell}, \eqref{Thermal} \\
\hline 
$\vartheta_{\alpha,\ell}$ & Legendre multipole functions of $\Theta_\alpha$ & \eqref{Thetaexp}, \eqref{CN-S-ell} \\
\hline 
\end{tabular}
\end{table}

\begin{table}[!ht]
\centering
\caption{Guided symbols in this paper 2.}
\begin{tabular}{|c|l|c|}
\hline
\textbf{Symbol} & \textbf{Meaning} & \textbf{Reference Eqn} \\
\hline
\hline 
${\cal P}_\ell(\upmu)$ & Legendre polynomials & \makecell[c]{\eqref{Thetaexp}, \eqref{CN-S-ell}, \eqref{Thermal}, \\ \eqref{ctilde_1}} \\
\hline
$f_{g,\ell}$ & \makecell[l]{first-order, gauge-dependent contribution to \\ the Boltzmann hierarchy equations} & \makecell[c]{\eqref{CN-S-ell}, \eqref{fg}, \eqref{f_g} \\
\eqref{HBE-CN-S}, \eqref{BHE-0-CN-S}}\\
\hline
$k$ & wavenumber of Fourier mode & \eqref{CN-S-ell}, \eqref{fg}, \eqref{BHE-0-CN-S} \\
\hline
$q_i$ & $= a p_i$ & -- \\
\hline
$\epsilon$ & $= (q^2 + a^2 m^2)^{1/2}$ & -- \\
\hline
\makecell[c]{$\phi$ \\ $\psi$} & transformed CN metric scalar perturbations & \eqref{fg} \\
\hline
\makecell[c]{$h$ \\ $\eta$} & \makecell[l]{transformed synchronous metric scalar \\ perturbations} & \eqref{f_g} \\
\hline 
$\tilde{C}_\ell$ & \makecell[l]{Legendre coefficient for the  collision term \\ at first order} & \eqref{C_ell}, \eqref{BH-CN-S} \\
\hline
$\nu_{\ell}$ & neutrino multipole moments & \makecell[c]{\eqref{vartheta_ell.expand}, \eqref{nu-ell}, \eqref{BH-CN-S}, \\ \eqref{ctilde_1}, \eqref{HBE-CN-S}} \\
\hline
$\nu_i$ & neutrino field (used at the Lagrangian level) &  \eqref{L-int-gen}, \eqref{L.Dirac}\\
\hline
$\tilde{F}^{(1)}$ & \makecell[l]{linear order on the perturbations of distribution \\ factor in collision term for $2 \to 2$ process} &  \eqref{C-first}, \eqref{F1-eslastic} \\
\hline
$f_{\nu_j}^{(0)}$ & \makecell[l]{unperturbed phase space distribution function \\
for neutrino with $i$ mass eigenstate} & \eqref{Psi_s} \\
\hline
$\Psi_s$ & \makecell[l]{Global factor in $\tilde{F}^{(1)}$ due to local \\ energy conservation} & \eqref{F1-eslastic}, \eqref{Psi_s}, \eqref{Psi_S} \\
\hline
$\alpha_r$ & \makecell[l]{determined by the ratio between couplings with \\
respect to $|g_{ii}|=g$} & \makecell[c]{\eqref{M2} \\ (see Appendix~\ref{Amplitudes})}\\
\hline
$T_r$ & contribution to the scattering amplitudes & \makecell[l]{\eqref{M2}, \eqref{Tr}, \eqref{Tmix} \\(see Appendix~\ref{Amplitudes})} \\
\hline
$\tilde{C}^{(1)}_{\nu_i}$ & \makecell[l]{contribution of one neutrino-neutrino scattering \\ to the collision term} & \eqref{ctilde_1} \\
\hline
${\cal J}_n$ & \makecell[l]{integral associated with the reference particle \\
with momentum ${\bf p}_n$, $n=1,2,3,4$} & \eqref{Gen.Jn} \\
\hline
${\cal K}^{ij}(z_1)$ & \makecell[l]{contribution of one elastic scattering process to \\ the collision term $\nu_i-\nu_j$} & \eqref{K-ell-ij} \\
\hline
\makecell[c]{$J_1, J_{1\ell}, J_{2\ell}, J_{3\ell}$} & \makecell[l]{reduced angular integrals of ${\cal J}_1, {\cal J}_2, {\cal J}_3, {\cal J}_4$}  & \eqref{J-integrals}, \eqref{J-integrals-g}  \\
\hline
${\cal C}^{ij}_\ell(z_1)$ & multipoles of the collision (elastic scatterings) & \eqref{HBE-CN-S}, \eqref{collision_cases}, \eqref{Collison-Deg}\\
\hline
${\rm g}_s$ & degeneracy factor per particle & \eqref{collision_cases}, \eqref{Collison-Deg} \\
\hline
${\cal F}_\ell$ & \makecell[l]{neutrino multipole moments for effectively \\ massless neutrinos} & \eqref{BHE-0-CN-S} \\
\hline
$\Gamma_\ell$ & effectively rate of interaction & \eqref{collision-term-MvsD} \\
\hline
$\beta_\ell, \beta_\ell^{j}$ & \makecell[l]{effective contributions arising from the collision \\ term averaged over the energy density of the \\ reference neutrino with negligible mass} & \makecell[c]{\eqref{collision-term-MvsD}, \eqref{Beta-ell-j}, \eqref{Beta-ell}, \\ \eqref{Xi-ell}} \\
\hline
$\xi_\ell$ & \makecell[l]{thermal correction in the light-mediator scenario} & \eqref{Xi-ell} \\
\hline
\end{tabular}
\end{table}

\newpage

\bibliographystyle{JHEP}
\bibliography{references} 

@article{Gelmini1981left,
  title={Left-handed neutrino mass scale and spontaneously broken lepton number},
  author={Gelmini, Graciela B and Roncadelli, Marco},
  journal={Phys. Lett. B},
  volume={99},
  number={5},
  pages={411--415},
  year={1981},
  publisher={Elsevier},
  doi = {10.1016/0370-2693(81)90559-1},
  url = {https://www.sciencedirect.com/science/article/abs/pii/0370269381905591?via%3Dihub}
}

@article{Chikashige1981there,
  title={Are there real goldstone bosons associated with broken lepton number?},
  author={Chikashige, Y and Mohapatra, Rabindra N and Peccei, Roberto D},
  journal={Phys. Lett. B},
  volume={98},
  number={4},
  pages={265--268},
  year={1981},
  publisher={Elsevier},
  doi = {10.1016/0370-2693(81)90011-3},
  url = {https://doi.org/10.1016/0370-2693(81)90011-3}
}

@article{Miranda_2015,
  doi = {10.1088/1367-2630/17/9/095002},
  url = {https://dx.doi.org/10.1088/1367-2630/17/9/095002},
  year = {2015},
  month = {sep},
  publisher = {IOP Publishing},
  volume = {17},
  number = {9},
  pages = {095002},
  author = {Miranda, O G and Nunokawa, H},
  title = {Non-standard neutrino interactions: current status and future prospects},
  journal = {New J. Phys.},
  eprint = {1505.06254},
  archivePrefix = "arXiv",
}

@article{Chao_2019_dm,
  doi = {10.1088/1475-7516/2019/08/010},
  url = {https://dx.doi.org/10.1088/1475-7516/2019/08/010},
  year = {2019},
  month = {aug},
  publisher = {},
  volume = {2019},
  number = {08},
  pages = {010},
  author = {Chao, Wei and Jiang, Jian-Guo and Wang, Xuan and Zhang, Xing-Yu},
  title = {Direct detections of dark matter in the presence of non-standard neutrino interactions},
  journal = {JCAP},
  eprint = {1904.11214},
  archivePrefix = "arXiv",
}

@Article{Bhupal2019status,
  title={{Neutrino non-standard interactions: A status report}},
  author={P. S. Bhupal Dev and K. S. Babu and Peter B. Denton and Pedro A. N. Machado and Carlos A. Argüelles and Joshua L. Barrow and Sabya Sachi Chatterjee and Mu-Chun Chen and André de Gouvêa and Bhaskar Dutta and Dorival Gonçalves and Tao Han and Matheus Hostert and Sudip Jana and Kevin J. Kelly and Shirley Weishi Li and Ivan Martinez-Soler and Poonam Mehta and Irina Mocioiu and Yuber F. Perez-Gonzalez and Jordi Salvado and Ian M. Shoemaker and Michele Tammaro and Anil Thapa and Jessica Turner and Xun-Jie Xu},
  journal={SciPost Phys. Proc.},
  eprint = {1907.00991},
  archivePrefix = "arXiv",
  pages={001},
  year={2019},
  publisher={SciPost},
  doi={10.21468/SciPostPhysProc.2.001},
  url={https://scipost.org/10.21468/SciPostPhysProc.2.001},
}

@Article{Babu2020_mass_models,
  author={Babu, K. S.
  and Dev, P.S. Bhupal
  and Jana, Sudip
  and Thapa, Anil},
  title={Non-standard interactions in radiative neutrino mass models},
  journal={JHEP},
  eprint = {1907.09498},
  archivePrefix = "arXiv",
  year={2020},
  month={Mar},
  day={02},
  volume={2020},
  number={3},
  pages={6},
  issn={1029-8479},
  doi={10.1007/JHEP03(2020)006},
  url={https://doi.org/10.1007/JHEP03(2020)006}
}

@Article{VanLoi2020,
  author={Van Loi, Duong
  and Van Dong, Phung
  and Van Soa, Dang},
  title={Neutrino mass and dark matter from an approximate {B-L} symmetry},
  journal={JHEP},
  eprint = {1911.04902},
  archivePrefix = "arXiv",
  year={2020},
  month={May},
  day={20},
  volume={2020},
  number={5},
  pages={90},
  issn={1029-8479},
  doi={10.1007/JHEP05(2020)090},
  url={https://doi.org/10.1007/JHEP05(2020)090}
}

@article{ELLIOTT2024,
  title = {The gallium anomaly},
  journal = {Prog. Part. Nucl. Phys.},
  eprint = {2306.03299},
  archivePrefix = "arXiv",
  volume = {134},
  pages = {104082},
  year = {2024},
  issn = {0146-6410},
  doi = {https://doi.org/10.1016/j.ppnp.2023.104082},
  url = {https://www.sciencedirect.com/science/article/pii/S0146641023000637},
  author = {S.R. Elliott and V.N. Gavrin and W.C. Haxton},
}

@article{Blinov2019,
  title = {Constraining the Self-Interacting Neutrino Interpretation of the Hubble Tension},
  author = {Blinov, Nikita and Kelly, Kevin J. and Krnjaic, Gordan and McDermott, Samuel D.},
  journal = {Phys. Rev. Lett.},
  eprint={1905.02727},
  archivePrefix = "arXiv",
  volume = {123},
  issue = {19},
  pages = {191102},
  numpages = {7},
  year = {2019},
  month = {Nov},
  publisher = {American Physical Society},
  doi = {10.1103/PhysRevLett.123.191102},
  url = {https://link.aps.org/doi/10.1103/PhysRevLett.123.191102}
}

@article{Berryman2023neutrino,
  title={Neutrino self-interactions: A white paper},
  author={Berryman, Jeffrey M and Blinov, Nikita and Brdar, Vedran and Brinckmann, Thejs and Bustamante, Mauricio and Cyr-Racine, Francis-Yan and Das, Anirban and de Gouv{\^e}a, Andr{\'e} and Denton, Peter B and Dev, PS Bhupal and others},
  journal={Phys. Dark Univ.},
  eprint={2203.01955},
  archivePrefix = "arXiv",
  volume={42},
  pages={101267},
  year={2023},
  publisher={Elsevier},
  doi={10.1016/j.dark.2023.101267},
  url={https://doi.org/10.1016/j.dark.2023.101267}
}

@article{Bell2006,
  title = {Cosmological signatures of interacting neutrinos},
  author = {Bell, Nicole F. and Pierpaoli, Elena and Sigurdson, Kris},
  journal = {Phys. Rev. D},
  volume = {73},
  issue = {6},
  pages = {063523},
  numpages = {17},
  year = {2006},
  month = {Mar},
  publisher = {American Physical Society},
  doi = {10.1103/PhysRevD.73.063523},
  url = {https://link.aps.org/doi/10.1103/PhysRevD.73.063523}
}

@article{CyrRacine2014,
  title = {Limits on neutrino-neutrino scattering in the early Universe},
  author = {Cyr-Racine, Francis-Yan and Sigurdson, Kris},
  eprint={1306.1536},
  archivePrefix = "arXiv",
  journal = {Phys. Rev. D},
  volume = {90},
  issue = {12},
  pages = {123533},
  numpages = {8},
  year = {2014},
  month = {Dec},
  publisher = {American Physical Society},
  doi = {10.1103/PhysRevD.90.123533},
  url = {https://link.aps.org/doi/10.1103/PhysRevD.90.123533}
}

@article{Archidiacono2014,
  year = 2014,
  month = {jul},
  publisher = {{IOP} Publishing},
  volume = {2014},
  number = {07},
  pages = {046--046},
  author = {Maria Archidiacono and Steen Hannestad},
  title = {Updated constraints on non-standard neutrino interactions from Planck},
  eprint={1311.3873},
  archivePrefix = "arXiv",
  journal = {JCAP},
  doi = {10.1088/1475-7516/2014/07/046},
  url = {https://doi.org/10.1088\%2F1475-7516\%2F2014\%2F07\%2F046},
}

@article{Oldengott2015,
	author = {Isabel M. Oldengott and Cornelius Rampf and Yvonne Y.Y. Wong},
	doi = {10.1088/1475-7516/2015/04/016},
	journal = {JCAP},
	month = {apr},
	number = {04},
	pages = {016},
	publisher = {},
	title = {{Boltzmann hierarchy for interacting neutrinos I: formalism}},
    eprint = "1409.1577",
    archivePrefix = "arXiv",
    url = {https://dx.doi.org/10.1088/1475-7516/2015/04/016},
    volume = {2015},
  year = {2015}
}

@article{Lancaster2017,
  doi = {10.1088/1475-7516/2017/07/033},
  url = {https://doi.org/10.1088\%2F1475-7516\%2F2017\%2F07\%2F033},
  year = 2017,
  month = {jul},
  publisher = {{IOP} Publishing},
  volume = {2017},
  number = {07},
  pages = {033--033},
  author = {Lachlan Lancaster and Francis-Yan Cyr-Racine and Lloyd Knox and Zhen Pan},
  title = {A tale of two modes: neutrino free-streaming in the early universe},
  journal = {JCAP},
  eprint = {1704.06657},
  archivePrefix = "arXiv",
}

@article{Oldengott_2017,
  author = {Isabel M. Oldengott and Thomas Tram and Cornelius Rampf and Yvonne Y.Y. Wong},
  doi = {10.1088/1475-7516/2017/11/027},
  journal = {JCAP},
  month = {nov},
  number = {11},
  pages = {027},
  publisher = {},
  title = {Interacting neutrinos in cosmology: exact description and constraints},
  eprint = {1706.02123},
  archivePrefix = "arXiv",
  url = {https://dx.doi.org/10.1088/1475-7516/2017/11/027},
  volume = {2017},
  year = {2017}
}

@article{Park2019,
  title = {$\mathrm{\ensuremath{\Lambda}}\mathrm{CDM}$ or self-interacting neutrinos: How {CMB} data can tell the two models apart},
  author = {Park, Minsu and Kreisch, Christina D. and Dunkley, Jo and Hadzhiyska, Boryana and Cyr-Racine, Francis-Yan},
  eprint = "1904.02625",
  archivePrefix = "arXiv",
  journal = {Phys. Rev. D},
  volume = {100},
  issue = {6},
  pages = {063524},
  numpages = {7},
  year = {2019},
  month = {Sep},
  publisher = {American Physical Society},
  doi = {10.1103/PhysRevD.100.063524},
  url = {https://link.aps.org/doi/10.1103/PhysRevD.100.063524}
}

@article{Kreisch2020,
  title = {Neutrino puzzle: Anomalies, interactions, and cosmological tensions},
  eprint = {1902.00534},
  archivePrefix = "arXiv",
  author = {Kreisch, Christina D. and Cyr-Racine, Francis-Yan and Dor\'e, Olivier},
  journal = {Phys. Rev. D},
  volume = {101},
  issue = {12},
  pages = {123505},
  numpages = {34},
  year = {2020},
  month = {Jun},
  publisher = {American Physical Society},
  doi = {10.1103/PhysRevD.101.123505},
  url = {https://link.aps.org/doi/10.1103/PhysRevD.101.123505}
}

@article{Choudhury2021,
  author = "Roy Choudhury, Shouvik and Hannestad, Steen and Tram, Thomas",
  title = "{Updated constraints on massive neutrino self-interactions from cosmology in light of the $H_0$ tension}",
  doi = "10.1088/1475-7516/2021/03/084",
  journal = "JCAP",
  volume={2021},
  number={03},
  pages = "084",
  year = "2021",
  eprint = "2012.07519",
  archivePrefix = "arXiv",
  primaryClass = "astro-ph.CO"
}

@article{brinckmann2020self,
  title = {{Self-interacting neutrinos, the Hubble parameter tension, and the cosmic microwave background}},
  author = {Brinckmann, Thejs and Chang, Jae Hyeok and LoVerde, Marilena},
  journal = {Phys. Rev. D},
  eprint = "2012.11830",
  archivePrefix = "arXiv",
  volume = {104},
  issue = {6},
  pages = {063523},
  numpages = {40},
  year = {2021},
  month = {Sep},
  publisher = {American Physical Society},
  doi = {10.1103/PhysRevD.104.063523},
  url = {https://link.aps.org/doi/10.1103/PhysRevD.104.063523}
}

@article{mazumdar2022flavour,
  doi = {10.1088/1475-7516/2022/10/011},
  url = {https://dx.doi.org/10.1088/1475-7516/2022/10/011},
  year = {2022},
  month = {oct},
  publisher = {IOP Publishing},
  volume = {2022},
  number = {10},
  pages = {011},
  author = {Arindam Mazumdar and Subhendra Mohanty and Priyank Parashari},
  title = {{Flavour specific neutrino self-interaction: $H_0$ tension and IceCube}},
  journal = {JCAP},
  eprint = "2011.13685",
  archivePrefix = "arXiv",
}

@article{Kreisch:2022zxp,
    author = "Kreisch, Christina D. and others",
    title = "{Atacama Cosmology Telescope: The persistence of neutrino self-interaction in cosmological measurements}",
    eprint = "2207.03164",
    archivePrefix = "arXiv",
    primaryClass = "astro-ph.CO",
    doi = "10.1103/PhysRevD.109.043501",
    journal = "Phys. Rev. D",
    volume = "109",
    number = "4",
    pages = "043501",
    year = "2024"
}

@article{RoyChoudhury2022,
    author = "Roy Choudhury, Shouvik and Hannestad, Steen and Tram, Thomas",
    title = "{Massive neutrino self-interactions and inflation}",
    eprint = "2207.07142",
    archivePrefix = "arXiv",
    primaryClass = "astro-ph.CO",
    doi = "10.1088/1475-7516/2022/10/018",
    journal = "JCAP",
    volume = "10",
    pages = "018",
    year = "2022"
}

@article{Das:2023npl,
    author = "Das, Anirban and Ghosh, Subhajit",
    title = "{The magnificent ACT of flavor-specific neutrino self-interaction}",
    eprint = "2303.08843",
    archivePrefix = "arXiv",
    primaryClass = "astro-ph.CO",
    reportNumber = "SLAC-PUB-17708",
    doi = "10.1088/1475-7516/2023/09/042",
    journal = "JCAP",
    volume = "09",
    pages = "042",
    year = "2023"
}

@article{Camarena2023,
  title = {Confronting self-interacting neutrinos with the full shape of the galaxy power spectrum},
  author = {Camarena, David and Cyr-Racine, Francis-Yan and Houghteling, John},
  eprint={2309.03941},
  archivePrefix = "arXiv",
  journal = {Phys. Rev. D},
  volume = {108},
  issue = {10},
  pages = {103535},
  numpages = {20},
  year = {2023},
  month = {Nov},
  publisher = {American Physical Society},
  doi = {10.1103/PhysRevD.108.103535},
  url = {https://link.aps.org/doi/10.1103/PhysRevD.108.103535}
}

@article{he2024self,
  title={Self-interacting neutrinos in light of large-scale structure data},
  author={He, Adam and An, Rui and Ivanov, Mikhail M and Gluscevic, Vera},
  journal={Phys. Rev. D},
  eprint = "2309.03956",
  archivePrefix = "arXiv",
  volume={109},
  number={10},
  pages={103527},
  year={2024},
  publisher={APS},
  doi= {10.1103/PhysRevD.109.103527},
  url= {https://link.aps.org/doi/10.1103/PhysRevD.109.103527}
}

@article{camarena2025strong,
  title={Strong constraints on a simple self-interacting neutrino cosmology},
  author={Camarena, David and Cyr-Racine, Francis-Yan},
  journal={Phys. Rev. D},
  eprint = "2403.05496",
  archivePrefix = "arXiv",
  volume={111},
  number={2},
  pages={023504},
  year={2025},
  publisher={APS},
  doi={10.1103/PhysRevD.111.023504},
  url={https://link.aps.org/doi/10.1103/PhysRevD.111.023504}
}

@article{abellan2026neutrino,
  title={Neutrino decays as a natural explanation of the neutrino mass tension},
  author={Abell{\'a}n, Guillermo Franco},
  journal={},
  year={2026},
  eprint = "2601.04312",
  archivePrefix = "arXiv",
}

@article{Poudou:2025qcx,
  author = "Poudou, Ad\`ele and Simon, Th\'eo and Montandon, Thomas and Teixeira, Elsa M. and Poulin, Vivian",
  title = "{Self-interacting neutrinos in light of recent CMB and LSS data}",
  eprint = "2503.10485",
  archivePrefix = "arXiv",
  primaryClass = "astro-ph.CO",
  month = "3",
  year = "2025",
  journal = {Phys. Rev. D},
  doi={10.1103/mljb-42fm},
  url={https://link.aps.org/10.1103/mljb-42fm}
}

@article{He:2025jwp,
  author = "He, Adam and Ivanov, Mikhail M. and Bird, Simeon and An, Rui and Gluscevic, Vera",
  title = "{A Fresh Look at Neutrino Self-Interactions With the Lyman-$\alpha$ Forest: Constraints from EFT and PRIYA}",
  eprint = "2503.15592",
  archivePrefix = "arXiv",
  primaryClass = "astro-ph.CO",
  reportNumber = "MIT-CTP/5854",
  month = "3",
  year = "2025",
  journal = {Phys. Rev. D},
  doi={10.1103/wzpy-p7w8},
  url={https://link.aps.org/10.1103/wzpy-p7w8}
}

@article{Montefalcone:2025ibh,
  title="{Directly probing neutrino interactions through CMB phase shift measurements}",
  author={Montefalcone, Gabriele and Ghosh, Subhajit and Boddy, Kimberly K and Ho, Daven Wei Ren and Tsai, Yuhsin},
  journal={Phys. Rev. D},
  volume={113},
  number={2},
  pages={023540},
  year={2026},
  publisher={APS},
  doi={10.1103/lylf-3xyq},
  url={https://journals.aps.org/prd/abstract/10.1103/lylf-3xyq},
  eprint = "2509.20363",
  archivePrefix = "arXiv",
  primaryClass = "astro-ph.CO",
}

@article{Forastieri:2015paa,
    author = "Forastieri, Francesco and Lattanzi, Massimiliano and Natoli, Paolo",
    title = "{Constraints on secret neutrino interactions after Planck}",
    doi = "10.1088/1475-7516/2015/07/014",
    journal = "JCAP",
    volume = "07",
    pages = "014",
    year = "2015",
    eprint = "1504.04999",
    archivePrefix = "arXiv",
    primaryClass = "astro-ph.CO"
}

@article{Forastieri:2019cuf,
    author = "Forastieri, Francesco and Lattanzi, Massimiliano and Natoli, Paolo",
    title = "{Cosmological constraints on neutrino self-interactions with a light mediator}",
    eprint = "1904.07810",
    archivePrefix = "arXiv",
    primaryClass = "astro-ph.CO",
    doi = "10.1103/PhysRevD.100.103526",
    journal = "Phys. Rev. D",
    volume = "100",
    number = "10",
    pages = "103526",
    year = "2019"
}

@article{Venzor2022,
  title = {Massive neutrino self-interactions with a light mediator in cosmology},
  author = {Venzor, Jorge and Garcia-Arroyo, Gabriela and P\'erez-Lorenzana, Abdel and De-Santiago, Josue},
  journal = {Phys. Rev. D},
  eprint = "2202.09310",
  archivePrefix = "arXiv",
  volume = {105},
  issue = {12},
  pages = {123539},
  numpages = {10},
  year = {2022},
  month = {Jun},
  publisher = {American Physical Society},
  doi = {10.1103/PhysRevD.105.123539},
  url = {https://link.aps.org/doi/10.1103/PhysRevD.105.123539}
}

@article{venzor2023resonant,
  title={Resonant neutrino self-interactions and the {$H_0$} tension},
  author={Venzor, Jorge and Garcia-Arroyo, Gabriela and De-Santiago, Josue and P{\'e}rez-Lorenzana, Abdel},
  eprint = {2303.12792},
  archivePrefix = "arXiv",
  journal={Phys. Rev. D},
  volume={108},
  number={4},
  pages={043536},
  year={2023},
  publisher={APS},
  doi= {10.1103/PhysRevD.108.043536},
  url={https://link.aps.org/doi/10.1103/PhysRevD.108.043536}
}

@article{Noriega2025resonant,
  title = {Resonant neutrino self-interactions: Insights from the full shape galaxy power spectrum},
  author = {Noriega, Hern\'an E. and De-Santiago, Josue and Garcia-Arroyo, Gabriela and Venzor, Jorge and P\'erez-Lorenzana, Abdel},
  journal = {Phys. Rev. D},
  eprint = {2506.07994},
  archivePrefix = "arXiv",
  volume = {112},
  issue = {6},
  pages = {063509},
  numpages = {12},
  year = {2025},
  month = {Sep},
  publisher = {American Physical Society},
  doi = {10.1103/b9x4-hnqn},
  url = {https://link.aps.org/doi/10.1103/b9x4-hnqn}
}

@article{Beacom2004,
  title = {Neutrinoless Universe},
  author = {Beacom, John F. and Bell, Nicole F. and Dodelson, Scott},
  journal = {Phys. Rev. Lett.},
  volume = {93},
  issue = {12},
  pages = {121302},
  numpages = {4},
  year = {2004},
  month = {Sep},
  publisher = {American Physical Society},
  doi = {10.1103/PhysRevLett.93.121302},
  url = {https://link.aps.org/doi/10.1103/PhysRevLett.93.121302}
}

@article{Hannestad2005,
  title={Structure formation with strongly interacting neutrinos—implications for the cosmological neutrino mass bound},
  author={Hannestad, Steen},
  journal={JCAP},
  volume={2005},
  number={02},
  pages={011},
  year={2005},
  publisher={IOP Publishing},
  doi={10.1088/1475-7516/2005/02/011},
  url={https://arxiv.org/abs/astro-ph/0411475}
}

@article{Heurtier2017,
  doi = {10.1088/1475-7516/2017/02/042},
  url = {https://doi.org/10.1088\%2F1475-7516\%2F2017\%2F02\%2F042},
  year = 2017,
  month = {feb},
  publisher = {{IOP} Publishing},
  volume = {2017},
  number = {02},
  pages = {042--042},
  author = {Lucien Heurtier and Yongchao Zhang},
  title = {Supernova constraints on massive (pseudo)scalar coupling to neutrinos},
  journal = {JCAP},
  eprint={1609.05882},
  archivePrefix = "arXiv"
}

@article{Huang2018,
  title = {Observational constraints on secret neutrino interactions from big bang nucleosynthesis},
  author = {Huang, Guo-yuan and Ohlsson, Tommy and Zhou, Shun},
  journal = {Phys. Rev. D},
  eprint={1712.04792},
  archivePrefix = "arXiv",
  volume = {97},
  issue = {7},
  pages = {075009},
  numpages = {12},
  year = {2018},
  month = {Apr},
  publisher = {American Physical Society},
  doi = {10.1103/PhysRevD.97.075009},
  url = {https://link.aps.org/doi/10.1103/PhysRevD.97.075009}
}

@article{Berryman2018,
  title = {Lepton-number-charged scalars and neutrino beamstrahlung},
  author = {Berryman, Jeffrey M. and de Gouv\^ea, Andr\'e and Kelly, Kevin J. and Zhang, Yue},
  journal = {Phys. Rev. D},
  eprint={1802.00009},
  archivePrefix = "arXiv",
  volume = {97},
  issue = {7},
  pages = {075030},
  numpages = {16},
  year = {2018},
  month = {Apr},
  publisher = {American Physical Society},
  doi = {10.1103/PhysRevD.97.075030},
  url = {https://link.aps.org/doi/10.1103/PhysRevD.97.075030}
}

@article{BLUM2018,
  title = {Neutrinoless double-beta decay with massive scalar emission},
  journal = {Phys. Lett. B},
  eprint={1802.08019},
  archivePrefix = "arXiv",
  volume = {785},
  pages = {354-361},
  year = {2018},
  issn = {0370-2693},
  doi = {https://doi.org/10.1016/j.physletb.2018.08.022},
  url = {https://www.sciencedirect.com/science/article/pii/S0370269318306282},
  author = {Kfir Blum and Yosef Nir and Michal Shavit}
}

@article{Brune2019,
  author = {Brune, Tim and P\"as, Heinrich},
  doi = {10.1103/PhysRevD.99.096005},
  issn = {2470-0029},
  journal = {Phys. Rev. D},
  number = {9},
  pages = {96005},
  publisher = {American Physical Society},
  title = {{Massive Majorons and constraints on the Majoron-neutrino coupling}},
  eprint={1808.08158},
  archivePrefix = "arXiv",
  url = {https://doi.org/10.1103/PhysRevD.99.096005},
  volume = {99},
  year = {2019}
}

@article{Barenboim2019inflation,
  title = {Constraints on inflation with an extended neutrino sector},
  author = {Barenboim, Gabriela and Denton, Peter B. and Oldengott, Isabel M.},
  journal = {Phys. Rev. D},
  eprint={1903.02036},
  archivePrefix = "arXiv",
  volume = {99},
  issue = {8},
  pages = {083515},
  numpages = {9},
  year = {2019},
  month = {Apr},
  publisher = {American Physical Society},
  doi = {10.1103/PhysRevD.99.083515},
  url = {https://link.aps.org/doi/10.1103/PhysRevD.99.083515}
}

@article{Schoneberg2019,
  doi = {10.1088/1475-7516/2019/10/029},
  url = {https://doi.org/10.1088/1475-7516/2019/10/029},
  year = 2019,
  month = {oct},
  publisher = {{IOP} Publishing},
  volume = {2019},
  number = {10},
  pages = {029--029},
  author = {Nils Schöneberg and Julien Lesgourgues and Deanna C. Hooper},
  title = {{The BAO+BBN take on the Hubble tension}},
  eprint={1907.11594},
  archivePrefix = "arXiv",
  journal = {JCAP}
}

@article{Escudero2020_CMB_search,
  author = {Escudero, Miguel and Witte, Samuel J},
  doi = {10.1140/epjc/s10052-020-7854-5},
  issn = {1434-6052},
  journal = {Eur. Phys. J. C},
  number = {4},
  pages = {294},
  title = {{A CMB search for the neutrino mass mechanism and its relation to the Hubble tension}},
  url = {https://doi.org/10.1140/epjc/s10052-020-7854-5},
  volume = {80},
  year = {2020},
  eprint={1909.04044},
  archivePrefix = "arXiv"
}

@Article{Chacko2020,
  author={Chacko, Zackaria
  and Dev, Abhish
  and Du, Peizhi
  and Poulin, Vivian
  and Tsai, Yuhsin},
  title={Cosmological limits on the neutrino mass and lifetime},
  journal={JHEP},
  eprint={1909.05275},
  archivePrefix = "arXiv",
  year={2020},
  month={Apr},
  day={03},
  volume={2020},
  number={4},
  pages={20},
  issn={1029-8479},
  doi={10.1007/JHEP04(2020)020},
  url={https://doi.org/10.1007/JHEP04(2020)020}
}

@article{Bustamante2020bounds,
  title={Bounds on secret neutrino interactions from high-energy astrophysical neutrinos},
  author={Bustamante, Mauricio and Rosenstr{\o}m, Charlotte and Shalgar, Shashank and Tamborra, Irene},
  eprint={2001.04994},
  archivePrefix = "arXiv",
  journal={Phys. Rev. D},
  volume={101},
  number={12},
  pages={123024},
  year={2020},
  publisher={APS},
  doi={10.1103/PhysRevD.101.123024},
  url={https://link.aps.org/10.1103/PhysRevD.101.123024}
}

@article{Brdar2020,
  title = {Revisiting neutrino self-interaction constraints from {$Z$} and $\ensuremath{\tau}$ decays},
  author = {Brdar, Vedran and Lindner, Manfred and Vogl, Stefan and Xu, Xun-Jie},
  journal = {Phys. Rev. D},
  eprint={2003.05339},
  archivePrefix = "arXiv",
  volume = {101},
  issue = {11},
  pages = {115001},
  numpages = {13},
  year = {2020},
  month = {Jun},
  publisher = {American Physical Society},
  doi = {10.1103/PhysRevD.101.115001},
  url = {https://link.aps.org/doi/10.1103/PhysRevD.101.115001}
}

@article{Deppisch2020,
  title = {Neutrino self-interactions and double beta decay},
  author = {Deppisch, Frank F. and Graf, Lukas and Rodejohann, Werner and Xu, Xun-Jie},
  journal = {Phys. Rev. D},
  eprint={2004.11919},
  archivePrefix = "arXiv",
  volume = {102},
  issue = {5},
  pages = {051701},
  numpages = {6},
  year = {2020},
  month = {Sep},
  publisher = {American Physical Society},
  doi = {10.1103/PhysRevD.102.051701},
  url = {https://link.aps.org/doi/10.1103/PhysRevD.102.051701}
}

@Article{Escudero2020relaxing,
  author={Escudero, Miguel
  and Lopez-Pavon, Jacobo
  and Rius, Nuria
  and Sandner, Stefan},
  title={Relaxing cosmological neutrino mass bounds with unstable neutrinos},
  journal={JHEP},
  eprint={2007.04994},
  archivePrefix = "arXiv",
  year={2020},
  month={Dec},
  day={18},
  volume={2020},
  number={12},
  pages={119},
  issn={1029-8479},
  doi={10.1007/JHEP12(2020)119},
  url={https://doi.org/10.1007/JHEP12(2020)119}
}

@article{Shalgar2021,
  title = {Core-collapse supernova stymie secret neutrino interactions},
  author = {Shalgar, Shashank and Tamborra, Irene and Bustamante, Mauricio},
  journal = {Phys. Rev. D},
  eprint = "1912.09115",
  archivePrefix = "arXiv",
  volume = {103},
  issue = {12},
  pages = {123008},
  numpages = {9},
  year = {2021},
  month = {Jun},
  publisher = {American Physical Society},
  doi = {10.1103/PhysRevD.103.123008},
  url = {https://link.aps.org/doi/10.1103/PhysRevD.103.123008}
}

@article{Lyu2021,
  title = {Self-interacting neutrinos: Solution to Hubble tension versus experimental constraints},
  author = {Lyu, Kun-Feng and Stamou, Emmanuel and Wang, Lian-Tao},
  journal = {Phys. Rev. D},
  eprint = "2004.10868",
  archivePrefix = "arXiv",
  volume = {103},
  issue = {1},
  pages = {015004},
  numpages = {11},
  year = {2021},
  month = {Jan},
  publisher = {American Physical Society},
  doi = {10.1103/PhysRevD.103.015004},
  url = {https://link.aps.org/doi/10.1103/PhysRevD.103.015004}
}

@article{Creque2021,
  title = {Resonant neutrino self-interactions},
  author = {Creque-Sarbinowski, Cyril and Hyde, Jeffrey and Kamionkowski, Marc},
  journal = {Phys. Rev. D},
  eprint={2005.05332},
  archivePrefix = "arXiv",
  volume = {103},
  issue = {2},
  pages = {023527},
  numpages = {11},
  year = {2021},
  month = {Jan},
  publisher = {American Physical Society},
  doi = {10.1103/PhysRevD.103.023527},
  url = {https://link.aps.org/doi/10.1103/PhysRevD.103.023527}
}

@article{Venzor2021,
  title = {Bounds on neutrino-scalar nonstandard interactions from big bang nucleosynthesis},
  author = {Venzor, Jorge and P\'erez-Lorenzana, Abdel and De-Santiago, Josue},
  journal = {Phys. Rev. D},
  eprint = "2009.08104",
  archivePrefix = "arXiv",
  volume = {103},
  issue = {4},
  pages = {043534},
  numpages = {12},
  year = {2021},
  month = {Feb},
  publisher = {American Physical Society},
  doi = {10.1103/PhysRevD.103.043534},
  url = {https://link.aps.org/doi/10.1103/PhysRevD.103.043534}
}

@article{Suliga2021,
  title = {Astrophysical constraints on nonstandard coherent neutrino-nucleus scattering},
  author = {Suliga, Anna M. and Tamborra, Irene},
  journal = {Phys. Rev. D},
  eprint = "2010.14545",
  archivePrefix = "arXiv",
  volume = {103},
  issue = {8},
  pages = {083002},
  numpages = {20},
  year = {2021},
  month = {Apr},
  publisher = {American Physical Society},
  doi = {10.1103/PhysRevD.103.083002},
  url = {https://link.aps.org/doi/10.1103/PhysRevD.103.083002}
}

@article{Barenboim2021invisible,
  title={Invisible neutrino decay in precision cosmology},
  author={Barenboim, Gabriela and Chen, Joe Zhiyu and Hannestad, Steen and Oldengott, Isabel M and Tram, Thomas and Wong, Yvonne YY},
  eprint = "2011.01502",
  archivePrefix = "arXiv",
  journal={JCAP},
  volume={2021},
  number={03},
  pages={087},
  year={2021},
  publisher={IOP Publishing},
  doi ={10.1088/1475-7516/2021/03/087},
  url ={https://iopscience.iop.org/article/10.1088/1475-7516/2021/03/087}
}

@article{Seto2021,
  title = {{Comparing early dark energy and extra radiation solutions to the Hubble tension with BBN}},
  author = {Seto, Osamu and Toda, Yo},
  journal = {Phys. Rev. D},
  eprint = "2101.03740",
  archivePrefix = "arXiv",
  volume = {103},
  issue = {12},
  pages = {123501},
  numpages = {12},
  year = {2021},
  month = {Jun},
  publisher = {American Physical Society},
  doi = {10.1103/PhysRevD.103.123501},
  url = {https://link.aps.org/doi/10.1103/PhysRevD.103.123501}
}

@article{Esteban2021,
  doi = {10.1088/1475-7516/2021/05/036},
  url = {https://doi.org/10.1088/1475-7516/2021/05/036},
  year = 2021,
  month = {may},
  publisher = {{IOP} Publishing},
  volume = {2021},
  number = {05},
  pages = {036},
  author = {Ivan Esteban and Jordi Salvado},
  title = {Long range interactions in cosmology: implications for neutrinos},
  journal={JCAP},
  eprint={2101.05804},
  archivePrefix = "arXiv",
}

@article{Das2021,
  doi = {10.1088/1475-7516/2021/07/038},
  url = {https://doi.org/10.1088/1475-7516/2021/07/038},
  year = 2021,
  month = {jul},
  publisher = {{IOP} Publishing},
  volume = {2021},
  number = {07},
  pages = {038},
  author = {Anirban Das and Subhajit Ghosh},
  title = {Flavor-specific interaction favors strong neutrino self-coupling in the early universe},
  journal = {JCAP},
  eprint={2011.12315},
  archivePrefix = "arXiv",
}

@article{Escrihuela:2021mud,
  author = "Escrihuela, F. J. and Flores, L. J. and Miranda, O. G. and Rend\'on, Javier",
  title = "{Global constraints on neutral-current generalized neutrino interactions}",
  eprint = "2105.06484",
  archivePrefix = "arXiv",
  primaryClass = "hep-ph",
  doi = "10.1007/JHEP07(2021)061",
  journal = "JHEP",
  volume = "07",
  pages = "061",
  year = "2021"
}

@article{Cerdeno2021,
  title = {Medium effects in supernovae constraints on light mediators},
  author = {Cerde\~no, David G. and Cerme\~no, Marina and P\'erez-Garc\'{\i}a, M. \'Angeles and Reid, Elliott},
  journal = {Phys. Rev. D},
  eprint = "2106.11660",
  archivePrefix = "arXiv",
  volume = {104},
  issue = {6},
  pages = {063013},
  numpages = {16},
  year = {2021},
  month = {Sep},
  publisher = {American Physical Society},
  doi = {10.1103/PhysRevD.104.063013},
  url = {https://link.aps.org/doi/10.1103/PhysRevD.104.063013}
}

@article{Esteban2021_probing,
  title = {Probing secret interactions of astrophysical neutrinos in the high-statistics era},
  author = {Esteban, Ivan and Pandey, Sujata and Brdar, Vedran and Beacom, John F.},
  journal = {Phys. Rev. D},
  eprint = "2107.13568",
  archivePrefix = "arXiv",
  volume = {104},
  issue = {12},
  pages = {123014},
  numpages = {17},
  year = {2021},
  month = {Dec},
  publisher = {American Physical Society},
  doi = {10.1103/PhysRevD.104.123014},
  url = {https://link.aps.org/doi/10.1103/PhysRevD.104.123014}
}

@article{Ge:2021lur,
  author = "Ge, Shao-Feng and Pasquini, Pedro",
  title = "{Probing light mediators in the radiative emission of neutrino pair}",
  eprint = "2110.03510",
  archivePrefix = "arXiv",
  primaryClass = "hep-ph",
  doi = "10.1140/epjc/s10052-022-10131-4",
  journal = "Eur. Phys. J. C",
  volume = "82",
  number = "3",
  pages = "208",
  year = "2022"
}

@article{Medhi:2021wxj,
  author = "Medhi, Abinash and Dutta, Debajyoti and Devi, Moon Moon",
  title = "{Exploring the effects of scalar non standard interactions on the CP violation sensitivity at DUNE}",
  eprint = "2111.12943",
  archivePrefix = "arXiv",
  primaryClass = "hep-ph",
  doi = "10.1007/JHEP06(2022)129",
  journal = "JHEP",
  volume = "06",
  pages = "129",
  year = "2022"
}

@Article{Abellan2022,
  author={Abell{\'a}n, Guillermo Franco
  and Chacko, Zackaria
  and Dev, Abhish
  and Du, Peizhi
  and Poulin, Vivian
  and Tsai, Yuhsin},
  title={Improved cosmological constraints on the neutrino mass and lifetime},
  journal={JHEP},
  eprint = {2112.13862},
  archivePrefix = "arXiv",
  year={2022},
  month={Aug},
  day={04},
  volume={2022},
  number={8},
  pages={76},
  issn={1029-8479},
  doi={10.1007/JHEP08(2022)076},
  url={https://doi.org/10.1007/JHEP08(2022)076}
}

@article{Chen2022weaker,
  title={Weaker yet again: mass spectrum-consistent cosmological constraints on the neutrino lifetime},
  eprint = {2203.09075},
  archivePrefix = "arXiv",
  author={Chen, Joe Zhiyu and Oldengott, Isabel M and Pierobon, Giovanni and Wong, Yvonne YY},
  journal={Eur. Phys. J. C},
  volume={82},
  number={7},
  pages={640},
  year={2022},
  publisher={Springer},
  doi = {10.1140/epjc/s10052-022-10518-3},
  url = {https://doi.org/10.1140/epjc/s10052-022-10518-3}
}

@article{Kumar_2022,
  title = "{Updating non-standard neutrinos properties with Planck-CMB data and full-shape analysis of BOSS and eBOSS galaxies}",
  journal = {JCAP},
  year = {2022},
  month = {sep},
  publisher = {IOP Publishing},
  volume = {2022},
  number = {09},
  pages = {060},
  author = {Kumar, Suresh and Nunes, Rafael C. and Yadav, Priya},
  doi = {10.1088/1475-7516/2022/09/060},
  url = {https://dx.doi.org/10.1088/1475-7516/2022/09/060},
  eprint = {2205.04292},
  archivePrefix = "arXiv"
}

@Article{Akita2022,
  author={Akita, Kensuke
  and Im, Sang Hui
  and Masud, Mehedi},
  title={Probing non-standard neutrino interactions with a light boson from next galactic and diffuse supernova neutrinos},
  journal={JHEP},
  eprint = {2206.06852},
  archivePrefix = "arXiv",
  year={2022},
  month={Dec},
  day={09},
  volume={2022},
  number={12},
  pages={50},
  issn={1029-8479},
  doi={10.1007/JHEP12(2022)050},
  url={https://doi.org/10.1007/JHEP12(2022)050}
}

@article{chang2022towards,
  title = {Toward Powerful Probes of Neutrino Self-Interactions in Supernovae},
  author = {Chang, Po-Wen and Esteban, Ivan and Beacom, John F. and Thompson, Todd A. and Hirata, Christopher M.},
  journal = {Phys. Rev. Lett.},
  eprint = {2206.12426},
  archivePrefix = "arXiv",
  volume = {131},
  issue = {7},
  pages = {071002},
  numpages = {8},
  year = {2023},
  month = {Aug},
  publisher = {American Physical Society},
  doi = {10.1103/PhysRevLett.131.071002},
  url = {https://link.aps.org/doi/10.1103/PhysRevLett.131.071002}
}

@article{Fiorillo2022,
  title = {Strong Supernova 1987A Constraints on Bosons Decaying to Neutrinos},
  author = {Fiorillo, Damiano F. G. and Raffelt, Georg G. and Vitagliano, Edoardo},
  journal = {Phys. Rev. Lett.},
  eprint = {2209.11773},
  archivePrefix = "arXiv",
  volume = {131},
  issue = {2},
  pages = {021001},
  numpages = {7},
  year = {2023},
  month = {Jul},
  publisher = {American Physical Society},
  doi = {10.1103/PhysRevLett.131.021001},
  url = {https://link.aps.org/doi/10.1103/PhysRevLett.131.021001}
}

@Article{Dutta:2023fdt,
  author={Dutta, Bhaskar
  and Ghosh, Sumit
  and Li, Tianjun
  and Thompson, Adrian
  and Verma, Ankur},
  title={Non-standard neutrino interactions in light mediator models at reactor experiments},
  journal={JHEP},
  eprint = {2209.13566},
  archivePrefix = "arXiv",
  year={2023},
  month={Mar},
  day={22},
  volume={2023},
  number={3},
  pages={163},
  issn={1029-8479},
  doi={10.1007/JHEP03(2023)163},
  url={https://doi.org/10.1007/JHEP03(2023)163}
}

@article{cerdeno2023constraints,
  title = {Constraints from the duration of supernova neutrino burst on on-shell light gauge boson production by neutrinos},
  author = {Cerde\~no, David and Cerme\~no, Marina and Farzan, Yasaman},
  journal = {Phys. Rev. D},
  eprint = {2301.00661},
  archivePrefix = "arXiv",
  volume = {107},
  issue = {12},
  pages = {123012},
  numpages = {15},
  year = {2023},
  month = {Jun},
  publisher = {American Physical Society},
  doi = {10.1103/PhysRevD.107.123012},
  url = {https://link.aps.org/doi/10.1103/PhysRevD.107.123012}
}

@article{Doring_2024,
  doi = {10.1088/1475-7516/2024/07/015},
  url = {https://dx.doi.org/10.1088/1475-7516/2024/07/015},
  year = {2024},
  month = {jul},
  publisher = {IOP Publishing},
  volume = {2024},
  number = {07},
  pages = {015},
  author = {Döring, Christian and Vogl, Stefan},
  title = {Testing secret interaction with astrophysical neutrino point sources},
  journal = {JCAP},
  eprint = "2304.08533",
  archivePrefix = "arXiv"
}

@Article{Sandner2023,
  author={Sandner, Stefan
  and Escudero, Miguel
  and Witte, Samuel J.},
  title={{Precision CMB constraints on eV-scale bosons coupled to neutrinos}},
  journal={Eur. Phys. J. C},
  eprint = {2305.01692},
  archivePrefix = "arXiv",
  year={2023},
  month={Aug},
  day={09},
  volume={83},
  number={8},
  pages={709}, 
  issn={1434-6052},
  doi={10.1140/epjc/s10052-023-11864-6},
  url={https://doi.org/10.1140/epjc/s10052-023-11864-6}
}

@article{Fiorillo2024_theoretical,
  title = {Supernova emission of secretly interacting neutrino fluid: Theoretical foundations},
  author = {Fiorillo, Damiano F. G. and Raffelt, Georg G. and Vitagliano, Edoardo},
  journal = {Phys. Rev. D},
  eprint = "2307.15122",
  archivePrefix = "arXiv",
  volume = {109},
  issue = {2},
  pages = {023017},
  numpages = {21},
  year = {2024},
  month = {Jan},
  publisher = {American Physical Society},
  doi = {10.1103/PhysRevD.109.023017},
  url = {https://link.aps.org/doi/10.1103/PhysRevD.109.023017}
}

@article{Fiorillo2024_small_impact,
  title = {Large Neutrino Secret Interactions Have a Small Impact on Supernovae},
  author = {Fiorillo, Damiano F. G. and Raffelt, Georg G. and Vitagliano, Edoardo},
  journal = {Phys. Rev. Lett.},
  eprint = "2307.15115",
  archivePrefix = "arXiv",
  volume = {132},
  issue = {2},
  pages = {021002},
  numpages = {7},
  year = {2024},
  month = {Jan},
  publisher = {American Physical Society},
  doi = {10.1103/PhysRevLett.132.021002},
  url = {https://link.aps.org/doi/10.1103/PhysRevLett.132.021002}
}

@article{Bostan_2024,
  doi = {10.1088/1475-7516/2024/07/032},
  url = {https://dx.doi.org/10.1088/1475-7516/2024/07/032},
  year = {2024},
  month = {jul},
  publisher = {IOP Publishing},
  volume = {2024},
  number = {07},
  pages = {032},
  author = {Bostan, Nilay and Choudhury, Shouvik Roy},
  title = {First constraints on non-minimally coupled Natural and Coleman-Weinberg inflation and massive neutrino self-interactions with Planck+BICEP/Keck},
  journal = {JCAP},
  eprint = "2310.01491",
  archivePrefix = "arXiv"
}

@article{akita2024limits,
  title="{Limits on heavy neutral leptons, Z' bosons and majorons from high-energy supernova neutrinos}",
  author={Akita, Kensuke and Im, Sang Hui and Masud, Mehedi and Yun, Seokhoon},
  journal={JHEP},
  eprint = "2411.11749",
  archivePrefix = "arXiv",
  volume={2024},
  number={7},
  pages={1--38},
  year={2024},
  publisher={Springer},
  doi={10.1007/JHEP07(2024)057},
  url={https://link.springer.com/article/10.1007/JHEP07(2024)057}
}

@article{Medhi:2023ebi,
  author = "Medhi, Abinash and Sarker, Arnab and Devi, Moon Moon",
  title = "{Scalar NSI: a unique tool for constraining absolute neutrino masses via neutrino oscillations}",
  eprint = "2307.05348",
  archivePrefix = "arXiv",
  primaryClass = "hep-ph",
  doi = "10.1140/epjc/s10052-025-14089-x",
  journal = "Eur. Phys. J. C",
  volume = "85",
  number = "4",
  pages = "380",
  year = "2025"
}

@article{dev2025new,
  title={New laboratory constraints on neutrinophilic mediators},
  author={Dev, PS Bhupal and Kim, Doojin and Sathyan, Deepak and Sinha, Kuver and Zhang, Yongchao},
  journal={Phys. Lett. B},
  pages={139765},
  year={2025},
  publisher={Elsevier},
  eprint = "2407.12738",
  archivePrefix = "arXiv",
  doi={10.1016/j.physletb.2025.139765},
  url={https://doi.org/10.1016/j.physletb.2025.139765}
}

@article{Denton:2024upc,
    author = "Denton, Peter B. and Giarnetti, Alessio and Meloni, Davide",
    title = "{Solar neutrinos and the strongest oscillation constraints on scalar NSI}",
    eprint = "2409.15411",
    archivePrefix = "arXiv",
    primaryClass = "hep-ph",
    doi = "10.1007/JHEP01(2025)097",
    journal = "JHEP",
    volume = "01",
    pages = "097",
    year = "2025"
}

@article{zhang2025neutrino,
  title={Neutrino self-interaction and weak mixing-angle measurements},
  author={Zhang, Yue},
  journal={Phys. Rev. D},
  eprint = "2411.05070",
  archivePrefix = "arXiv",
  volume={112},
  number={3},
  pages={035027},
  year={2025},
  publisher={APS},
  doi={10.1103/7y8g-56fz},
  url={https://journals.aps.org/prd/abstract/10.1103/7y8g-56fz}
}

@article{De2025bounds,
  title="{Bounds on new neutrino interactions from the first CE$\nu$NS data at direct detection experiments}",
  author={De Romeri, Valentina and Papoulias, Dimitrios K and Ternes, Christoph A},
  journal={JCAP},
  eprint = "2411.11749",
  archivePrefix = "arXiv",
  volume={2025},
  number={05},
  pages={012},
  year={2025},
  publisher={IOP Publishing},
  doi={10.1088/1475-7516/2025/05/012},
  url={https://iopscience.iop.org/article/10.1088/1475-7516/2025/05/012}
}

@article{foroughi2025enabling,
  title="{Enabling Strong Neutrino Self-Interaction with an Unparticle Mediator}",
  author={Foroughi-Abari, Saeid and Kelly, Kevin J and Rai, Mudit and Zhang, Yue},
  journal={Phys. Rev. Lett.},
  volume={134},
  number={18},
  pages={181001},
  year={2025},
  publisher={APS},
  eprint = "2501.02049",
  archivePrefix = "arXiv",
  doi={10.1103/PhysRevLett.134.181001},
  url={https://doi.org/10.1103/PhysRevLett.134.181001}
}

@article{Wang:2025qap,
  title = {Probing a New Regime of Neutrino Self-Interactions with Astrophysical Neutrinos and the Relativistic Cosmic Neutrino Background},
  author = {Wang, Isaac R. and Xu, Xun-Jie and Zhou, Bei},
  journal = {Phys. Rev. Lett.},
  volume = {135},
  issue = {18},
  pages = {181002},
  numpages = {8},
  year = {2025},
  month = {Oct},
  publisher = {American Physical Society},
  eprint = "2501.07624",
  archivePrefix = "arXiv",
  doi = {10.1103/9ddp-j1z9},
  url = {https://link.aps.org/doi/10.1103/9ddp-j1z9}
}

@article{Machado2025widen,
  title="{Widen the Resonance at Ultra-High Energies: Novel Probes of Neutrino Self-interactions in the High-Mass Regime}",
  author={Machado, Pedro AN and Wang, Isaac R and Xu, Xun-Jie and Zhou, Bei},
  journal={},
  year={2025},
  eprint = "2512.00165",
  archivePrefix = "arXiv",
}

@article{Das2025impostor,
  title = "{Impostor among Neutrinos: Dark Radiation Masquerading as Self-Interacting Neutrinos}",
  author = {Das, Anirban and Dev, P. S. Bhupal and Gao, Christina and Ghosh, Subhajit and Kim, Taegyun},
  journal = {Phys. Rev. Lett.},
  volume = {136},
  issue = {13},
  pages = {131003},
  numpages = {10},
  year = {2026},
  month = {Apr},
  publisher = {American Physical Society},
  doi = {10.1103/jprg-jll6},
  url = {https://link.aps.org/doi/10.1103/jprg-jll6},
  eprint={2506.08085},
  archivePrefix={arXiv},
  primaryClass={hep-ph}
}

@article{Foroughi-Abari:2025mhj,
  author = "Foroughi-Abari, Saeid and Kelly, Kevin J. and Zhang, Yue",
  title = "{Radiative Correction from Secret Neutrino Interactions and Implications for Neutrino-Scattering Experiments}",
  eprint = "2510.15023",
  archivePrefix = "arXiv",
  primaryClass = "hep-ph",
  reportNumber = "MI-HET-868",
  month = "10",
  year = "2025",
  journal={}
}

@article{Whitford:2025dmq,
  title="{Limits on self-interacting neutrinos from the BAO and CMB phase shift}",
  author={Whitford, Abb{\'e} M and Howlett, Cullan and Davis, Tamara M and Camarena, David and Cyr-Racine, Francis-Yan},
  journal={JCAP},
  volume={2026},
  number={03},
  pages={064},
  year={2026},
  publisher={IOP Publishing},
  doi={10.1088/1475-7516/2026/03/064},
  url={https://iopscience.iop.org/article/10.1088/1475-7516/2026/03/064},
  eprint = "2511.00800",
  archivePrefix = "arXiv",
  primaryClass = "astro-ph.CO"
}

@article{kang1992cosmological,
  title={Cosmological constraints on neutrino degeneracy},
  author={Kang, Ho-Shik and Steigman, Gary},
  journal={Nucl. Phys. B},
  volume={372},
  number={1-2},
  pages={494--520},
  year={1992},
  publisher={Elsevier},
  doi={10.1016/0550-3213(92)90329-A},
  url={https://doi.org/10.1016/0550-3213(92)90329-A}
}

@article{Esposito2000standard,
  title={The standard and degenerate primordial nucleosynthesis versus recent experimental data},
  author={Esposito, Salvatore and Mangano, Gianpiero and Miele, Gennaro and Pisanti, Ofelia},
  journal={JHEP},
  volume={2000},
  number={09},
  pages={038},
  year={2000},
  publisher={IOP Publishing},
  doi={10.1088/1126-6708/2000/09/038},
  url={https://iopscience.iop.org/article/10.1088/1126-6708/2000/09/038}
}

@article{Hansen2001constraining,
  title={Constraining neutrino physics with big bang nucleosynthesis and cosmic microwave background radiation},
  author={Hansen, SH and Mangano, G and Melchiorri, A and Miele, Gennaro and Pisanti, Ofelia},
  journal={Phys. Rev. D},
  volume={65},
  number={2},
  pages={023511},
  year={2001},
  publisher={APS},
  doi={10.1103/PhysRevD.65.023511},
  url={https://link.aps.org/doi/10.1103/PhysRevD.65.023511}
}

@article{Serpico2005lepton,
  title={Lepton asymmetry and primordial nucleosynthesis in the era of precision cosmology},
  author={Serpico, Pasquale D and Raffelt, Georg G},
  journal={Phys. Rev. D},
  volume={71},
  number={12},
  pages={127301},
  year={2005},
  publisher={APS},
  doi={10.1103/PhysRevD.71.127301},
  url={https://link.aps.org/doi/10.1103/PhysRevD.71.127301}
}

@article{Bardeen1980gauge,
  title={Gauge-invariant cosmological perturbations},
  author={Bardeen, James M},
  journal={Phys. Rev. D},
  volume={22},
  number={8},
  pages={1882},
  year={1980},
  publisher={APS},
  doi = {10.1103/PhysRevD.22.1882},
  url = {https://link.aps.org/doi/10.1103/PhysRevD.22.1882}
}

@incollection{Durrer20042,
  title={{2} {C}osmological Perturbation Theory},
  
  author={Durrer, Ruth},
  booktitle={The Physics of the Early Universe},
  pages={31--69},
  year={2004},
  publisher={Springer},
}

@article{Ma1995,
  title = {Cosmological perturbation theory in the synchronous and conformal newtonian gauges},
  author = {Ma, Chung-Pei and Bertschinger, Edmund},
  journal = {Astrophys. J.},
  volume = {455},
  pages = {7-25},
  year = {1995},
  month = {Dec},
  publisher = {American Astronomical Society},
  doi = {10.1086/176550},
  url = {https://ui.adsabs.harvard.edu/abs/1994astro.ph..1007M/abstract}
}

@book{Dodelson2024modern,
  title={Modern cosmology},
  author={Dodelson, Scott and Schmidt, Fabian},
  year={2024},
  publisher={Elsevier}
}

@article{Bardin1970nu,
  title={On the $\nu$-$\nu$ interaction},
  author={Bardin, D Yu and Bilenky, SM and Pontecorvo, B},
  journal={Phys. Lett. B},
  volume={32},
  number={2},
  pages={121--124},
  year={1970},
  publisher={Elsevier},
  doi={10.1016/0370-2693(70)90602-7},
  url={https://www.sciencedirect.com/science/article/pii/0370269370906027?via%3Dihub}
}

@article{Choi199117,
  title={17 keV neutrino in a singlet-triplet majoron model},
  author={Choi, Kiwoon and Santamaria, A},
  journal={Phys. Lett. B},
  volume={267},
  number={4},
  pages={504--508},
  year={1991},
  publisher={Elsevier},
  doi = {10.1016/0370-2693(91)90900-B},
  url = {https://doi.org/10.1016/0370-2693(91)90900-B}
}

@article{Audren2014strongest,
  title={Strongest model-independent bound on the lifetime of Dark Matter},
  author={Audren, Benjamin and Lesgourgues, Julien and Mangano, Gianpiero and Serpico, Pasquale Dario and Tram, Thomas},
  journal={JCAP},
  volume={2014},
  number={12},
  pages={028},
  year={2014},
  publisher={IOP Publishing},
  doi={10.1088/1475-7516/2014/12/028},
  url={https://iopscience.iop.org/article/10.1088/1475-7516/2014/12/028},
  eprint={1407.2418},
  archivePrefix = "arXiv",
}

@article{Acker1992neutrino,
  title={A neutrino decay model, solar antineutrinos and atmospheric neutrinos},
  author={Acker, Andy and Joshipura, Anjan and Pakvasa, Sandip},
  journal={Phys. Lett. B},
  volume={285},
  number={4},
  pages={371--375},
  year={1992},
  publisher={Elsevier},
  doi = {10.1016/0370-2693(92)91520-J},
  url = {https://doi.org/10.1016/0370-2693(92)91520-J}
}

@article{Acker1992decaying,
  title={Decaying Dirac neutrinos},
  author={Acker, A and Pakvasa, S and Pantaleone, J},
  journal={Phys. Rev. D},
  volume={45},
  number={1},
  year={1992},
  publisher={APS},
  doi = {10.1103/PhysRevD.45.R1},
  url = {https://link.aps.org/doi/10.1103/PhysRevD.45.R1}
}

@article{Pal2011dirac,
  title={Dirac, majorana, and weyl fermions},
  author={Pal, Palash B},
  journal={Am. J. Phys.},
  eprint = {1006.1718},
  archivePrefix = "arXiv",
  volume={79},
  number={5},
  pages={485--498},
  year={2011},
  publisher={AIP Publishing},
  doi={10.1119/1.3549729},
  url={https://doi.org/10.1119/1.3549729}
}

@article{Kayser1982distinguishing,
  title={Distinguishing between Dirac and Majorana neutrinos in neutral-current reactions},
  author={Kayser, Boris and Shrock, Robert E},
  journal={Phys. Lett. B},
  volume={112},
  number={2},
  pages={137--142},
  year={1982},
  publisher={Elsevier},
  doi = {10.1016/0370-2693(82)90314-8},
  url = {https://doi.org/10.1016/0370-2693(82)90314-8}
}

@article{de20212020,
  title={2020 global reassessment of the neutrino oscillation picture},
  author={de Salas, Pablo F and Forero, DV and Gariazzo, S and Mart{\'\i}nez-Mirav{\'e}, P and Mena, O and Ternes, CA and T{\'o}rtola, Mariam and Valle, JWF},
  eprint={2006.11237},
  archivePrefix = "arXiv",
  journal={JHEP},
  volume={2021},
  number={2},
  pages={1--36},
  year={2021},
  publisher={Springer},
  doi={ 10.1007/JHEP02(2021)071},
  url={https://link.springer.com/article/10.1007/JHEP02(2021)071}
}

@article{Ng2014cosmic,
  title={Cosmic neutrino cascades from secret neutrino interactions},
  author={Ng, Kenny CY and Beacom, John F},
  eprint = {1404.2288},
  archivePrefix = "arXiv",
  journal={Phys. Rev. D},
  volume={90},
  number={6},
  pages={065035},
  year={2014},
  publisher={APS},
  doi= {10.1103/PhysRevD.90.065035},
  url={https://link.aps.org/doi/10.1103/PhysRevD.90.065035}
}

@article{Gross1951plasma,
  title={Plasma oscillations in a static magnetic field},
  author={Gross, Eugene P},
  journal={Phys. Rev.},
  volume={82},
  number={2},
  pages={232},
  year={1951},
  publisher={APS},
  doi = {10.1103/PhysRev.82.232},
  url = {https://doi.org/10.1103/PhysRev.82.232}
}

@article{Bhatnagar1954model,
  title={A model for collision processes in gases. I. Small amplitude processes in charged and neutral one-component systems},
  author={Bhatnagar, Prabhu Lal and Gross, Eugene P and Krook, Max},
  journal={Phys. Rev.},
  volume={94},
  number={3},
  pages={511},
  year={1954},
  publisher={APS},
  doi = {10.1103/PhysRev.94.511},
  url = {https://journals.aps.org/pr/abstract/10.1103/PhysRev.94.511}
}

@article{Hannestad2000self,
  title={Self-interacting warm dark matter},
  author={Hannestad, Steen and Scherrer, Robert J},
  journal={Phys. Rev. D},
  volume={62},
  number={4},
  pages={043522},
  year={2000},
  publisher={APS},
  doi = {10.1103/PhysRevD.62.043522},
  url = {https://link.aps.org/doi/10.1103/PhysRevD.62.043522}
}

@article{Atrio1997interacting,
  title={Interacting hot dark matter},
  author={Atrio-Barandela, Fernando and Davidson, Sacha},
  journal={Phys. Rev. D},
  volume={55},
  number={10},
  pages={5886},
  year={1997},
  publisher={APS},
  doi={10.1103/PhysRevD.55.5886},
  url={ https://doi.org/10.1103/PhysRevD.55.5886}
}

@article{Yunis2020boltzmann,
  title={Boltzmann hierarchies for self-interacting warm dark matter scenarios},
  author={Yunis, Rafael and Arg{\"u}elles, Carlos R and Nacir, Diana L{\'o}pez},
  journal={JCAP},
  volume={2020},
  number={09},
  pages={041},
  year={2020},
  publisher={IOP Publishing},
  doi={10.1088/1475-7516/2020/09/041},
  url={https://iopscience.iop.org/article/10.1088/1475-7516/2020/09/041},
  eprint = {2002.05778},
  archivePrefix = "arXiv"
}

@article{Yunis2022self,
  title={Self-interacting dark matter in cosmology: Accurate numerical implementation and observational constraints},
  author={Yunis, Rafael and Arg{\"u}elles, Carlos R and Sc{\'o}ccola, Claudia G and Nacir, Diana L{\'o}pez and Giordano, Gaston},
  journal={JCAP},
  volume={2022},
  number={02},
  pages={024},
  year={2022},
  publisher={IOP Publishing},
  doi={10.1088/1475-7516/2022/02/024},
  url={https://iopscience.iop.org/article/10.1088/1475-7516/2022/02/024},
  eprint = {2108.02657},
  archivePrefix = "arXiv"
}

@article{Pal_2025,
  title = {Exploring neutrino interactions in light of present and upcoming galaxy surveys},
  journal = {JCAP},
  year = {2025},
  month = {mar},
  publisher = {IOP Publishing},
  volume = {2025},
  number = {03},
  pages = {047},
  author = {Pal, Sourav and Samanta, Rickmoy and Pal, Supratik},
  doi = {10.1088/1475-7516/2025/03/047},
  url = {https://dx.doi.org/10.1088/1475-7516/2025/03/047},
  eprint = {2409.03712},
  archivePrefix = "arXiv",
}

@article{Libanore2025,
  title = {Joint 21-cm and CMB forecasts for constraining self-interacting massive neutrinos},
  author = {Libanore, Sarah and Ghosh, Subhajit and Kovetz, Ely D. and Boddy, Kimberly K. and Raccanelli, Alvise},
  journal = {Phys. Rev. D},
  volume = {112},
  issue = {6},
  pages = {063502},
  numpages = {28},
  year = {2025},
  month = {Sep},
  publisher = {American Physical Society},
  doi = {10.1103/tdms-6n76},
  url = {https://link.aps.org/doi/10.1103/tdms-6n76},
  eprint = {2504.15348},
  archivePrefix = "arXiv",
}

@article{He_2020,
  doi = {10.1088/1475-7516/2020/11/003},
  url = {https://dx.doi.org/10.1088/1475-7516/2020/11/003},
  year = {2020},
  month = {nov},
  publisher = {},
  volume = {2020},
  number = {11},
  pages = {003},
  author = {Hong-Jian He and Yin-Zhe Ma and Jiaming Zheng},
  title = {Resolving Hubble tension by self-interacting neutrinos with Dirac seesaw},
  journal = {JCAP},
  eprint = {2003.12057},
  archivePrefix = "arXiv"
}

@article{Berbig2020,
  title = {The Hubble tension and a renormalizable model of gauged neutrino self-interactions},
  author = {Berbig, Maximilian and Jana, Sudip and Trautner, Andreas},
  journal = {Phys. Rev. D},
  volume = {102},
  issue = {11},
  pages = {115008},
  numpages = {10},
  year = {2020},
  month = {Dec},
  publisher = {American Physical Society},
  doi = {10.1103/PhysRevD.102.115008},
  url = {https://link.aps.org/doi/10.1103/PhysRevD.102.115008},
  eprint = {2004.13039},
  archivePrefix = "arXiv"
}

@article{Denner1992feynman,
  title={Feynman rules for fermion-number-violating interactions},
  author={Denner, Ansgar and Eck, H and Hahn, O and K{\"u}blbeck, J},
  journal={Nucl. Phys. B},
  volume={387},
  number={2},
  pages={467--481},
  year={1992},
  publisher={Elsevier},
  doi = {10.1016/0550-3213(92)90169-C},
  url = {https://doi.org/10.1016/0550-3213(92)90169-C}
}

@book{tables,
title={Table of Integrals, Series, and Products},
author={Gradshteyn, I.S. and  Ryzhik, I.M.},
year={2014},
Note={Eigth Ed.},
publisher={Elsevier Academic Press},
address={London. U.K.},
ISBN={978-0-12-294760-5}
}

@article{Yueh1976scattering,
  title={Scattering functions for neutrino transport},
  author={Yueh, William R and Buchler, J Robert},
  journal={Astrophysics and Space Science},
  volume={39},
  number={2},
  pages={429--435},
  year={1976},
  publisher={Springer},
  doi={10.1007/BF00648341},
  url={https://link.springer.com/article/10.1007/BF00648341}
}

@article{Hannestad1995neutrino,
  title={Neutrino decoupling in the early universe},
  author={Hannestad, Steen and Madsen, Jes},
  journal={Phys. Rev. D},
  volume={52},
  number={4},
  pages={1764},
  year={1995},
  publisher={APS},
  doi={10.1103/PhysRevD.52.1764},
  url={https://link.aps.org/doi/10.1103/PhysRevD.52.1764}
}

@article{Abenza2020precision,
  title={Precision early universe thermodynamics made simple: {$N_{\rm eff}$} and neutrino decoupling in the Standard Model and beyond},
  author={Abenza, Miguel Escudero},
  journal={JCAP},
  volume={2020},
  number={05},
  pages={048},
  year={2020},
  publisher={IOP Publishing},
  doi={10.1088/1475-7516/2020/05/048},
  url={https://iopscience.iop.org/article/10.1088/1475-7516/2020/05/048},
  eprint = {2001.04466},
  archivePrefix = "arXiv",
}

@article{PETERLEPAGE1978192,
  title = {A new algorithm for adaptive multidimensional integration},
  journal = {J. Comput. Phys},
  volume = {27},
  number = {2},
  pages = {192-203},
  year = {1978},
  issn = {0021-9991},
  doi = {https://doi.org/10.1016/0021-9991(78)90004-9},
  url = {https://www.sciencedirect.com/science/article/pii/0021999178900049},
  author = {G {Peter Lepage}},
}

@article{Lepage2021adaptive,
  title={Adaptive multidimensional integration: VEGAS enhanced},
  author={Lepage, G Peter},
  journal={J. Comput. Phys},
  volume={439},
  pages={110386},
  year={2021},
  publisher={Elsevier},
  doi={10.1016/j.jcp.2021.110386},
  url={https://doi.org/10.1016/j.jcp.2021.110386},
  eprint = {2009.05112},
  archivePrefix = "arXiv",
}

\end{document}